\documentclass[journal]{IEEEtran}

\usepackage[pdftex]{graphicx}
\graphicspath{{./figures/}{./photos/}{./bio/}}
\DeclareGraphicsExtensions{.pdf,.jpg,.jpeg,.png}

\usepackage[utf8]{inputenc}
\usepackage[T1]{fontenc} 
\usepackage[cmintegrals]{newtxmath}
\usepackage{bm} 
\usepackage{amsmath}

\usepackage{hyperref}

\usepackage{array}
\usepackage{tabularx}
\usepackage[table]{xcolor}

\usepackage{bookmark}
\usepackage{booktabs}
\usepackage{multirow}

\usepackage{datetime}

\usepackage{todonotes}

\usepackage{pgfplots}
\usepackage{pgfplotstable}
\pgfplotsset{compat=1.7}
\usepackage{tikz}
\usepackage{circuitikz}

\usetikzlibrary{positioning}
\usetikzlibrary{arrows,decorations.pathreplacing}

\newlength{\layersep}
\usepackage{url} 
\usepackage[english]{babel} 

\usepackage{threeparttable}

\usepackage{cite}
\usepackage[caption=false,font=footnotesize]{subfig}

\newcommand{\ra}[1]{\renewcommand{\arraystretch}{#1}}

\definecolor{blue}{RGB}{31, 119, 180}
\definecolor{orange}{RGB}{255, 127, 14}
\definecolor{shadecolor}{RGB}{196,196,196}

\newcommand{\eg}{e.\,g.,\ }

\newcommand{\matr}[1]{\mathbf{#1}} 

\begin{document}


\title{Learning to Detect Anomalous Wireless Links in IoT Networks}
\date{2020}

\author{%
	Gregor Cerar$^{1,2}$,
	Halil Yetgin$^{1,3}$, 
	Bla\v{z} Bertalani\v{c}$^{1,4}$, and
	Carolina Fortuna$^{1}$\\
	$^{1}$Department of Communication Systems, Jo{\v z}ef Stefan Institute, SI-1000 Ljubljana, Slovenia.\\
	$^{2}$Jo\v{z}ef Stefan International Postgraduate School, Jamova 39, SI-1000 Ljubljana, Slovenia.\\
	$^{3}$Department of Electrical and Electronics Engineering, Bitlis Eren University, 13000 Bitlis, Turkey.\\
	$^{4}$Faculty of Electrical Engineering, University of Ljubljana, 1000 Ljubljana, Slovenia.\\
	\{gregor.cerar $\mid$ halil.yetgin $\mid$ blaz.bertalanic $\mid$ carolina.fortuna\}@ijs.si
	\thanks{This work was funded by the Slovenian Research Agency (Grant no. P2-0016 and J2-9232).}
}






\makeatletter

\def\ps@IEEEtitlepagestyle{%
	\def\@oddfoot{\mycopyrightnotice}%
	\def\@evenfoot{}%
}
\def\mycopyrightnotice{%
	{\begin{minipage}{2\linewidth}\footnotesize\bfseries \copyright 2020 IEEE.  Personal use of this material is permitted.  Permission from IEEE must be obtained for all other uses, in any current or future media, including reprinting/republishing this material for advertising or promotional purposes, creating new collective works, for resale or redistribution to servers or lists, or reuse of any copyrighted component of this work in other works.\hfill\end{minipage}}
	\gdef\mycopyrightnotice{}
	
}

\maketitle

\begin{abstract}
	After decades of research, Internet of Things (IoT) is finally permeating real-life and helps improve the efficiency of infrastructures and processes as well as our health. As massive number of IoT devices are deployed, they naturally incurs great operational costs to ensure intended operations. To effectively handle such intended operations in massive IoT networks, automatic detection of malfunctioning, namely anomaly detection, becomes a critical but challenging task. In this paper, motivated by a real-world experimental IoT deployment, we introduce four types of wireless network anomalies that are identified at the link layer. We study the performance of threshold- and machine learning (ML)-based classifiers to automatically detect these anomalies. We examine the relative performance of three supervised and three unsupervised ML techniques on both non-encoded and encoded (autoencoder) feature representations. Our results demonstrate that; i) selected supervised approaches are able to detect anomalies with F1 scores of above 0.98, while unsupervised ones are also capable of detecting the said anomalies with F1 scores of, on average, $0.90$, and ii) OC-SVM outperforms all the other unsupervised ML approaches reaching at F1 scores of $0.99$ for SuddenD, $0.95$ for SuddenR, $0.93$ for InstaD and $0.95$ for SlowD.  
\end{abstract}


\section{Introduction}
\label{sec:intro}

The Internet of Things (IoT) has received a plethora of attention from both industry and academia due to the market release of a variety of smart devices on a regular basis, e.g. the devices retrofitted in home appliances, wearables, healthcare, vehicles and industrial machinery, just to name a few~\cite{qiu2018can}. To this end, extensive research efforts have been put forward for their active deployment and development to enable increasingly efficient and more automated operations in manufacturing, agriculture, transportation and healthcare, but also due to their massive economic contributions~\cite{davies2020internet}.

Valid business cases~\cite{diaz2017business} and successful real-world large-scale IoT deployments are emerging as a way to improve existing business processes as well as enable new applications~\cite{davies2020internet}. However, once the network of sensors is deployed, it becomes part of the operational infrastructure of a business, and needs to be maintained and serviced similar to any other infrastructure, such as legacy IT infrastructure, robots and machines just to name a few. Minimizing maintenance costs while ensuring the reliability of IoT network~\cite{wetzker2016troubleshooting} becomes prohibitive when the number of sensors are in their thousands or tens of thousands. To efficiently manage such massive IoT networks, automatic IoT network monitoring~\cite{silva2019m4dn} and malfunction detection~\cite{sheth2006mojo} solutions that automatically report relevant malfunctions and filter them out without influencing the business process are required.

IoT network or node malfunctioning can also be referred to as network or node anomaly and to date, it has been defined in various ways, often from the perspective of monitored networking aspects. For instance, Sheth~\textit{et al.}~\cite{sheth2006mojo} define and identify anomalies from the IEEE~802.11 physical layer perspective, namely, hidden terminal, capture effect, noise and signal strength variation anomalies, whereas Gupta~\textit{et al.}~\cite{gupta2007andes} define anomalies from multihop networking perspective with the aspects, such as black hole, sink hole, selective forwarding and flooding. Alipour~\textit{et al.}~\cite{7109166} define the anomalies from IEEE~802.11 link layer security perspective with the focus on aspects, such as injection test, deauthentication attack, disassociation attack, association flood and authentication flood. Generally speaking, anomaly detection research in IoT networks can be found in the form of intrusion, fraud and fault detection, system health monitoring, event detection in sensor networks and detecting ecosystem disturbances~\cite{chandola2009anomaly}, where most studies mainly concerned with a certain type of anomaly within a specific scenario. 

In this paper, motivated by a real-world experimental IoT deployment, we define four types of IoT anomalies that can be identified at the link layer, namely \textit{sudden degradation}, \textit{sudden degradation with recovery}, \textit{instantaneous degradation} and \textit{slow degradation}. Rather than focusing on the cause of an anomaly as realized in~\cite{sheth2006mojo} and~\cite{gupta2007andes}, we focus our attention on the observable symptoms of link measurements, namely the changes in the expected received signal. Based on the type of anomaly, we identify possible root causes that may be related to hardware, firmware and the channel, and develop models for automatically classifying the introduced anomalies. By accurately detecting these four types of anomalies, a wireless network operator is able to quickly and proactively detect issues within the operation of the network without waiting to be explicitly alerted by users. Proactively detecting and mitigating malfunctions can increase user satisfaction, reduce churn and ultimately show significant improvements in business KPIs. Additionally, the detected and classified anomaly type can aid technical staff with the well-informed decisions so as to diagnose and resolve the issues. For instance, \textit{sudden degradation with recovery} is observed frequently after updating the firmware of devices in the network, which is highly likely related to the bugs of the firmware that prevent devices from working as intended and trigger the watchdog to reset. Therefore, discriminating between four of those types of anomalies and automatizing this process can speed up the real-time resolution of the network-related issues, in turn diminishing the allotted personnel and their efforts, and network-wide operational costs of mobile operators. The major contributions of this paper are as follows.

\begin{enumerate}
	\item We define four types of anomalies that can appear on wireless links and are representative for narrowing down the causes and enabling more efficient mitigation. Driven by a real-world operational wireless infrastructure, for each of the defined anomalies we identify their symptoms from the application perspective and potential underlying causes.
	\item We study the performance of standard manually-engineered features and a proposed autoencoder-based automatic feature generation approach, and show the performance improvement brought by the latter.
	\item We also analyse the relative performance of three supervised and three unsupervised ML techniques. More explicitly, we consider regression-based, tree-based and kernel-based methods as part of our supervised techniques, while nearest neighbours, tree- and kernel-based methods are leveraged as their unsupervised counterpart techniques.
\end{enumerate}

Additionally, minor contributions are outlined as follows:
\begin{enumerate}
	\item Based on the gained knowledge while operating the LOG-a-TEC wireless experimentation testbed~\cite{vucnik2018continuous}, we provide an analysis on real-world operational measurements that further stresses the need for automated anomaly detection in massive IoT networks. 
	\item We produce a publicly available anomaly detection tool-set\footnote{Script for the design and development of anomaly detection models: \\\url{https://gist.github.com/gcerar/0b03e55f41147a7b7230f45d1f1209d6}} including entire procedures, e.g., anomaly injection into trace-sets, feature generation out of data representations, and model training and development.
\end{enumerate}

This paper is structured as follows. Section~\ref{sec:related} summarizes the related work and Section~\ref{sec:motivation} presents an analysis of the real-world testbed measurements motivating our contributions, while Section~\ref{sec:anomaly} introduces the four types of IoT network anomalies. Then, Section~\ref{sec:representations} elaborates on various data representations that can be used to generate features for training the proposed ML models, whereas Section~\ref{sec:approaches} discusses the threshold-based approach as well as the selected supervised and unsupervised ML techniques. Section~\ref{sec:meth} describes the relevant methodological and experimental details, while Section~\ref{sec:evaluation} provides thorough analyses of the results and discusses the limitations. Finally, Section~\ref{sec:conclusions} concludes the paper.

\section{Related work}
\label{sec:related}
We provide related work to the main contributions of this paper as follows. First, we discuss related works that define anomalies in wireless and IoT networks, then we stress on the use of autoencoders for improving various aspects of wireless networks including anomaly detection, and finally, we focus on ML models that support for improved operations of wireless networks.

\subsection{Anomaly definitions in wireless networks}

Generally speaking, \textit{an anomaly} is defined as an outlier, a distant object, an exception, a surprise, an aberration or a peculiarity, depending on the domain, research community and specific application scenario~\cite{chandola2009anomaly, gogoi2011survey, zimek2012survey, aggarwal2013outlier, gupta2013outlier, xu2019recent}. A widely used classification of anomalies, including in wireless sensor network research is provided in~\cite{chandola2009anomaly,cook2019survey}, where three classes of anomalies are defined based on their nature; point anomalies, contextual anomalies and collective anomalies. In~\cite{gupta2013outlier}, Gupta~\textit{et al.} classify relevant studies on outlier detection for time series data, one of which is the point outlier as defined in~\cite{chandola2009anomaly}, and others are subsequence outliers, global and local outliers. More recently, Lavin~\textit{et al.}~\cite{lavin2015numenta} introduce a benchmark for anomaly detection, and target mainly at cloud networks and associated services, where they provide reference datasets to be used when evaluating the performance of anomaly detection algorithms. While they do not specifically define the type of anomalies, their benchmark datasets include several anomalies.

Due to the spatio-temporal nature of wireless sensor network monitoring and data collection, Jurdak~\textit{et al.}~\cite{jurdak2011} introduce temporal, spatial and spatio-temporal anomalies as well as node, network and data anomalies, followed by even finer grained anomalies, such as node resets, node failures, etc. A number of studies then introduce more focused and application specific anomalies. For instance, Sheth~\textit{et al.}~\cite{sheth2006mojo} define and identify anomalies from the IEEE 802.11 physical layer perspective namely; hidden terminal, capture effect, noise and signal strength variation anomalies. Moreover, Gupta~\textit{et al.}~\cite{gupta2007andes} define anomalies with the aspects of multihop networking, such as black hole, sink hole, selective forwarding and flooding, whereas Alipour~\textit{et al.}~\cite{7109166} define anomalies from IEEE 802.11 link layer security aspects, such as injection test, deauthentication attack, disassociation attack, association flood and authentication flood. For further details, motivated readers are referred to~\cite{jurdak2011} for the diagnosis and detection of wireless network anomalies.

\subsection{Autoencoders for improving wireless network operations and anomaly detection}
With the advent of deep learning, one class of techniques belonging to this class of ML, referred to as autoencoders, has been proven to be particularly useful at performing automatic feature engineering also for time series data~\cite{kieu2019outlier}. Autoencoders attempts to learn a lossless compression of the data and the code resulting from that compression represents a superior feature set.

Generally in wireless, autoencoders have been successfully applied by~\cite{oshea2016unsupervised} and their subsequent works, such as~\cite{oshea2017deep} to accurately reconstruct physical layer signals and~\cite{zhang2019dualband} signal denoising for more accurate localization. For anomaly detection in wireless and IoT networks, Wang~\textit{et al.}~\cite{wang2018anomaly} proposed autoencoders for more accurate identification of faulty parts of WSNs, as well as faulty antennas in antenna arrays, whereas Shahid~\textit{et al.} and Chen~\textit{et al.}~\cite{8935007,chen2018autoencoder} proposed autoencoders for identifying anomalies in wireless and IoT networks based on transport layer traces, and recently, Yin~\textit{et al.}~\cite{yin2020anomaly} proposed recurrent autoencoders for time series anomaly detection for IoT networks. However, they used a synthetic dataset with metrics derived from several Yahoo services.  Unlike the state-of-the-art, this work proposes autoencoders as an automatic feature generation method for link layer anomaly detection and uses a real-world wireless dataset in which the introduced four types of anomalies are synthetically injected.

\subsection{ML techniques for wireless and IoT network anomaly detection}
In the literature, it is often a good practice that when a ML solution to a specific problem is considered, several counterpart ML models are evaluated against each other for performance analyses. For instance, Kieu~\textit{et al.}~\cite{kieu2019outlier} compare the performance of ten different ML techniques, such as Support Vector Machines, Local Outlier Factor, Isolation Forest, just to name a few, on six different datasets that are suitable for anomaly detection.

With respect to wireless and IoT network anomalies, Thing~\textit{et al.}~\cite{thing2017ieee} evaluate the relative performance of four deep learning and one decision tree models for anomaly detection and attack classification in IEEE 802.11 networks, whereas Chen~\textit{et al.}~\cite{chen2018autoencoder} evaluate the relative performance of principal component analysis, standard and convolutional autoencoder for detecting anomalies in transport layer traces, i.e., TCP, UDP and ICMP of wireless networks. Moreover, Ran~\textit{et al.}~\cite{ran2019semi} evaluate the relative performance of their proposed semi-supervised approach of IEEE.802.11 anomaly detection, and similarly Salem~\textit{et al.}~\cite{salem2014anomaly} evaluate the relative performance of five ML techniques, i.e., SVM, decision trees (J48), logistic regression, Na\"ive Bayes, and Decision Table for anomaly detection in WSNs. Additionally, the previous authors~\cite{salem2013sensor} also evaluate the performance of their proposed algorithm against selected three ML techniques, namely linear regression, additive regression, and J48 decision tree for anomaly detection in WSNs. However, in most of the ML-based network anomaly detection research discussed in this section as well as in~\cite{alsheikh2014machine} provide only limited relative performance evaluation results. To the best of our knowledge, this paper is the first attempt to provide relative comparisons between three supervised and three unsupervised ML techniques based on various data representations and their encoded counterpart features.

\section{Motivation}
\label{sec:motivation}

\begin{figure*}[tbp]
	\centering
	\subfloat[Sudden degradation with no recovery between Node 5 and Node 3.\label{fig:anomaly:step-tb}]{
		\includegraphics[width=.47\linewidth]{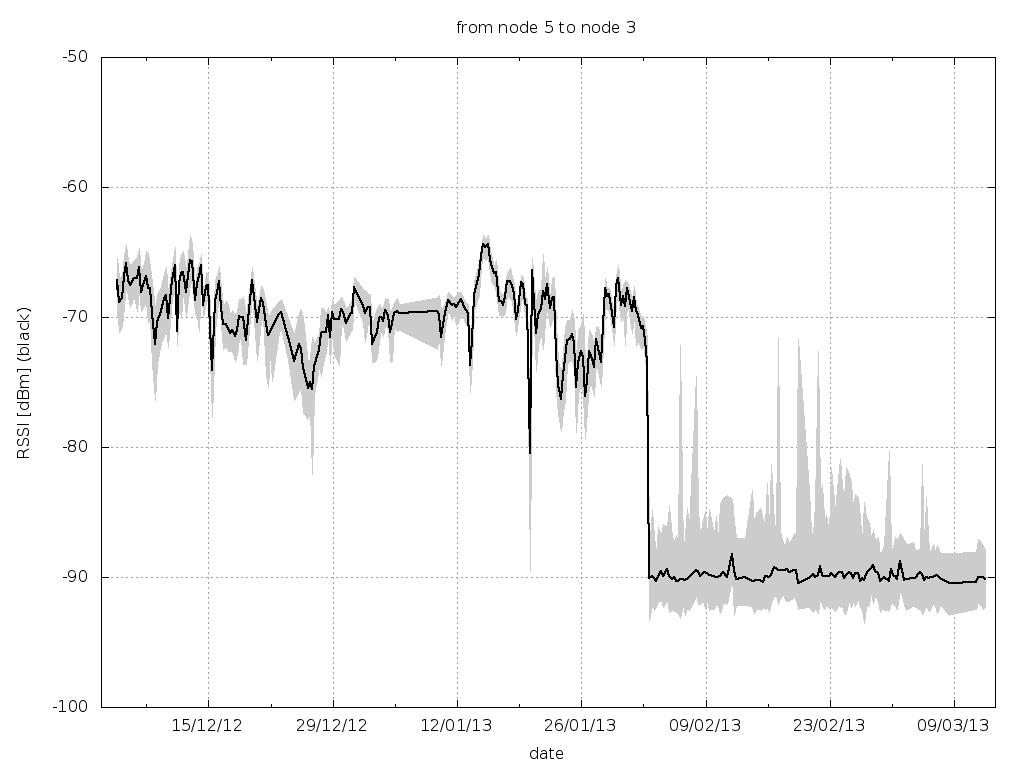}
	}
	\subfloat[Sudden degradation with recovery between Node 6 and Node 13.\label{fig:anomaly:step-recovery-tb}]{
		\includegraphics[width=.47\linewidth]{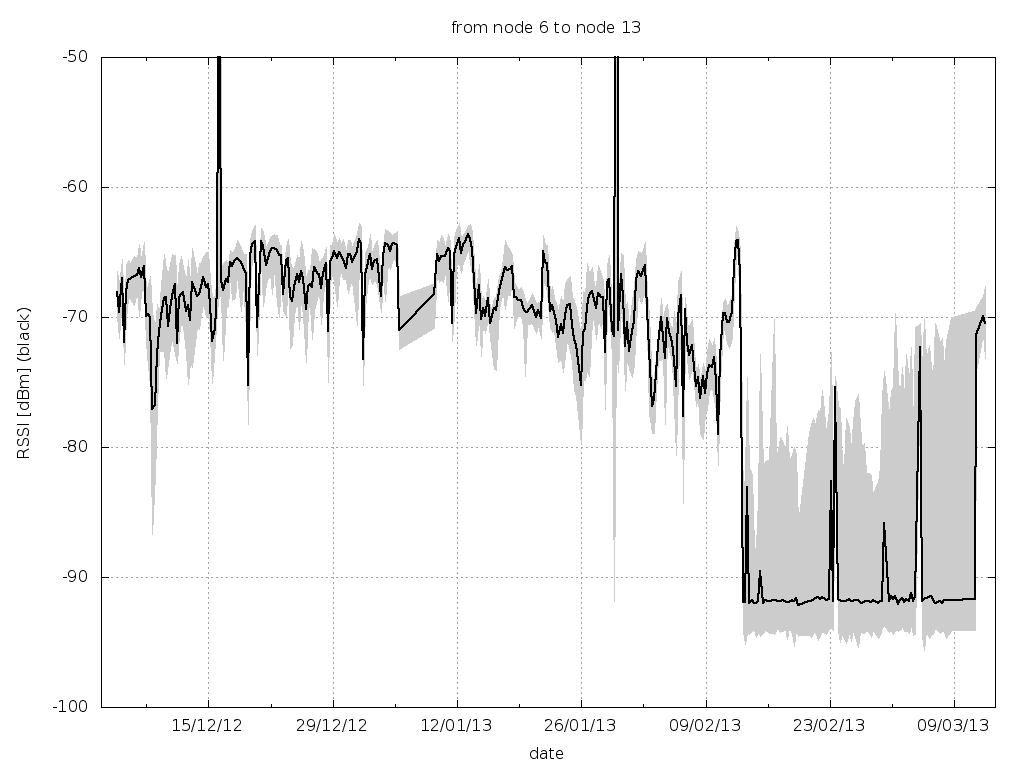}
	}

	\subfloat[Spike-like instantaneous degradation between Node 13 and Node 15.\label{fig:anomaly:spikes-tb}]{
		\includegraphics[width=.47\linewidth]{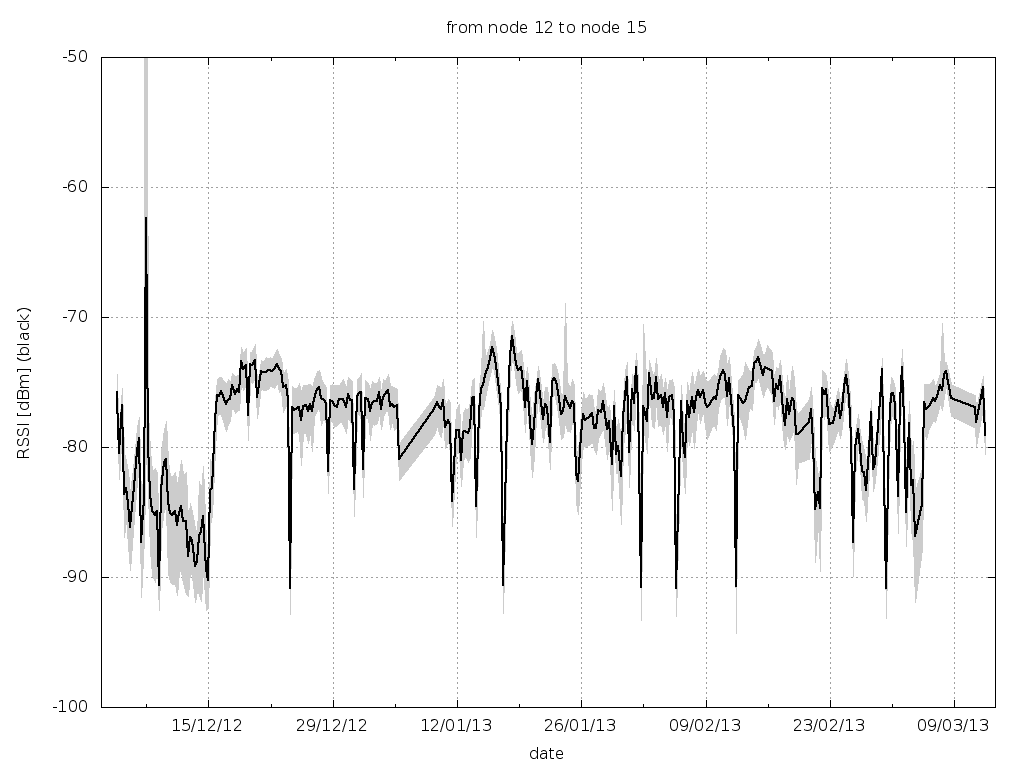}
	}
	\subfloat[Slow degradation between Node 4 and Node 26.\label{fig:anomaly:slow-tb}]{
		\includegraphics[width=.47\linewidth]{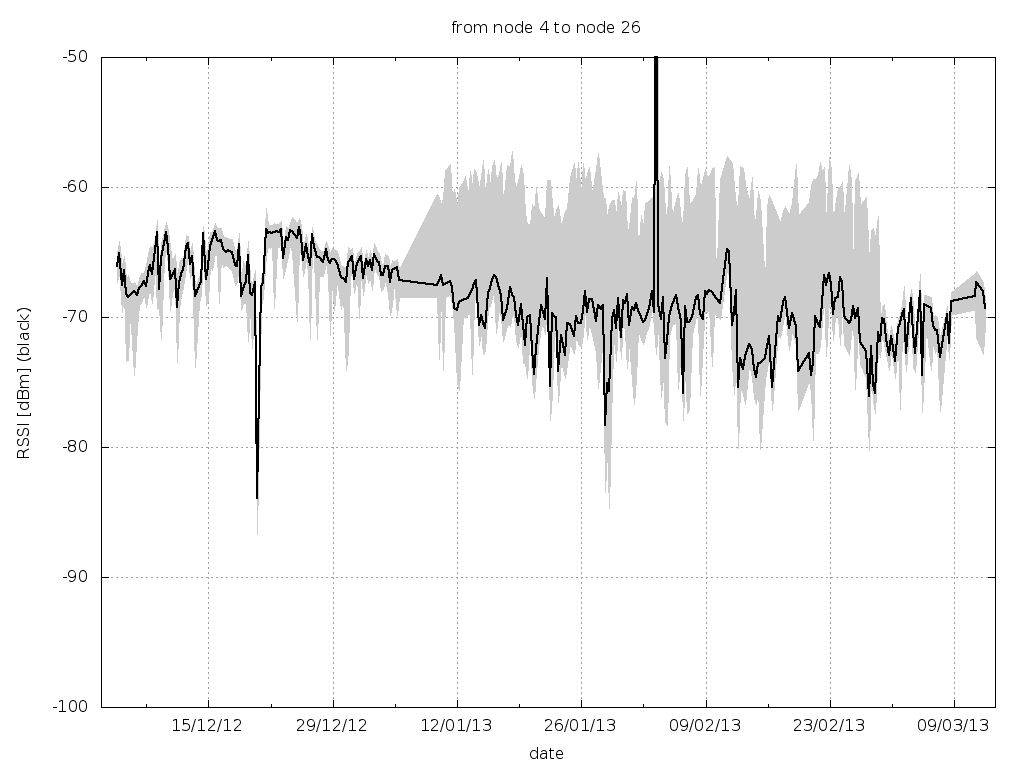}
	}
	\caption{Anomalies observed in operational environment, where solid black lines represent average RSSI and greyed areas show maximum/minimum values.}
	\label{fig:operational-anomalies}
\end{figure*}

Our lab runs the LOG-a-TEC \footnote{LOG-a-TEC testbed with sensor platforms~\url{http://log-a-tec.eu}} testbed that has empowered wireless experimentation for more than ten years. The first version of the testbed comprised of our custom embedded platform~\cite{vsolc2015low} was mounted on public light poles in a small municipality of Slovenia~\cite{vsolc2013network}. It included more than fifty nodes, most of which were situated in hard-to-reach locations. A sensor management system~\cite{vucnik2018continuous} is used to keep the record of each node for its hardware and software versions, configurations, and locations. This system also performs a number of management and diagnosis related tasks to monitor the operation of the devices.

Over time, the users of the testbed had difficulties in reaching some of the nodes or noticed unexplainable measurements collected during their testbed experimentation. For instance, the transceivers on some of the nodes were degraded significantly for their receiver sensitivity and transmit power performances, and in some cases to such a degree that they became inoperative. As depicted in Figure~\ref{fig:anomaly:step-tb}, third node (ID-3) sensed transmissions from fifth node with received signal strength indicator (RSSI) of about -70 [dBm] on average till 2nd February of 2013. Following that, either fifth node's transmit power or third node's receiver sensitivity was degraded significantly, which was reduced to about -90 [dBm] on average. After investing a good amount of time and effort in understanding and reproducing the anomaly, the fifth node was diagnosed with a hardware failure, and it could only be restored to normal operation by replacing the integrated circuit for transceiver (TI CC2500).

Similarly, another anomaly type is experienced in Figure~\ref{fig:anomaly:step-recovery-tb} with a sudden degradation and there were several recovery attempts between February 15th and March 9th 2013. In this particular case, we figured out that the sixth node was accidentally downgraded in February to an older version of the firmware that had a bug in the spectrum sensing code, which directly affected the operations of the sixth node and degraded its transmit power. Figure~\ref{fig:anomaly:spikes-tb} presents several spike-like instantaneous degradation anomalies between nodes 12 and 15. We were not able to discover anything technically wrong with these respective nodes. Therefore, we assumed that these anomalies were probably due to weather and/or large objects moving around the radios, since these two devices were mounted in an industrial zone, where moving large trucks and massive long-term standing objects were not an uncommon occurrence, which can indeed incur spikes due to the instantaneous non-line-of-sight channels experienced. Finally, Figure~\ref{fig:anomaly:slow-tb} also exhibits two distinguishable rapid drops and climbs, but most importantly, on average, shows a slightly degrading performance in sensitivity and/or transmit power between nodes 4 and 26 after December 2012. We were not able to readily justify such behaviour of the device, but ageing of electronic components may induce such behaviour, which is a well-known issue~\cite{Mann2016}. 

\section{Wireless Network Anomalies}
\label{sec:anomaly}

\begin{figure}[htbp]
\centering

\subfloat[Sudden degradation\label{fig:anomaly:step}]{
	\resizebox{.47\linewidth}{!}{%
		\begin{tikzpicture}
		\begin{axis}[ymin = 0, ymax = 1.1, axis lines = middle, ticks=none, label style={font=\Large}, x label style={anchor=north}, xlabel= time, y label style={anchor=south,rotate=90}, ylabel=value]
		\addplot [domain=0:10, no markers, black!80, ultra thick, dashed] { x < 5 ? 0.95 : 0.05 };
		\end{axis}
		\end{tikzpicture}
}}
\subfloat[Sudden degradation with recovery\label{fig:anomaly:step-recovery}]{
	\resizebox{.47\linewidth}{!}{%
		\begin{tikzpicture}
		\begin{axis}[ymin = 0, ymax = 1.1, axis lines = center, ticks=none, label style={font=\Large}, x label style={anchor=north}, xlabel= time, y label style={anchor=south,rotate=90}, ylabel=value]
		\addplot [domain=0:10, no markers, black!80, ultra thick, dashed] { x < 3 ? 0.95 : (x < 6 ? 0.05 : 0.95) };
		\end{axis}Algorithms and neural network certainly have some room to improve, since we do not perform any hyper-parametrization. We went with default parameters with some minor tweaks. We made arbitrary decision for going with four latent connection. However, this would certainly need further investigation under NN hyper-parametrization.
		\end{tikzpicture}
}}

\subfloat[Instantaneous degradation\label{fig:anomaly:spikes}]{
	\resizebox{.47\linewidth}{!}{%
		\begin{tikzpicture}
		\begin{axis}[ymin = 0, ymax = 1.1, axis lines = center, ticks=none, label style={font=\Large}, x label style={anchor=north}, xlabel= time, y label style={anchor=south,rotate=90}, ylabel=value]
		\addplot [domain=0:10, no markers, black!80, ultra thick, dashed] { x == 5 ? 0.05 : 0.95};
		\end{axis}
		\end{tikzpicture}
}}
\subfloat[Slow degradation\label{fig:anomaly:slow}]{
	\resizebox{.47\linewidth}{!}{%
		\begin{tikzpicture}
		\begin{axis}[ymin = 0, ymax = 1.1, axis lines = center, ticks=none, label style={font=\Large}, x label style={anchor=north}, xlabel= time, y label style={anchor=south,rotate=90}, ylabel=value]
		\addplot [domain=0:10, no markers, black!80, ultra thick, dashed] {0.95 - x/50};
		\end{axis}
		\end{tikzpicture}
}}

\caption{Visual representation of anomalies abbreviated as; a) SuddenD, b) SuddenR, c) InstaD, d) SlowD.}
\label{fig:anomaly-types}
\end{figure}
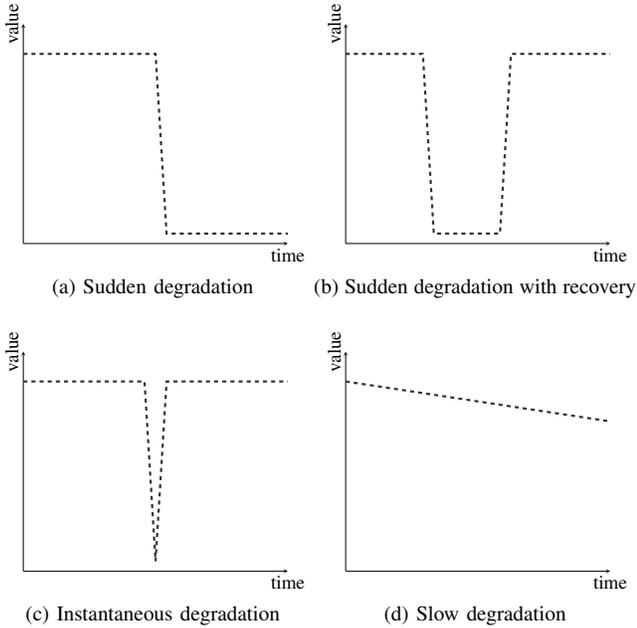

\begin{figure*}[thbp]
	\centering
	\subfloat[Time-value perpective\label{fig:synthetic:step:ts}]{
		\includegraphics[width=.35\linewidth]{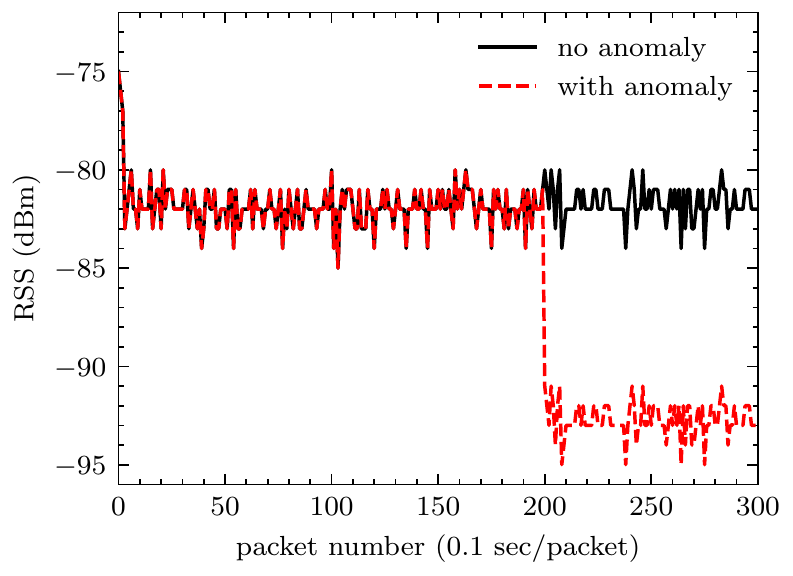}
	}%
	\subfloat[Aggregated features perspective\label{fig:synthetic:step:dist}]{
		\includegraphics[width=.35\linewidth]{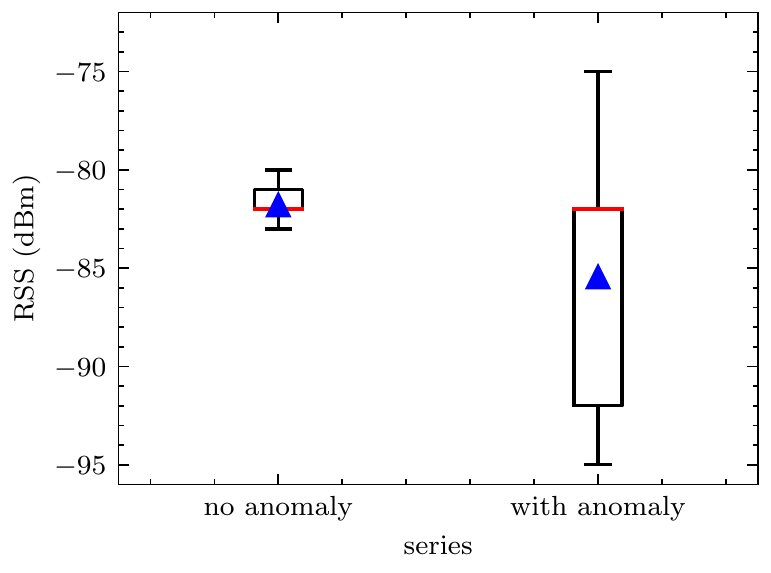}
	}

	\subfloat[Histogram perspective\label{fig:synthetic:step:hist}]{
		\includegraphics[width=.35\linewidth]{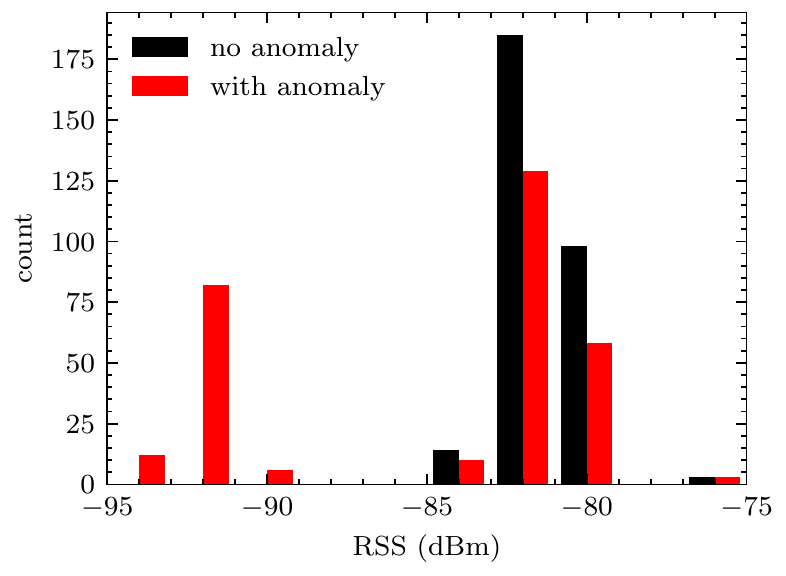}
	}%
	\subfloat[Frequency perspective (FFT)\label{fig:synthetic:step:fft}]{
		\includegraphics[width=.35\linewidth]{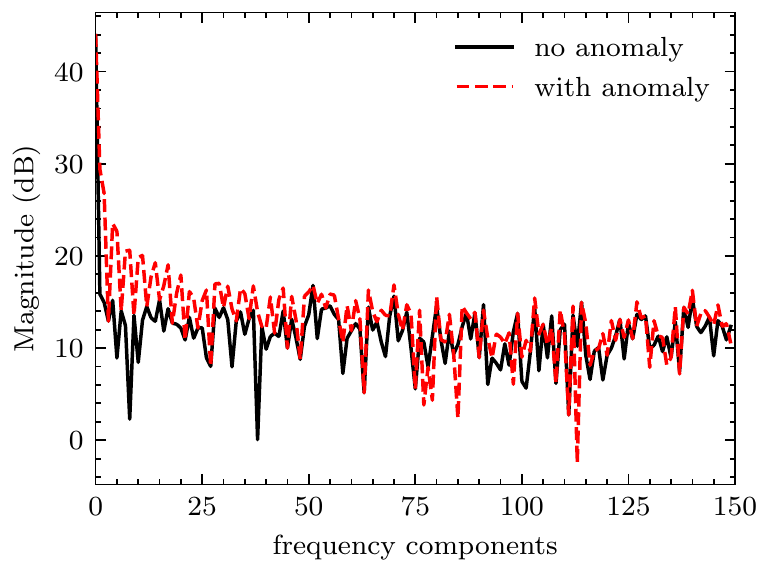}
	}%
	\caption{Distinct representations of the data for sudden degradation anomaly (SuddenD).}
	\label{fig:synthetic:step}
\end{figure*}
\begin{figure*}[thbp]
	\centering
	\subfloat[Time-value perpective\label{fig:synthetic:step-recovery:ts}]{
		\includegraphics[width=.35\linewidth]{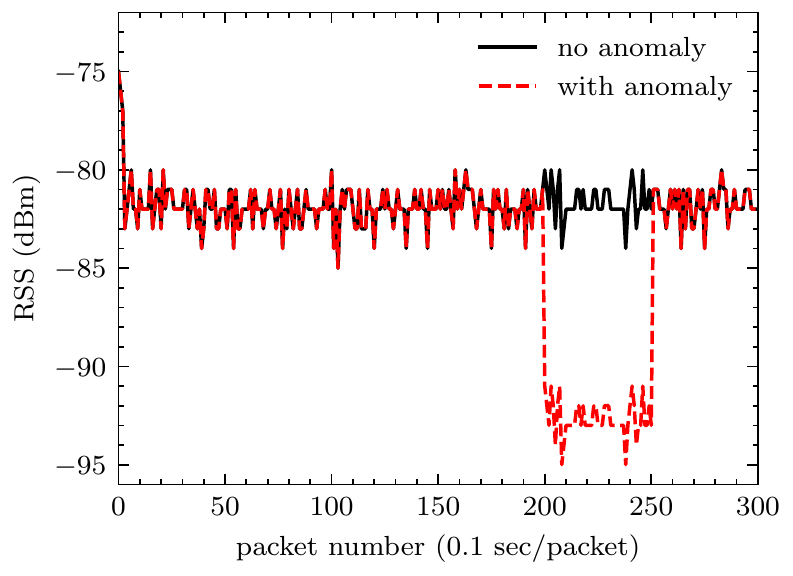}
	}%
	\subfloat[Aggregated features perspective\label{fig:synthetic:step-recovery:dist}]{
		\includegraphics[width=.35\linewidth]{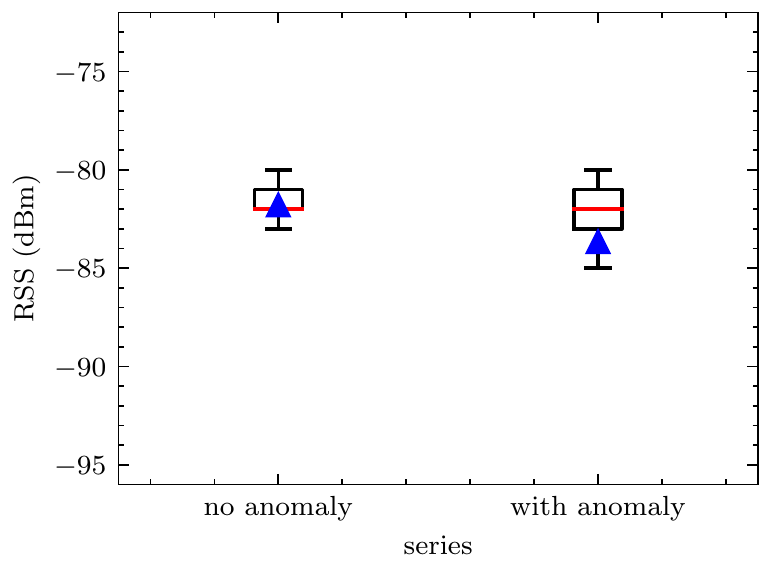}
	}

	\subfloat[Histogram perspective\label{fig:synthetic:step-recovery:hist}]{
		\includegraphics[width=.35\linewidth]{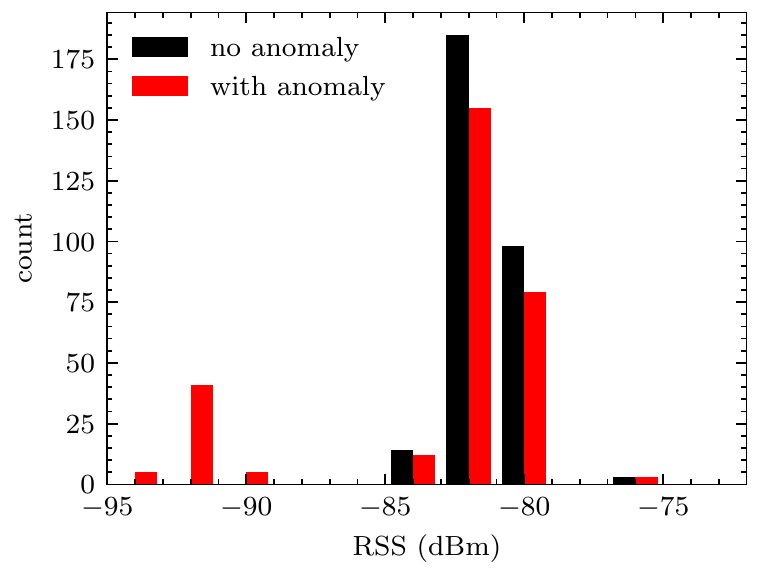}
	}%
	\subfloat[Frequency perspective (FFT)\label{fig:synthetic:step-recovery:fft}]{
		\includegraphics[width=.35\linewidth]{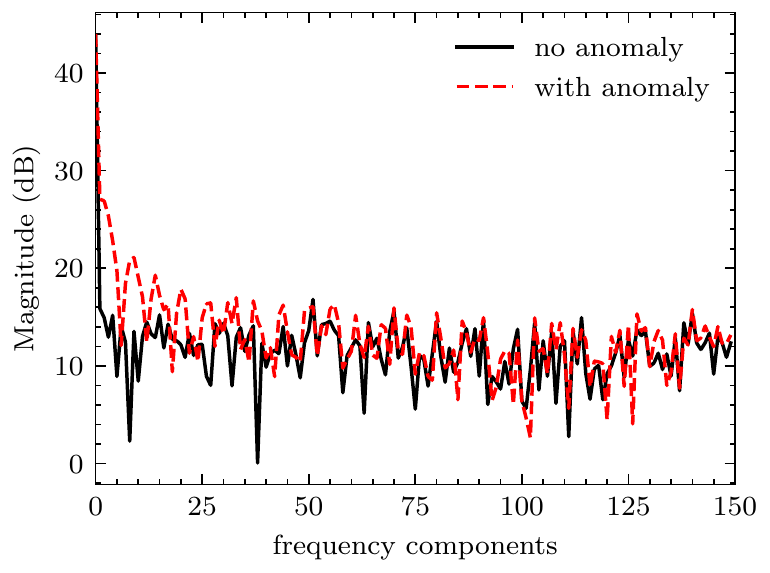}
	}%
	\caption{Distinct representations of the data for sudden degradation with recovery anomaly (SuddenR).}
	\label{fig:synthetic:step-recovery}
\end{figure*}

\begin{figure*}[thbp]
	\centering
	\subfloat[Time-value perpective\label{fig:synthetic:spikes:ts}]{
		\includegraphics[width=.35\linewidth]{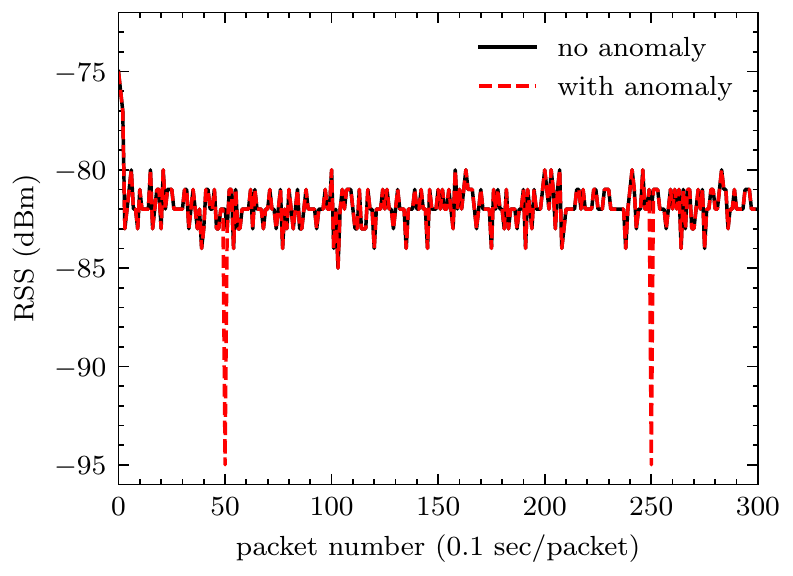}
	}%
	\subfloat[Aggregated features perspective\label{fig:synthetic:spikes:dist}]{
		\includegraphics[width=.35\linewidth]{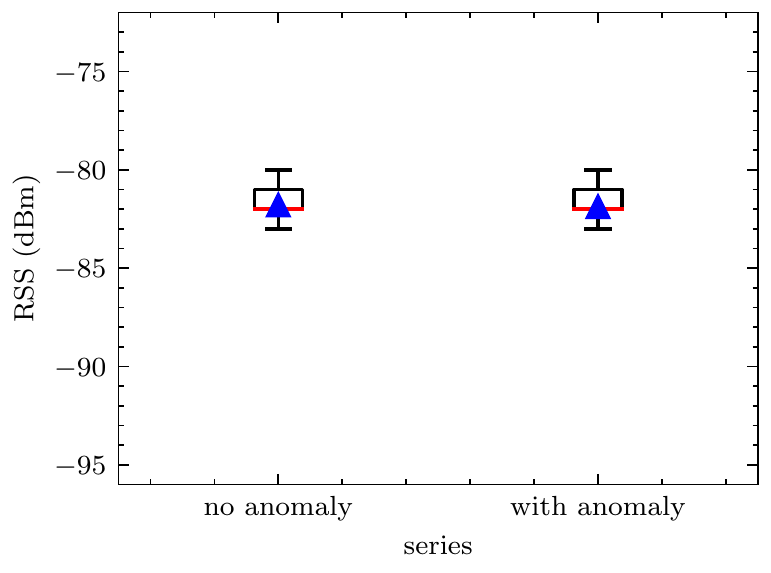}
	}

	\subfloat[Histogram perspective\label{fig:synthetic:spikes:hist}]{
		\includegraphics[width=.35\linewidth]{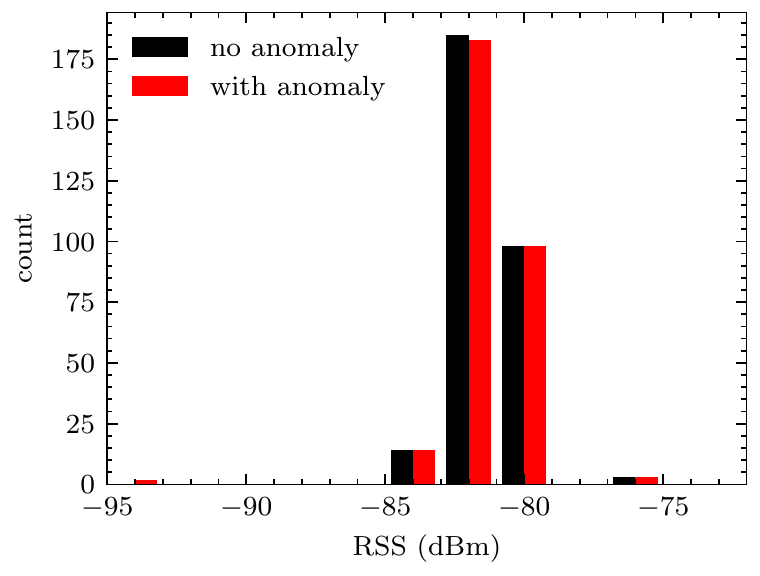}
	}%
	\subfloat[Frequency perspective (FFT)\label{fig:synthetic:spikes:fft}]{
		\includegraphics[width=.35\linewidth]{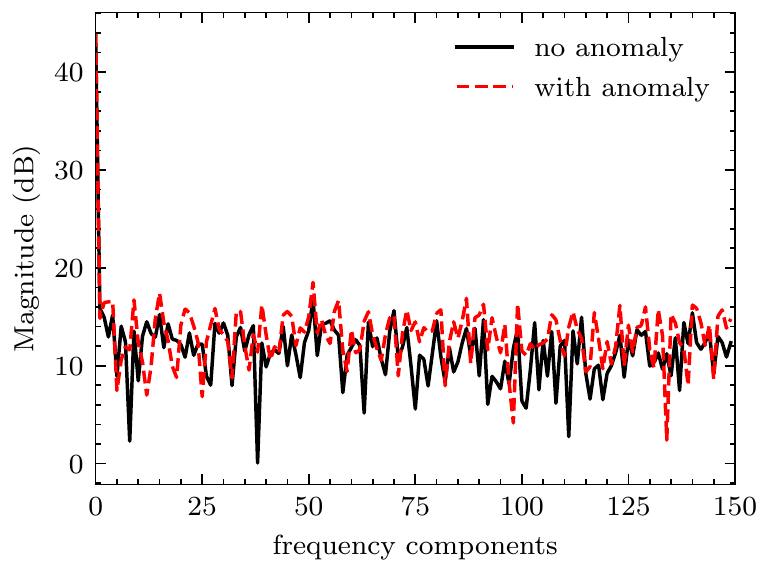}
	}%
	\caption{Distinct representations of the data for spike-like instantaneous degradation anomaly (InstaD).}
	\label{fig:synthetic:spikes}
\end{figure*}
\begin{figure*}[thbp]
	\centering
	\subfloat[Time-value perpective\label{fig:synthetic:slow:ts}]{
		\includegraphics[width=.35\linewidth]{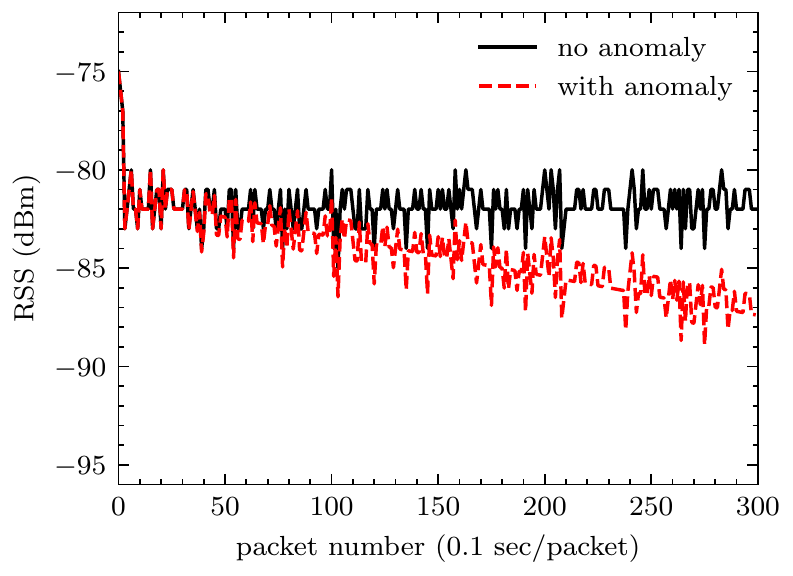}
	}%
	\subfloat[Aggregated features perspective\label{fig:synthetic:slow:dist}]{
		\includegraphics[width=.35\linewidth]{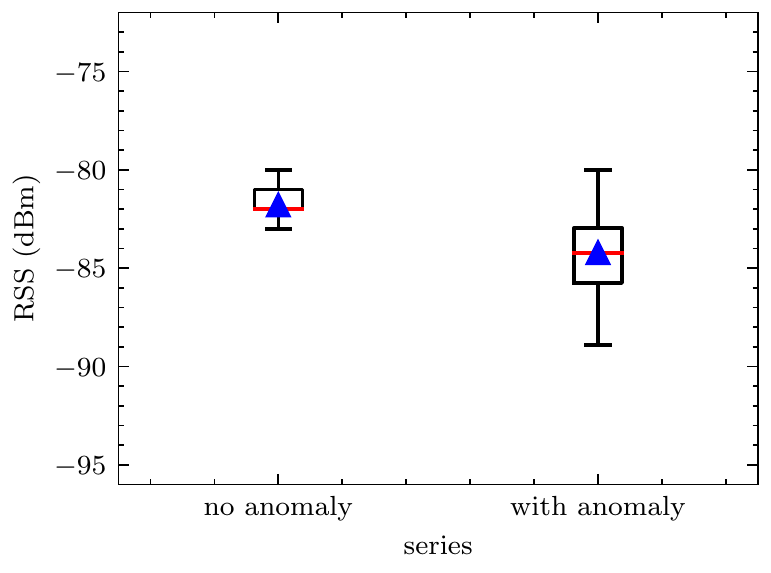}
	}

	\subfloat[Histogram perspective\label{fig:synthetic:slow:hist}]{
		\includegraphics[width=.35\linewidth]{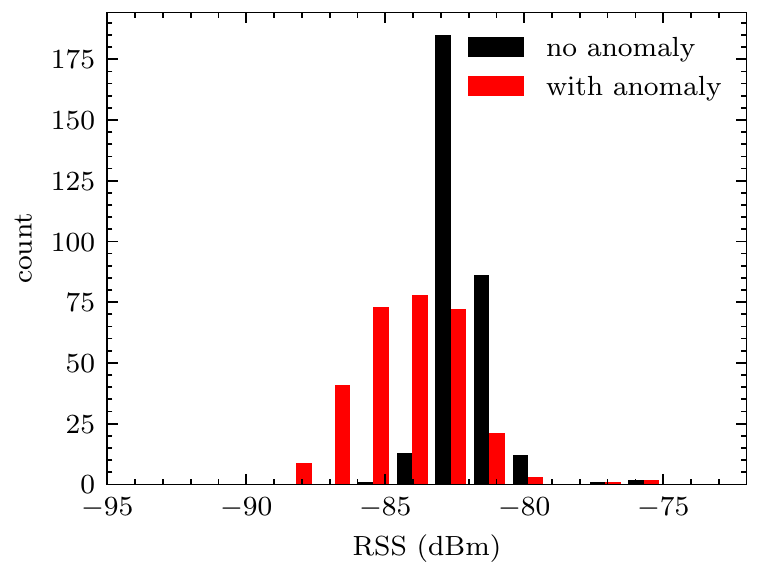}
	}%
	\subfloat[Frequency perspective (FFT)\label{fig:synthetic:slow:fft}]{
		\includegraphics[width=.35\linewidth]{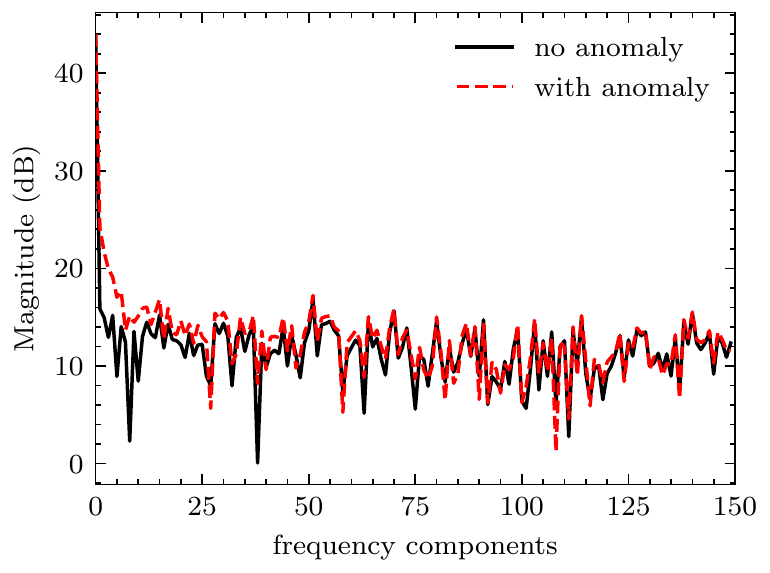}
	}%
	\caption{Distinct representations of the data for slow degradation anomaly (SlowD).}
	\label{fig:synthetic:slow}
\end{figure*}

Wireless networks are designed to exchange data between two communicating parties, e.g., video, voice and sensor measurements. As long as the network remains functional and is not interrupted, all the devices within the network are considered ordinarily operable. When the devices are compromised as exemplified in Section~\ref{sec:motivation}, then a degradation in the service quality is experienced. The way how anomalies affect the user's service quality experience is stringently associated with the type of anomaly. Therefore, in this section, we introduce four types of anomalies that can be observed in communication links of wireless networks, which were mainly discovered in our evaluation of a real-world experimentation, as discussed in Section~\ref{sec:motivation}: a) sudden degradation, b) sudden degradation with recovery, c) spike-like instantaneous degradation and d) slow degradation.

\paragraph{\textbf{Sudden degradation (SuddenD)}} The sudden degradation anomaly can be mathematically represented by a step function with decreasing slope, as depicted in Figure~\ref{fig:anomaly:step}. In our case, this represents a sudden persistent change in the state of a link. While this sudden change with an increasing slope is also possible in theory, typically it will only lead to a more reliable link, therefore they are not accounted as an anomaly. 

\textit{Symptom:} From the perspective of a user, services may become unavailable, offline and unreachable. From the perspective of a network, either the transmitter stops generating electromagnetic field or the receiver is unable to receive data. 

\textit{Possible causes:} Such sudden degradation can be induced by a transceiver failure as discussed in Section~\ref{sec:motivation} and depicted in Figure~\ref{fig:anomaly:step-tb}, a significant and sudden change in the position of one or both of the communicating parties leading them to remain disconnected, moving from line-of-sight to a non-line-of-sight environment with obstacles preserving electromagnetic shielding materials, and a significant hardware or software failure where built-in recovery mechanisms, such was watchdogs cannot be triggered.

\paragraph{\textbf{Sudden degradation with recovery (SuddenR)}} The sudden degradation with recovery anomaly can be mathematically represented by a step function with decreasing slope, as depicted in Figure~\ref{fig:anomaly:step-recovery}. In this case, the state of a link suddenly changes, stays in the new state for a longer period of time and ultimately returns to the previous state. In sudden degradation with recovery, communication is interrupted for a certain period of time. 

\textit{Symptom:} From user's perspective, provided services may become sluggish and unavailable for a certain period of time and later resume back to their regular operations. From the perspective of the network, in the case of sudden degradation with recovery, either transmitter temporarily stops generating electromagnetic field or the receiver temporarily is unable to receive it. 

\textit{Possible causes:} This type of degradation can be caused by buffer congestion and software bug, as discussed in Section~\ref{sec:motivation} and depicted in Figure~\ref{fig:anomaly:step-recovery-tb}, where watchdog performs reboot after a certain timeout, a radio  remaining in excessive active state and requiring recalibration, an obstacle blocking the communication for some time, and a signal jammer equipped on a military vehicle that is passing by.

\paragraph{\textbf{Instantaneous degradation (InstaD)}} The instantaneous degradation anomaly can be mathematically represented by a step function with steeply decreasing slope, forming a sudden spike, as depicted in Figure~\ref{fig:anomaly:spikes}. In this case, the state of the link changes suddenly, but instantaneously returns to its previous state. The instantaneous degradation anomaly may appear as an information loss.

\textit{Symptoms:} From user's perspective, a real-time service may experience instant lags, while other non-real-time services may work unaffected. From the perspective of the network, either transmitter experiences a deep fading instance or the receiver becomes unable to receive data due to an instant exposure to excessive noise or interference.

\textit{Possible causes:} This type of degradation can be caused by an instant interference, collision, quantization errors, value reading errors or sudden saturations in the transceiver's electronic components, as discussed in Section~\ref{sec:motivation} and depicted in Figure~\ref{fig:anomaly:spikes-tb}, where anomaly can be stringently induced by the issues related to the propagation environment, such as an external device communicating on the same frequency, excessive background noise and multipath fading, just to name a few.

\paragraph{\textbf{Slow degradation (SlowD)}} The slow degradation anomaly can be mathematically represented as a normalized linear function with slightly decreasing slope, as depicted in Figure~\ref{fig:anomaly:slow}. In this case, the state of the link undertakes slight and unnoticeable changes for a longer period of time and it may never resume to its original state. The slow degradation anomaly may commence triggering information loss and interruptions after a certain amount of time.

\textit{Symptom:} Slow degradation anomaly could go unnoticed for a very long time, where users may not even notice any difference in service quality immediately. When relevant thresholds are triggered, users commence experiencing deteriorated service quality. After employed compensation methods are exhausted (\eg buffers, queues, bandwidth preservation strategies), communication may be interrupted and intended services may become unavailable. From the perspective of the network, either transmitter gradually stops generating sufficient electromagnetic field to satisfy a received signal-to-noise ratio threshold or the receiver is not able detect or collect enough electromagnetic radiation to decode the information, which can also be induced by the aging of electronic components.

\textit{Possible causes:} This type of degradation may be caused by easier aging of electronic components in extreme working conditions (\eg high moisture and heat) as it is discussed in Section~\ref{sec:motivation} and depicted in Figure~\ref{fig:anomaly:slow-tb}, where it reflects a gradual but permanent impairment to the hardware or, slowly increasing obstacle such as a building being slowly built or vegetation growing. 

\section{Data representation}
\label{sec:representations}
Sections~\ref{sec:motivation} and~\ref{sec:anomaly} provided real-world anomaly examples and formalized wireless link anomalies, respectively. In the following, we provide five distinct ways to represent data that can be used as features while training the machine learning model.

\paragraph{Time-value representation} The anomalies appearing in time series of RSSI values and in Figures~\ref{fig:operational-anomalies} and~\ref{fig:anomaly-types} are recorded as raw time-ordered values, thus forming a time series. We refer to this time-ordered values as \textit{time-value} representation. In Figures~\ref{fig:synthetic:step:ts},~\ref{fig:synthetic:step-recovery:ts},~\ref{fig:synthetic:spikes:ts} and~\ref{fig:synthetic:slow:ts}, the time-value representation of an ordinary link is depicted with solid black lines and its anomaly injected counterpart, as per the definition from Section~\ref{sec:anomaly} is depicted with dashed red lines.

However, through mathematical transformations, time series can be represented in other domains that, in some cases may be more suitable for the analysis of anomaly or pattern recognition. Motivated readers are referred to~\cite{lin2012pattern} for a comprehensive taxonomy of time series representation. In addition to the time-value representation, in this study, we also consider an aggregated representation, a histogram representation, a frequency domain representation and an automatically encoded representation.

\begin{figure}[tbh]
	\centering
	\resizebox{.95\linewidth}{!}{%
	\begin{tikzpicture}[shorten >=1pt,->,draw=black!60, node distance=\layersep]
	\tikzstyle{every pin edge}=[stealth-,shorten <=1pt]
	\tikzstyle{neuron}=[circle,fill=black!25,minimum size=17pt,inner sep=0pt]
	\tikzstyle{input neuron}=[neuron, fill=green!50];
	\tikzstyle{output neuron}=[neuron, fill=red!50];
	\tikzstyle{hidden neuron}=[neuron, fill=blue!50];
	\tikzstyle{compression neuron}=[neuron, fill=violet!50];
	\tikzstyle{annot} = [text width=4em, text centered]
	\tikzstyle{conn} = [-stealth];
	
	
	\foreach \name / \y in {1,...,4}
		\node[input neuron, pin=left: $x_{\name}$] (I-\name) at (0,-\y) {};
	
	\node[] (I-dots) at (0,-5 + 0.2) {$\vdots$};
	\node[input neuron, pin=left: $x_n$] (I-n) at (0,-6 + 0.2) {};
	
	\foreach \name / \y in {1,...,5}
		\path[yshift=-0.35cm] node[hidden neuron] (E-\name) at (\layersep,-\y cm) {};

	\foreach \name / \y in {1,...,2}
		\path[yshift=-1.85cm] node[compression neuron] (C-\name) at (2*\layersep,-\y cm) {};
	
	\foreach \name / \y in {1,...,5}
		\path[yshift=-0.35cm] node[hidden neuron] (D-\name) at (3*\layersep,-\y cm) {};
	
	\foreach \name / \y in {1,...,4}
		\node[output neuron, pin={[pin edge={conn}]right:$\hat{x}_\name$}] (O-\name) at (4*\layersep,-\y) {};
	
	\node[] (O-dots) at (4*\layersep,-5 + 0.2) {$\vdots$};
	\node[output neuron, , pin={[pin edge={conn}]right:$\hat{x}_n$}] (O-n) at (4*\layersep,-6 + 0.2) {};

	\foreach \source in {1,...,4,n}
		\foreach \dest in {1,...,5}
			\path (I-\source) edge[conn] (E-\dest);
	
	\foreach \source in {1,...,5}
		\foreach \dest in {1,...,2}
			\path (E-\source) edge[conn] (C-\dest);
	
	\foreach \source in {1,...,2}
		\foreach \dest in {1,...,5}
			\path (C-\source) edge[conn] (D-\dest);
	
	\foreach \source in {1,...,5}
		\foreach \dest in {1,...,4,n}
			\path (D-\source) edge[conn] (O-\dest);
	
	\node[annot,above of=I-1, node distance=1cm] (il) {Input\\$\matr{x}$};
	\node[annot,above of=O-1, node distance=1cm] (ol) {Output\\$\matr{\hat{x}}$};
	\node[annot,above of=E-1, node distance=1cm] (el) {Encoder\\$\matr{h} = f(\matr{x})$};
	\node[annot,above of=C-1, node distance=1.6cm] (cl) {Code\\$\matr{h}$};
	\node[annot,above of=D-1, node distance=1cm] (dl) {Decoder\\$\matr{\hat{x}} = g(\matr{h})$};
	\end{tikzpicture}}
	\caption{Illustration of autoencoder configuration during training process.}
	\label{fig:autoencoder}
\end{figure}
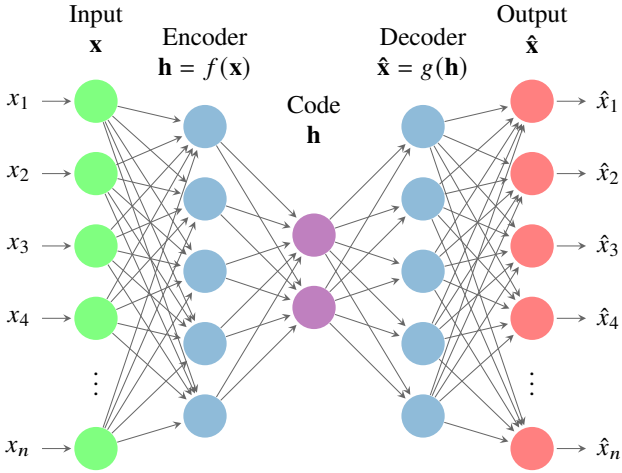

\begin{figure*}[bthp]
	\centering
	\subfloat[Sudden degradation\label{fig:synthetic:step:autoencoder}]{
		\includegraphics[width=.35\linewidth]{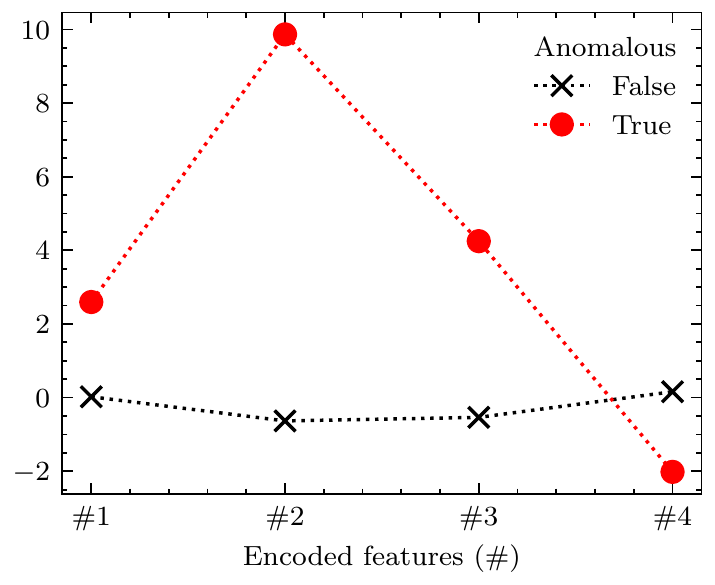}
	}%
	\subfloat[Sudden degradation with recovery\label{fig:synthetic:step-recovery:autoencoder}]{
		\includegraphics[width=.35\linewidth]{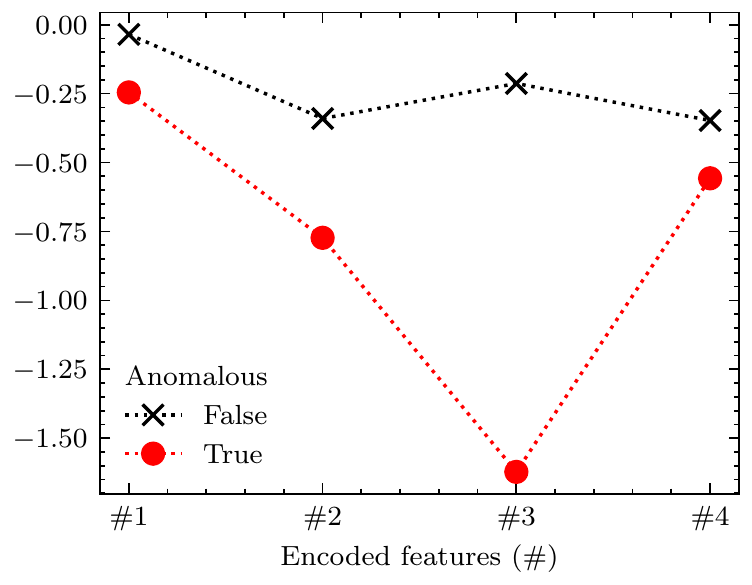}
	}

	\subfloat[Instantaneous degradation\label{fig:synthetic:spikes:autoencoder}]{
		\includegraphics[width=.35\linewidth]{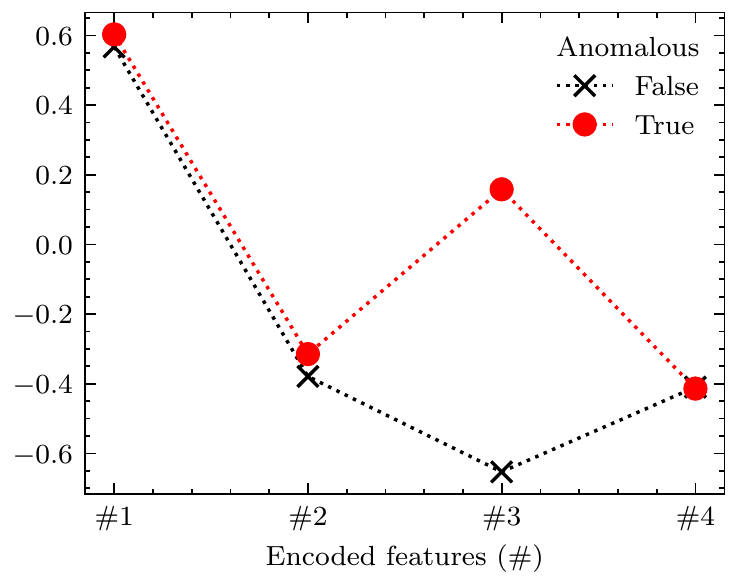}
	}%
	\subfloat[Slow degradation\label{fig:synthetic:slow:autoencoder}]{
		\includegraphics[width=.35\linewidth]{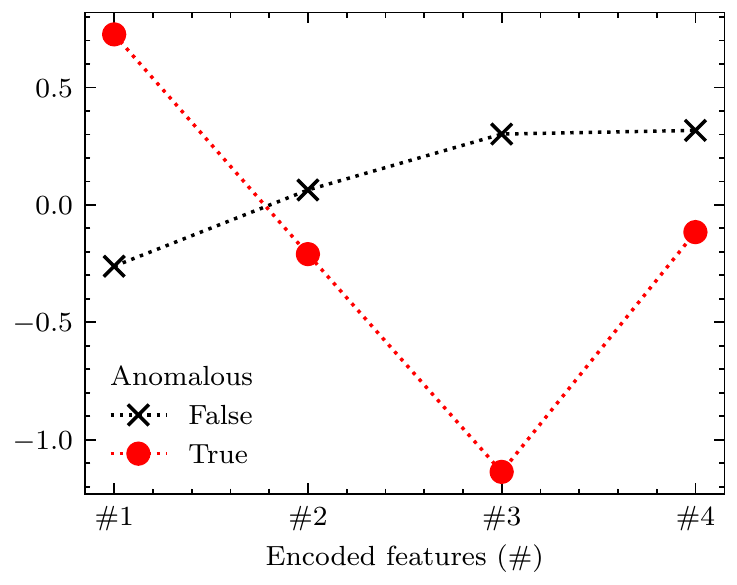}
	}%
	\caption{Automatically generated features (code) exemplified for time-value representations.}
	\label{fig:auto-features}
\end{figure*}

\paragraph{Aggregated representation} This representation contains seven statistical aggregates computed from the time-value representation, namely average, standard deviation, and all five quantile (Q) values, such as zeroth quantile (minimum), first quantile, second quantile (median), third quantile, and fourth quantile (maximum). This representation is depicted in Figures~\ref{fig:synthetic:step:dist},~\ref{fig:synthetic:step-recovery:dist},~\ref{fig:synthetic:spikes:dist} and~\ref{fig:synthetic:slow:dist} for each anomaly type, where they present values belonging to middle quantiles (Q1-Q3) as a box shape, first quantile (Q0-Q1) and third quantile (Q2-Q3) are marked as separate whiskers on top and the bottom, median value (Q2) is shown as a red bar within the box shape ({\color{red}--}), and finally, average is portrayed as a blue triangle shape ({\color{blue}$\blacktriangle$}).

\paragraph{Histogram representation} The histogram representation observed in Figures~\ref{fig:synthetic:step:hist},~\ref{fig:synthetic:step-recovery:hist},~\ref{fig:synthetic:spikes:hist} and~\ref{fig:synthetic:slow:hist} is performed via splitting the range between (global) minimum and maximum values into ten equally-sized bins. More explicitly, this representation exhibits the percentage of values allotted in each bin. 

\paragraph{FFT representation} The frequency domain representation provided in Figures~\ref{fig:synthetic:step:fft},~\ref{fig:synthetic:step-recovery:fft},~\ref{fig:synthetic:spikes:fft} and~\ref{fig:synthetic:slow:fft} utilizes absolute value of complex transformation, which is presented using log-scale for better contrasting "with anomaly" scenario against the "no anomaly" one. 

\paragraph{Encoded representation}
A recent revolution of deep learning techniques, namely autoencoders, exhibits great performance returns in a diverse set of problems. To contrast against the above-mentioned traditional representations, we propose automatically generated encoded (autoencoder) representations for all anomaly types introduced in Section~\ref{sec:anomaly}. 

Autoencoders~\cite{kramer1991autoencoder,goodfellow2016deep,cook2019survey} are neural networks which are trained to generate a representation from the reduced encoding that is very similar compared its original input. The middle layer of an autoencoder is depicted with the purple circles in Figure~\ref{fig:autoencoder} containing the reduced version of the input data and is referred to as a code $\textbf{h}$ whose size is expected to be smaller than the size of the input data. As portrayed in Figure~\ref{fig:autoencoder}, an autoencoder is composed of two parts; i) an encoder function $\textbf{h} = f(\textbf{x})$, and ii) a decoder function producing a reconstruction $\hat{\textbf{x}} = g(\textbf{h})$. The autoencoders thus learn to include only the most useful signals from the input data, while mitigating the unnecessary signal noise. 

An undercomplete autoencoder, where code size is smaller than input size, with nonlinear activation functions presents a generalized form of principal component analysis (PCA). Through the training process, the error between input $\matr{x}$ and output $\matr{\hat{x}}$ becomes negligible. Consequently, neural network learns a new representation of the input data, within a reduced feature-space. For example, in Figure~\ref{fig:synthetic:step:autoencoder} we transform time-value representation containing 300 dimensions into a newly encoded representation having only 4 dimensions. Figures~\ref{fig:synthetic:step:autoencoder},~\ref{fig:synthetic:step-recovery:autoencoder},~\ref{fig:synthetic:spikes:autoencoder}, and~\ref{fig:synthetic:slow:autoencoder} present scenarios for a link with both; i) ordinary (non-anomalous) data , ii) anomaly injected (anomalous) data for SuddenD, SuddenR, InstaD and SlowD anomalies, respectively. Non-anomalous link is depicted with a solid black line, whereas anomalous link is marked with a dashed red line.

\section{Approaches for the detection of anomalies}
\label{sec:approaches}

Considering the link anomalies defined in Section~\ref{sec:anomaly} and their corresponding representations depicted in Figures~\ref{fig:synthetic:step},~\ref{fig:synthetic:step-recovery},~\ref{fig:synthetic:spikes} and~\ref{fig:synthetic:slow}, it is clear that setting predefined thresholds for the investigated data would enable the detection of abnormal measurements and aid in treating them as an outlier. However, it has been proven that since fixed threshold-based approaches do not adapt to fluctuating behaviour of the data, selecting a threshold becomes consequential and thus may lead to poor performance, especially in real-time prediction applications~\cite{Alawe2018}. On the contrary, adaptive and proactive approaches, such as deep learning neural network (DNN) and recurrent neural network (RNN)~\cite{Alawe2018}, can learn from regular patterns of the data and accurately identify abnormal behaviours to enable more accurate anomaly detection.

\subsection{Threshold based detection}
\label{sub:threshold-based-detection}
Considering Figure~\ref{fig:anomaly:step}, detecting SuddenD requires the diagnosis of steep falling slopes that do not recover for a relatively long, possibly predefined,  period of time. Detecting SuddenR amounts to the identification of a sudden drop and later a boost in signal that resumes back to the original strength level within a predefined time window. SuddenR and InstaD are somewhat similar from application perspective. However, the distinction lies in the length of the time window at which the signal recovers back to its original levels within an instant of the time for InstaD. Detecting SlowD requires the diagnosis of a slowly but rather consistently falling slope for a relatively long, possibly predefined time window.

The time-value rules are a straightforward way to approach link-level anomaly detection. These rules may either be set based on an experienced arbitrary threshold or they can be identified using a theoretical or numerical method. However, as discussed in Section~\ref{sec:representations}, there are various possible ways to detect anomalies. For instance, it can be seen on Figures~\ref{fig:synthetic:step:dist},~\ref{fig:synthetic:step-recovery:dist} and~\ref{fig:synthetic:slow:dist} that RSS distribution of an average healthy link is significantly different than the RSS distribution of the same link when anomaly is injected, which is readily distinguishable for SuddenD, SuddenR and SlowD anomalies at a glance. More explicitly, the spread of RSS for the anomaly injected link is wider, and its mean and median values are overwritten accordingly. Similar conclusions can be made for the respective histograms in Figures~\ref{fig:synthetic:step:hist},~\ref{fig:synthetic:step-recovery:hist} and~\ref{fig:synthetic:slow:hist}. However, abnormal distributions in SlowD anomaly can only be detected with long-term observations. Moreover, sudden changes in time series can also be detected in frequency domain, which in our case, are readily observed for SuddenD and SuddenR anomalies as larger magnitudes at lower frequencies in Figures~\ref{fig:synthetic:step:dist} and~\ref{fig:synthetic:step-recovery:dist}, respectively. Changes due to injected anomalies are almost indistinguishable in the case of InstaD and SlowD while leveraging frequency domain.

Details of the threshold strategy are provided in Section~\ref{sec:evaluation}. For time-value perspective, we consider D'Agostino-Pearson's normality statistical test \cite{dagostino1971omnibus, dagostino1973tests}. The test assesses whether certain set of points come from normal distribution or not. If the $p$ value is below threshold, it is likely that the measurements do not come from normal distribution. Notice that Pearson's normality test is not sufficient condition for normality claims. Although, the approach may work fine for our limited line-of-sight scenario, it will not work for mobile or non line of sight scenario. For aggregated perspective, we consider for a link to have an anomaly two separate criteria. One criterion is based on the difference between mean and median values, which (if we assume normal distribution) are fairly close. The second criterion is how much can values deviate in standard deviation. Either of them has to be true for a link to be marked to have an anomaly. For histogram perspective, we define and arbitrary threshold. Anything below that is marked as an anomaly.

\subsection{Machine learning-based detection}
\label{sub:ml-based-detection}
A ML model is expected to distinguish between anomalous and ordinary behaviours of a link, thus requires to solve a binary classification problem. There are two ways to train a ML model to identify such distinctions. The first one is based on a supervised training approach where all anomaly data are labelled, although in many practical applications, producing a reliable training dataset is expensive and it can inevitably cover only the type of anomalies that are present in the training dataset, which then cannot cope with the abnormal link behaviours in a comprehensive manner. For this reason, training a ML model in an unsupervised way is more practical, where learning from patterns of the overall link operations so as to distinguish the abnormal behaviours of a link from the anticipated behaviours is provoked, which is referred to as the automated detection of an outlier~\cite{hodge2004survey} or an anomaly~\cite{cook2019survey} using ML models. 

In addition to baseline threshold-based approach discussed in Section~\ref{sub:threshold-based-detection}, we also consider three supervised and three unsupervised ML techniques as elaborated in the following sections.

\subsubsection{Supervised approaches}
To evaluate the performance of selected supervised ML techniques against each other and against the threshold-based approach, we opt for a set of candidate supervised approaches leveraging one representative technique from three different classes: i) Logistic Regression from Regression Analysis~\cite{malouf2002comparison}, ii) Random Forest from tree ensemble class~\cite{aggarwal2015outlier} and iii) Support Vector Machines~(SVM) from kernel-method class~\cite{aggarwal2015outlier}.

\textit{Logistic Regression}~\cite{malouf2002comparison} is a modified linear regression able to work on classification problems. In linear regression the goal is to fit a line to data samples and minimize loss. Similarly, logistic regression aims for fitting sigmoid function with the goal to minimize loss at predicting any two classes. Logistic regression also includes a generalized form suitable for high-dimensional input data and multi-class rather than binary classification.

\textit{Random Forests}~\cite{breiman2001random} is an ensemble method that uses a number of decision tree classifiers followed by a voting mechanisms to perform multi-class classification. The trees are learnt by randomly splitting a relatively large feature space into smaller subspaces. Each tree provides a class in which a specific data point falls into, the class corresponds to the "vote" of that tree. The final outcome of the classifier then uses a mechanism, such as majority voting to provide the final result. 

\textit{Support Vector Machine}~\cite{vapnik2013nature} is a learning algorithm that belongs to the family of kernel methods. Roughly speaking, SVMs attempt to learn a hyperplane that best splits a set of data into two classes. The shape of the hyperplane depends on the type of kernel function selected for the algorithm. When the kernel function is linear, so is the learnt hyperplane. When non-linear kernels are chosen, for instance RBF kernel~\cite{xiao2006optimization}, then the hyperplane is non-linear therefore better suited to approximate or discriminate non-linear random variables.

\subsubsection{Unsupervised approaches}
The cost of producing labels for supervised learning is discussed in Section~\ref{sub:ml-based-detection}. As a countermeasure, we also consider a set of candidate unsupervised approaches for developing anomaly detection models~\cite{aggarwal2015outlier}, where we leverage one representative technique from three different classes: i) Local Outlier Factor from Nearest Neighbour~(NN) class~\cite{aggarwal2015outlier}, ii) Isolation Forest from tree ensemble class~\cite{aggarwal2015outlier} and iii) one-class Support Vector Machines~(SVM) from kernel-method class~\cite{aggarwal2015outlier}.

\textit{Local Outlier Factor}~\cite{breunig2000lof} belongs to the k-Nearest Neighbour (kNN) family of algorithms, which rely on the computation of the distance between data points of the feature space. The feature vectors with smaller distance are alike and thus clustered together. One drawback for this family of algorithms is that as the dimensionality of the training data grows, the computational complexity evolves exponentially. However, there have been attempts in circumventing this exponential complexity, \eg~Ball Tree.

\textit{Isolation Forest}~\cite{liu2008isolation} belongs to tree-based ensemble methods, and works in a roughly similar way as Random Forests as described above. Essentially, it represents a Random Forest adapted so that it optimizes outlier detection rather than multi-class classification of majority of data it sees. Based on certain metrics and distinct criteria, the algorithm decides whether particular subspaces contain any abnormal samples, namely anomalies. 

\textit{Support Vector Machine}, as described at the end of supervised approaches, can also be used in an unsupervised mode for anomaly detection. In fact, most ML techniques can be used in both supervised and unsupervised mode. With this one-class approach, the model is expected to distinguish data as negative or positive instances. Then, the model can learn the boundaries of the data so as to detect the points that lie outside the boundary exposed as anomalies or outliers.

\section{Methodology and experimental details}
\label{sec:meth}

Before we proceed with the analysis of the relative performance of the wireless link anomaly detection approaches proposed in this paper, we  provide relevant methodological and experimental details.  

\subsection{Training dataset generation}\label{sub:injection}
For our experimental evaluation, we consider a real-world measurement dataset, i.e., Rutgers~\cite{rutgers-noise-20070420}, which contains measurements from 29 nodes at 5 different noise levels and each record has 300 measurements. Although every link is measured at five different noise levels, wee consider each recording as a different link and we assume that there is no correlation. On this existing real-world dataset we synthetically inject the four types of anomalies proposed in this paper as follows. First, we only pick the links without packet loss. This reduces our dataset from 4\,060 to 2\,123 ($\approx{52}\%$) of independent links. Second, by means of applying one anomaly type at a time, we randomly pick 33\% of these links, at which the anomaly is injected according to guidelines in Table~\ref{tab:injection-scenario}, while the remaining is left intact.

\begin{table}[htbp]
	\centering
	\ra{1.2}
	\begin{threeparttable}[b]
		\caption{Artificial anomaly injections for each anomaly scenario.}
		\label{tab:injection-scenario}
		\begin{tabularx}{\linewidth}{@{}lllll@{}}
			\toprule
			Type
			& Links
			& Affected
			& Appearance
			& Persistence
			\\\midrule
			
			SuddenD 
			& \multirow{4}{*}{2\,123}
			& \multirow{4}{*}{33\% (700)}
			& once, [200$^\textrm{th}$,~280$^\textrm{th}$] 
			& for $\infty$
			\\
			
			SuddenR 
			&
			&
			& once, [25$^\textrm{th}$,~275$^\textrm{th}$]
			& for [5,~20]
			\\
			
			InstaD 
			&
			&
			& on $\approx$1\% of a link
			& for 1 datapoint
			\\
			
			SlowD 
			&
			&
			& once, [1$^\textrm{st}$,~20$^\textrm{th}$]
			& for [150, 180]$^\dagger$
			\\
			
			\bottomrule
		\end{tabularx}
		\begin{tablenotes}
			\item[$\dagger$] RSSI$(x, \textrm{start})$ $\leftarrow$ RSSI$(x)$ + $\min(0, -\textrm{rand}(0.5, 1.5)\cdot(x-\textrm{start}))$
\end{tablenotes}	
	\end{threeparttable}
\end{table}

The suddenD anomaly, observed in Figure~\ref{fig:anomaly:step}, on the affected link appears arbitrarily between 200th and 280th packet and it persists indefinitely. In case of suddenR, observed in Figure~\ref{fig:anomaly:step-recovery}, anomaly applied on the link appears only once with a random start from 25th to 275th packet, where it persists for an arbitrary duration between 5 to 20 measurements. For InstaD of Figure~\ref{fig:anomaly:spikes}, the anomaly can appear anywhere in the entire series with 0.01 probability, which means that each anomaly on the affected link appears three times on average. Finally, SlowD anomaly of Figure~\ref{fig:anomaly:slow} appears arbitrarily between 1st and 20th measurements, where it commences with a random degrading pace of duration between 150 and 280 packets. In a nutshell, anomaly injection details are provided in Table~\ref{tab:injection-scenario}.

\begin{table}[htbp]
	\centering
	\ra{1.2}
	\footnotesize
	\begin{threeparttable}[b]
		\caption{Autoencoder configurations.}
		\label{tab:autoencoder-config}
		\begin{tabularx}{1\linewidth}{@{}lll@{}}
			\toprule
			Role & Layer & Notes \\
			\midrule
			\multirow{7}{*}{Encoder} & Input(*) & \\
			& Dense(128) & \\
			& BN + LeakyReLU($\alpha=0.2$) & \\
			& Dense(64) & \\
			& BN + LeakyReLU($\alpha=0.2$) & \\
			& Dense(32) & \\
			& BN + LeakyReLU($\alpha=0.2$) & \\
			& Dense(4) & no activation\\
			\midrule
			
			\multirow{7}{*}{Decoder} & Input(4) & \\
			& Dense(32) & \\
			& BN + LeakyReLU($\alpha=0.2$) & \\
			& Dense(64) & \\
			& BN + LeakyReLU($\alpha=0.2$) & \\
			& Dense(128) & \\
			& BN + LeakyReLU($\alpha=0.2$) & \\
			& Dense(*) & no activation\\
			\bottomrule

		\end{tabularx}
		\begin{tablenotes}
			\item[*] input/output size depends on feature vector
			\item[-] {Implementation of autoencoders in TensorFlow/Keras is available at: \\\url{https://gist.github.com/gcerar/5e4e53902493632a3cfb5cc06c3317b7}}
		\end{tablenotes}
	\end{threeparttable}
\end{table}

\subsection{Computing standard and encoded representations}
\label{sec:encodedRep}
Once anomalies are injected as specified in Table~\ref{tab:injection-scenario}, we compute four different data representations described in Section~\ref{sec:representations}. The first one, namely time-value representation of Section~\ref{sec:representations}-a, converts each link into a single feature vector containing 300 features. The second one, the so-called aggregated feature, summarizes each link with 7 features, which are described in Section~\ref{sec:representations}-b. The third one, namely histogram feature discussed in Section~\ref{sec:representations}-c, defines ten equally spaced bins, which are then presented to a model as a feature vector containing 10 features. The forth one, namely frequency feature elaborated in Section~\ref{sec:representations}-d, gives the model a large feature vector of frequency-domain representation summing up to nearly 150 features. As we compute four representations for each of the four types of anomalies, we generate 16 candidate datasets.

Next, we also consider autoencoders for each anomaly scenario and each of the four standard representations. As any other deep neural network, autoencoder also requires many iterations of training. To produce credible results with autoencoder, we build the generic model in two steps. In the first step, we split the dataset into training and test groups with a 60:40 ratio, respectively. In the second step, when the weights of the autoencoder are converged, we perform an end-to-end evaluation on the test group.  Relevant autoencoder configurations are provided in Table~\ref{tab:autoencoder-config}, where the layers and their required parameters are outlined for the encoder and the decoder. Although recent trends in DNNs go towards the use of convolutional layers, a convolution layer would make sense only in case of time-value and frequency perspective, due to their reasonable size and correlated neighbouring vector values. Therefore, our decision is to go with fully connected (dense) layers. For the activation part, we use batch normalization (BN) followed by Leaky Rectified Linear Unit (leaky ReLU, or LReLU) with $\alpha=0.2$ coefficient for negative values. While plain ReLU is most widely used non-linear activation function, its leaky version has shown several benefits and minor overall improvements~\cite{maas2013rectifier}. 

To produce the encoded representations, we feed the 16 datasets corresponding to the representation provided in Sections~\ref{sec:representations}-(a),(b),(c),(d) into the autoencoder, resulting in additional 16 candidate datasets. Therefore, to continue with the anomaly detection, we train both supervised and unsupervised ML models on a total of 32 datasets, 16 corresponding to the four standard representations of each anomaly and the other 16 corresponding to the encoded representations.

\subsection{Performing automatic anomaly detection}

Next, we compute the performance of the threshold, three supervised and three unsupervised ML techniques described in Section~\ref{sec:approaches} on the 32 generated datasets corresponding to the proposed anomalies and representations. Each approaches' output is compared to a label to identify whether the link actually contains anomalies or not. 

\begin{table}[htbp]
	\centering
	\ra{1.2}
	\footnotesize
	\begin{threeparttable}[b]
		\caption{Predetermined anomaly thresholds.}
		\label{tab:threshold-config}
		\footnotesize
		\begin{tabular}{ll}
			\toprule
			Features
			& Anomaly thresholds
			\\
			\midrule
			
			Time-series
			& Normality test~\cite{dagostino1971omnibus, dagostino1973tests}, when p < $10^{-3}$
			\\
			
			Aggregated
			& ($|$mean - median$|$ > 3\,dB) \textbf{OR} (2$\cdot$stdev > 2.5\,dB)
			\\
			
			Histogram
			& RSSI < -85\,dBm
			\\
			
			\bottomrule
			
		\end{tabular}
	\end{threeparttable}
\end{table}

\paragraph{Threshold approach} Descriptive details of leveraging certain thresholds for each anomaly can be found in Section~\ref{sub:threshold-based-detection}. The utilized experimental threshold parameters are listed in Table~\ref{tab:threshold-config}. The threshold for the time-series representation that uses the D'Agostino-Pearson's normality statistical test~\cite{dagostino1971omnibus, dagostino1973tests} is $p<10^{-3}$. The threshold for the aggregated representation assumes the absolute difference between mean and median is higher than $3dB$ or that the double of the standard deviation is higher than $2.5dB$. The threshold for the histogram representation is set at $RSSI < -85dBm$ while threshold selection for the FFT and encoded representations were infeasible to find using our trial-and-error approach. The differences in the FFT representation are not easily visible or detectable using simple methods while the encoded representations cannot be easily interpreted, therefore also deriving an appropriate threshold is not possible.

\begin{table*}[htbp]
	\centering
	\ra{1.2}
	\footnotesize
	\begin{threeparttable}[b]
		\caption{ML techniques and their relevant parameters.}
		\label{tab:alg}
		\footnotesize
		\begin{tabular}{llll}
			\toprule
			
			Approach
			& Technique
			& Implementation
			& Parameters and their range
			\\\midrule
			
			\multirow{6}{*}{Supervised} 
			&Logistic Regression 
			& LogisticRegression
			& penalty='l2',
			dual=False,
			tol=1e-4,
			C= (1e-3, 1e-2, 1e-1, 1.0, 10., 100.)
			fit\_intercept=True,
			\\
			
			&(LR)
			&  from sklearn
			& intercept\_scaling=1,
			class\_weight=None,
			solver='lbfgs',
			l1\_ratio=None	
			\\\cmidrule{2-4}
			
			&Random Forest 
			& BaggingClassifier
			& base\_estimator=None,
			n\_estimators=[10, 20, 30, 40, 50, 70, 100],
			max\_samples=1.0,
			\\

			& (RForest)
			&  from sklearn
			& 
			max\_features=1.0,
			oob\_score=False,
			intercept\_scaling=1,
			\\\cmidrule{2-4}
			
			&Support Vector 
			& SVC
			& C=(1e-3, 1e-2, 1e-1, 1.0, 10., 100.),
			kernel=('linear', 'rbf'),
			gamma=('auto', 'scale'),
			\\
			
			& Machine (SVM)
			&  from sklearn
			& 
			tol=1e-3,
			decision\_function\_shape='ovr',
			break\_ties=False
			\\\midrule
			
			\multirow{6}{*}{Unsupervised} 
			&Local Outlier 
			& LocalOutlierFactor
			& n\_neighbors=[5, 10, 20, 40, 50, 80],
			algorithm=['ball\_tree', 'kd\_tree', 'brute'],
			\\
			
			&Factor (LOF)
			& from sklearn
			& 
			leaf\_size=[10, 30, 50, 80],
			p=[1, 2]
			metric\_params=None,
			contamination="auto",
			\\\cmidrule{2-4}
			
			&Isolation Forest 
			& IsolationForest
			& n\_estimators=[10, 20, 30, 40, 50, 70, 100],
			max\_samples='auto',
			contamination='auto',
			\\
			
			&(IForest)
			&  from sklearn
			& 
			max\_features=1.0,
			bootstrap=False,
			\\\cmidrule{2-4}
			
			&Support Vector 
			& OneClassSVM
			& nu=[0.10, 0.3, 0.5, 0.70, 0.90, 1.0],
			kernel=('linear', 'rbf'),
			gamma=('auto', 'scale'),
			\\
			
			& Machine (OC-SVM)
			&  from sklearn
			& coef0=0.0,
			tol=1e-3,
			
			\\\bottomrule
			
		\end{tabular}
	\end{threeparttable}
\end{table*}

\begin{figure}[tbp]
	\centering
	\includegraphics[width=.8\linewidth]{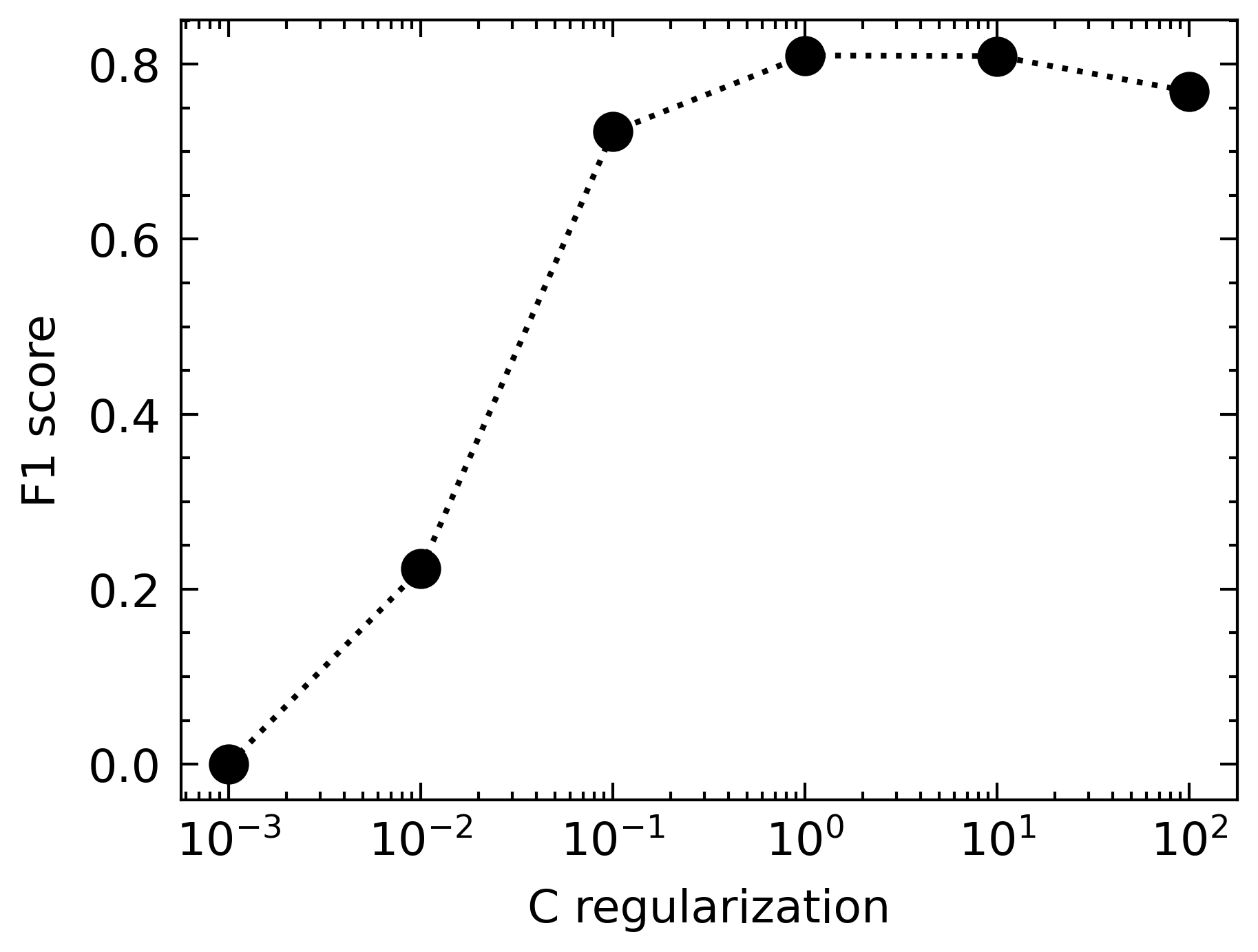}
	\caption{Regularization parameter ($C$) search for selecting the best performing model that is, for example, trained using LR on time-value representation for SuddenD anomalies and based on robust scaler.}
	\label{fig:optimization}
\end{figure}

\paragraph{Machine learning-based approaches} For each of the six selected ML techniques, we use standard ML cross-validation\footnote{Stratified K-Fold cross validation is implemented by using StratifiedKFold parameter in Python Scikit Learn toolbox \url{https://scikit-learn.org/stable/}}. We train the models using shuffled data split into training and test sets with a 80:20 ratio, respectively. Model is trained with the training set and evaluated using the test set in order to ensure credible results. We use standard metrics for evaluating classifiers: precision, recall and F1 score. Precision measures how many of the instances detected as class A actually belong to class A, expressed as; $\textrm{Precision} = \frac{\textrm{TP}}{\textrm{TP} + \textrm{FP}}$, whereas recall measures how many of the instances belonging to class A were actually detected, expressed as; $\textrm{Recall} = \frac{\textrm{TP}}{\textrm{TP} + \textrm{FN}}$, where TP, FP and FN stand for true positives, false positives and false negatives, respectively. F1 score is quantified by the harmonic mean of the precision and the recall, where larger values indicate better classifiers with balanced and higher precision and recall performances.

For each of the ML techniques selected in Section~\ref{sec:approaches}, Table~\ref{tab:alg} lists the respective implementations and parameters used in the experiments. For instance, for logistic regression we use the LogisticRegression implementation available in the Python Scikit Learn toolbox\footnote{\url{https://scikit-learn.org/stable/}}. As the LogisticRegression implementation enables setting 12 different parameters that influence the final model, we generally select standard values that have been proven to work on large number of cases and datasets by the ML community. However, we identify selected parameters that should be optimized, such as the regularization strength $C$ in this case. We search for the best configuration by adapting an array of possible values $C\in[10^{-3}, 10^{-2}, 10^{-1}, 10^{0}, 10^1, 10^2]$ and ultimately select the best performing regularization factor $C$ among them. For instance, Figure~\ref{fig:optimization} presents the scenario where a model is trained using LR on time-value representation for SuddenD anomalies and based on robust scaler. For this particular scenario, the best $F1$ score of this model is attained by means of setting $C$ to any value that is larger than $1$. For the results presented in the next sections, we only account for the best $F1$ scores obtained after searching for such near-optimal regularization parameter values.

The implementations chosen for the remaining algorithms also include over ten possible input parameters. For LOF, we vary the number of neighbours, algorithm and leaf size for finding the best performing model. For RForest and IForest, we vary the number of base estimators, whereas for SVM and OC-SVM, we vary the regularization factor $C$, the kernel and the kernel coefficient $gamma$ for the $rbf$ kernel, respectively. 

As some of the models are sensitive to scaling, we also consider training on data that is; i) not scaled, ii) scaled by using mean values, iii) scaled using mean and deviation, and iv) scaled using min-max. The entire procedure and parameters can be readily found and used in the existing public open source repository\footnote{Script for the design and development of anomaly detection models excluding data preprocessing is available at: \\\url{https://gist.github.com/gcerar/0b03e55f41147a7b7230f45d1f1209d6}}. Six selected ML techniques with the associated parameter tuning are trained over the 32 datasets, totalling at more than 40,000 anomaly detection models.

\section{Evaluation}\label{sec:evaluation}
In this section, we evaluate the relative performance of various data representations discussed in Section~\ref{sec:representations} and of approaches discussed in Section~\ref{sec:approaches} for detecting four types of anomalies introduced in Section~\ref{sec:anomaly}. The methodological and experimental details utilized for obtaining the results are elaborated in Section~\ref{sec:meth}.

\subsection{Performance analyses of data representations}
\label{sec:perf-repr}

\begin{figure*}[thbp]
	\centering
	\subfloat[Weights for time-value \label{fig:LG_norecovery_ts}]{
		\includegraphics[width=.35\linewidth]{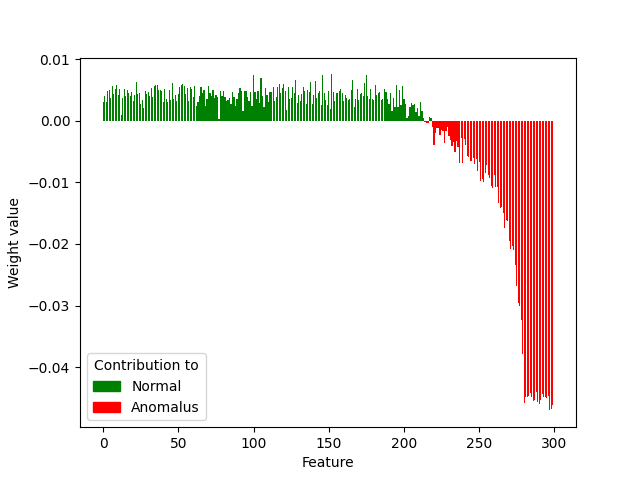}
	}%
	\subfloat[Weights for aggregated \label{fig:LG_norecovery_aggr}]{
		\includegraphics[width=.35\linewidth]{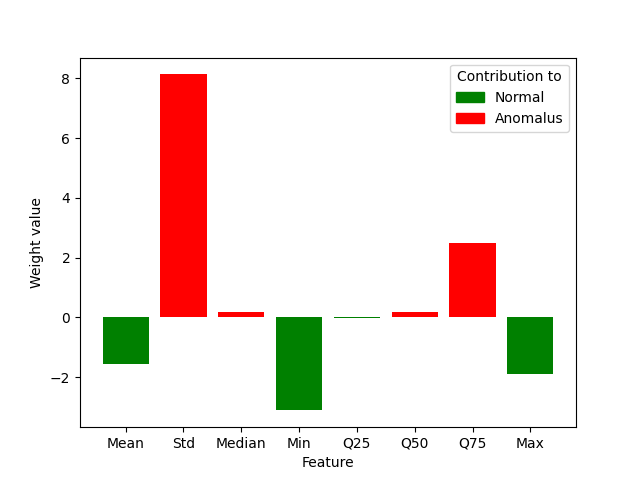}
	}

	\subfloat[Weights for histogram \label{fig:LG_norecovery_hist}]{
		\includegraphics[width=.35\linewidth]{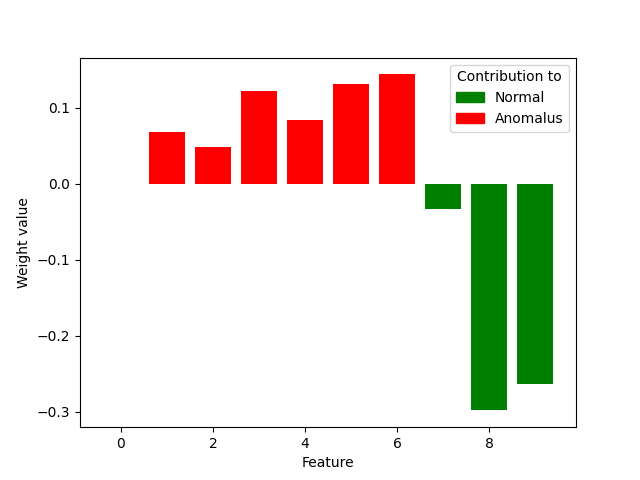}
	}%
	\subfloat[Weights for frequency (FFT)\label{fig:LG_norecovery_fft}]{
		\includegraphics[width=.35\linewidth]{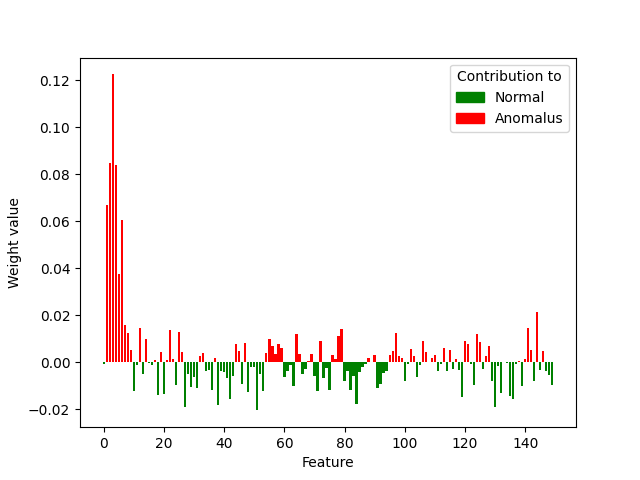}
	}%
	\caption{Learnt feature importance for distinct representations of the data for sudden degradation anomaly (SuddenD).}
	\label{fig:learnt:step}
\end{figure*}
\begin{figure*}[thbp]
	\centering
	\subfloat[Weights for time-value \label{fig:LG_recovery_ts}]{
		\includegraphics[width=.35\linewidth]{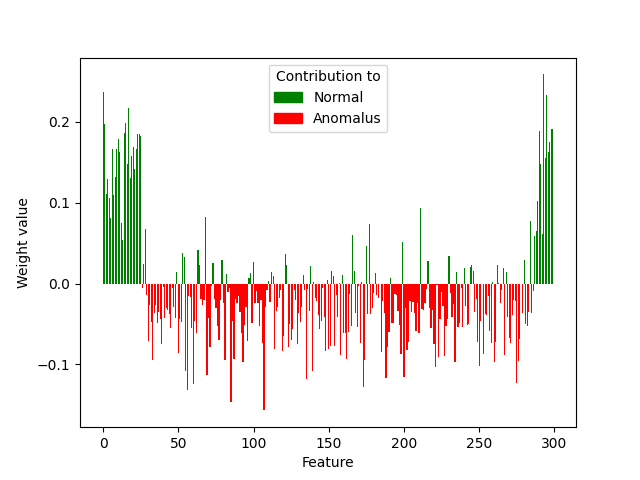}
	}%
	\subfloat[Weights for aggregated \label{fig:LG_recovery_aggr}]{
		\includegraphics[width=.35\linewidth]{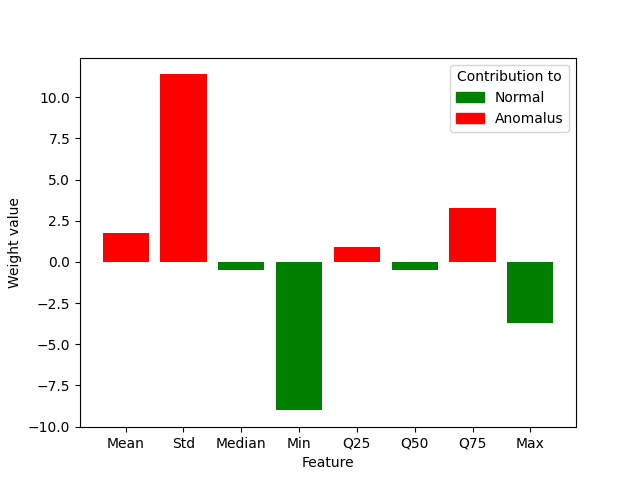}
	}

	\subfloat[Weights for histogram \label{fig:LG_recovery_hist}]{
		\includegraphics[width=.35\linewidth]{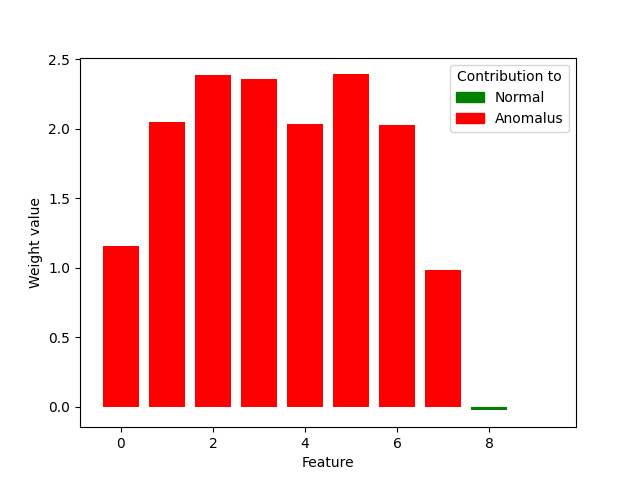}
	}%
	\subfloat[Weights for frequency (FFT)\label{fig:LG_recovery_fft}]{
		\includegraphics[width=.35\linewidth]{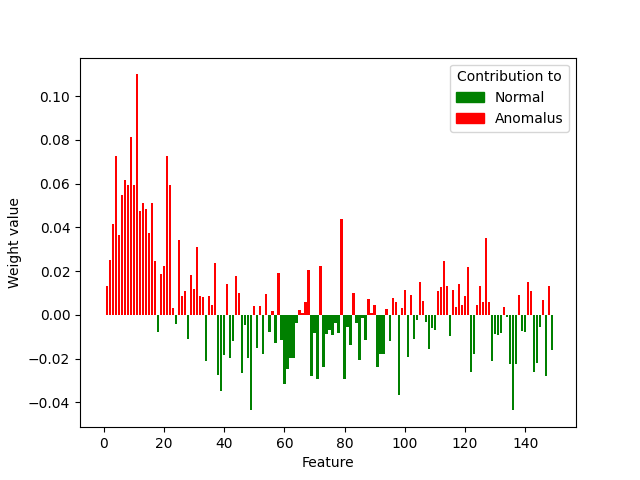}
	}%
	\caption{Learnt feature importance for distinct representations of the data for sudden degradation with recovery anomaly (SuddenR).}
	\label{fig:learnt:step-recovery}
\end{figure*}

\begin{figure*}[thbp]
	\centering
	\subfloat[Weights for time-value \label{fig:LG_spikes_ts}]{
		\includegraphics[width=.35\linewidth]{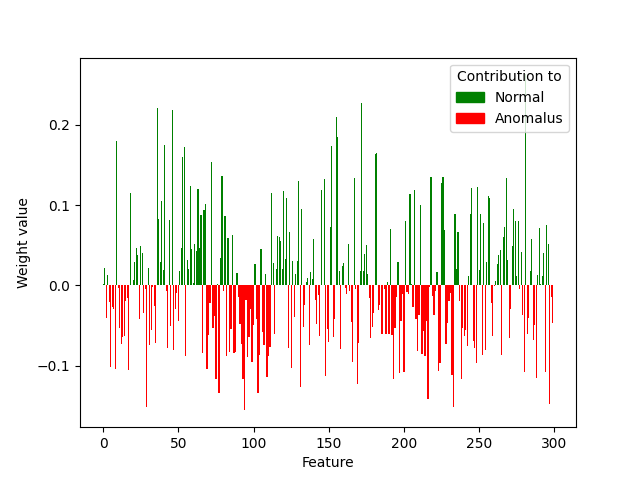}
	}%
	\subfloat[Weights for aggregated \label{fig:LG_spikes_aggr}]{
		\includegraphics[width=.35\linewidth]{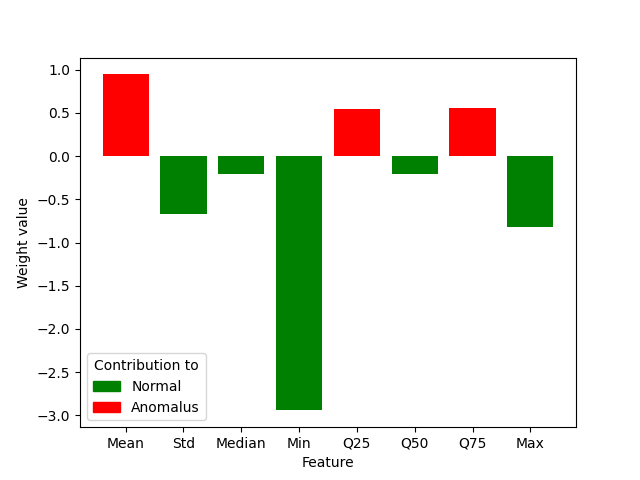}
	}

	\subfloat[Weights for histogram \label{fig:LG_spikes_hist}]{
		\includegraphics[width=.35\linewidth]{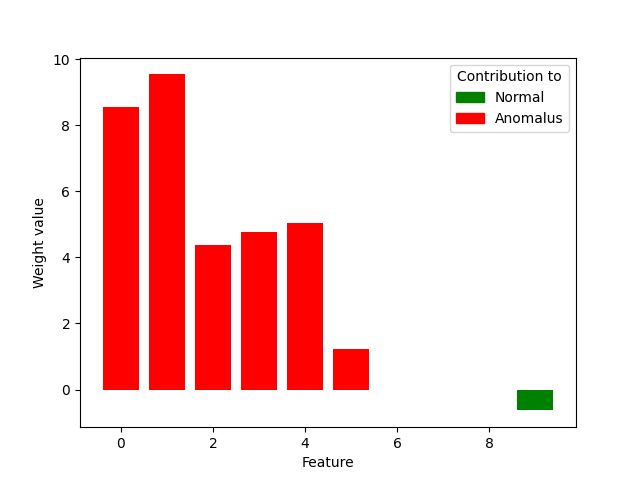}
	}%
	\subfloat[Weights for frequency  (FFT)\label{fig:LG_spikes_fft}]{
		\includegraphics[width=.35\linewidth]{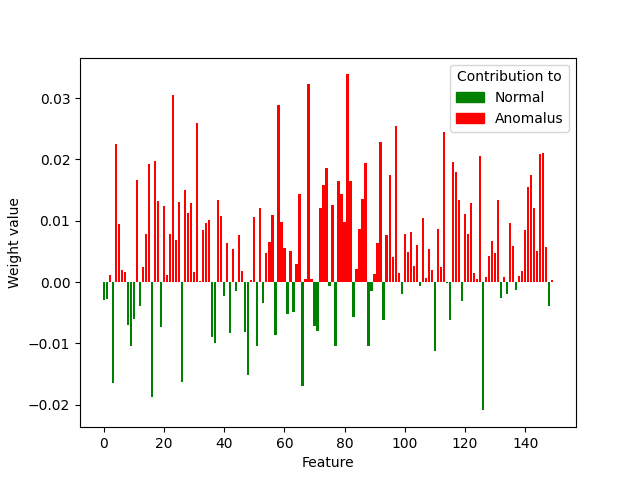}
	}%
	\caption{Learnt feature importance for distinct representations of the data for spike-like instantaneous degradation anomaly (InstaD).}
	\label{fig:learnt:spikes}
\end{figure*}
\begin{figure*}[thbp]
	\centering
	\subfloat[Weights for time-value \label{fig:LG_slow_ts}]{
		\includegraphics[width=.35\linewidth]{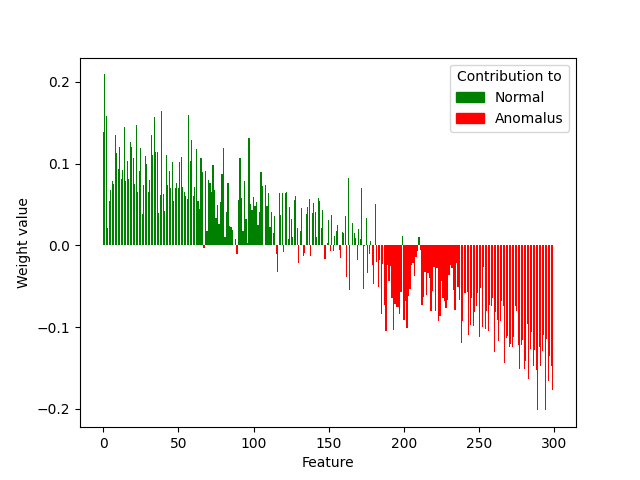}
	}%
	\subfloat[Weights for aggregated \label{fig:LG_slow_aggr}]{
		\includegraphics[width=.35\linewidth]{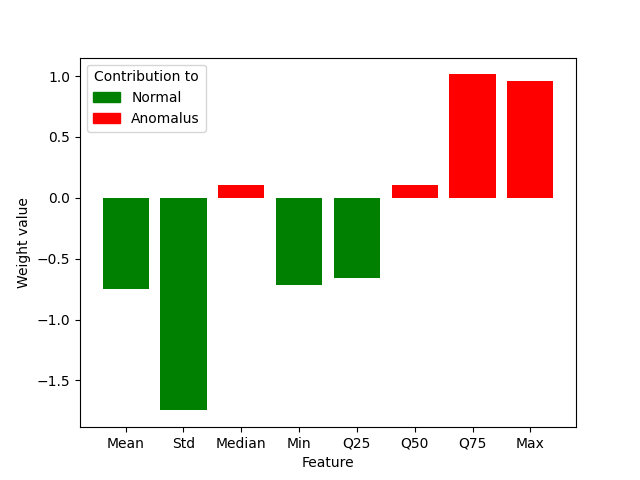}
	}

	\subfloat[Weights for histogram \label{fig:LG_slow_hist}]{
		\includegraphics[width=.35\linewidth]{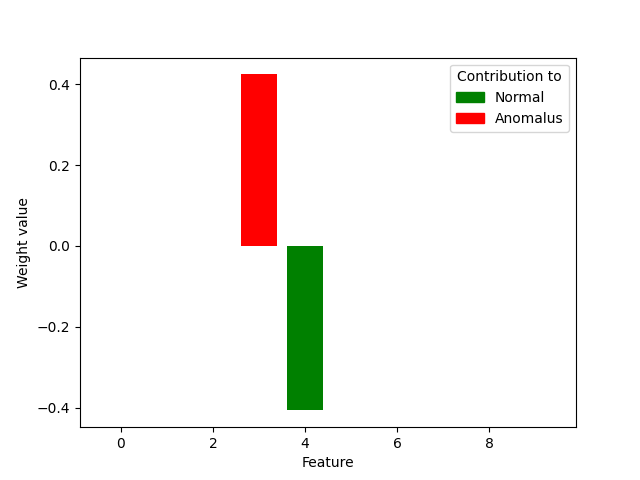}
	}%
	\subfloat[Weights for frequency  (FFT)\label{fig:LG_slow_fft}]{
		\includegraphics[width=.35\linewidth]{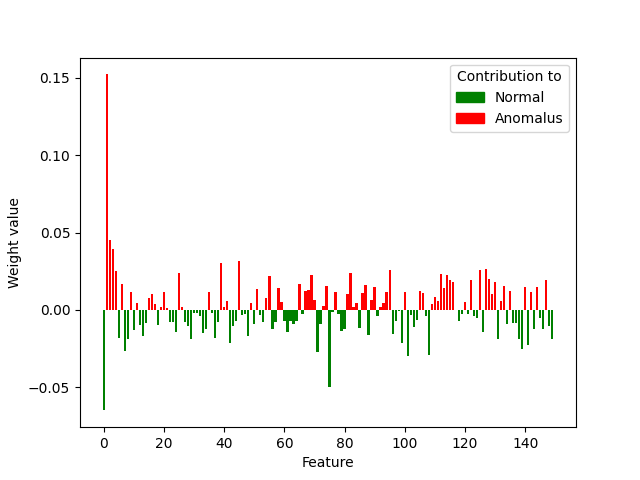}
	}%
	\caption{Learnt feature importance for distinct representations of the data for slow degradation anomaly (SlowD).}
	\label{fig:learnt:slow}
\end{figure*}

In this section, we first provide insight into how a model learns to classify by discussing the importance of various features resulting from the four manually generated and interpretable representations for discriminating the four types of anomalies defined in Section~\ref{sec:anomaly}. Next, we  discuss the influence of the five data representations, including those four manually generated ones and the automatically generated (autoencoder) one, as elaborated in Section~\ref{sec:representations}, on the performance of the learnt models. This entire subsection focuses on the influence of representations on the final models, while the influence of the ML approaches is analysed in Section~\ref{sub:MLappr}.

\subsubsection{Analysing the discriminative importance of features}
For analysing the discriminative power of the features in learning to classify the four anomaly types,  we choose LR for its simplicity and reasonable tractability. As explaining the meaning of the automatically generated features is infeasible, we exclude them from this part of the analysis, without loss of generality. 

Figures~\ref{fig:learnt:step},~\ref{fig:learnt:step-recovery},~\ref{fig:learnt:spikes} and~\ref{fig:learnt:slow} depict the weights learnt by the LR on the representations discussed in Section~\ref{sec:representations}. Each set of figures corresponds to an anomaly type, namely SuddenD, SuddenR, InstaD and SlowD. In the above-referred figures, the green weights depict the features that are important for identifying normal links, whereas the red weights are important for detecting the anomalous links. Using these learnt features, it is possible to look at the LR as a linear function with as many variables as the length of the feature vector, e.g., 300 for time-values representation and 8 for the aggregated. Each point in the feature vector has its corresponding weight with which it is multiplied. When all multiplications (weight * variable(n)) are summed up, a positive or a negative value corresponding to one of the two classes are obtained, i.e., normal or anomalous links.

For the case of the \textit{time-value representation} of the SuddenD anomaly from Figure~\ref{fig:synthetic:step:ts}, it can be seen that the points depicted with red, mostly starting from somewhere after feature 200 play a more important role when making the decision on whether an input feature vector contains an anomaly or not. The reason why LR learns that these features are the most important ones can be explained from the way the SuddenD anomaly is injected in the training dataset. According to Table~\ref{tab:injection-scenario}, SuddenD is injected randomly between packets 200 and 280. Therefore LR learns that those points are more discriminative for the anomalies. Simplistically, when multiplying the anomalous vector from Figure~\ref{fig:synthetic:step:ts} with the weights in Figure~\ref{fig:LG_norecovery_ts}, and subsequently summing up, the result will become positive, and hence the input will be classified as anomaly. On the other hand, when the normal vector from Figure~\ref{fig:synthetic:step:ts} is multiplied with the weights in Figure~\ref{fig:LG_norecovery_ts}, upon summing them up, the result will become negative, thus the vector will be classified as normal.

Similar discussions over time-value representations can be made for all the other anomalies. SuddenR anomaly is randomly injected between packets 25 and 275 of the time-value representation as per Table~\ref{tab:injection-scenario}, and it can be seen from Figure~\ref{fig:LG_recovery_ts} that the most important features for detecting the anomaly, represented with red, lie within this range. The importance of features for the spike anomaly that is quite random in nature and also occurs often in the data due to the nature of the wireless channel is depicted in Figure~\ref{fig:LG_spikes_ts}. Finally, the importance of the features for detecting SlowD is higher in the second half of the feature vector as depicted in Figure~\ref{fig:LG_slow_ts} since that's where the degradation becomes more evident.

Moving to \textit{aggregated representations}, it can be seen from Figure~\ref{fig:LG_norecovery_aggr} that standard deviation (Std) and the last quantile (Q75) are the most important features for detecting the anomaly, with minor contribution from the median and Q50. This is because standard deviation increases when SuddenD anomaly is present while the count of high RSSI values in the last quantile is smaller when this anomaly is present. Next, for SuddenR, the two main features remain the same as the shape is very similar to the SuddenD as can be seen in Figure~\ref{fig:LG_recovery_aggr}, albeit the duration differs leading to a more prominent influence of the mean for discrimination. For InstaD, that can be seen as a very narrow SuddenR randomly appearing on 1\% of the link, Std looses importance while the mean and two quantiles become more predictive as depicted in Figure~\ref{fig:LG_spikes_aggr}. For SlowD, the model learns that features which inform about the slope that appears and increases, therefore Q75 counting high RSSI values and the maximum (max) are predictive. The median and Q50 that capture the intermediate values of the slowly increasing slope also add minor discriminative power, as portrayed in Figure~\ref{fig:LG_slow_aggr}.

In the case of \textit{histogram representation}, the first bins where \textit{cumulated} low RSSI values corresponding to SuddenD, SuddenR and InstaD anomalies are the most important ones according to Figures~\ref{fig:LG_norecovery_hist},~\ref{fig:LG_recovery_hist} and~\ref{fig:LG_spikes_hist}. For the case of SlowD presented in Figure~\ref{fig:LG_slow_hist}, one of the middle bins that capture intermediate values is the most discriminative while the other bins seem to not contribute to either class. 

Finally, the importance of features in the case of \textit{frequency representation} presents a similar line of reasoning as for the other representations. For SuddenD and SuddenR anomaly amplitudes at low frequencies that introduce a major shift in the mean are the most important features, as portrayed in Figures~\ref{fig:LG_norecovery_fft} and~\ref{fig:LG_recovery_fft}. For InstaD there is no clear importance pattern as shown in Figure~\ref{fig:LG_spikes_fft}, whereas for SlowD the feature amplitudes around 0 are the most prominent ones as illustrated in Figure~\ref{fig:LG_slow_fft}.

\begin{table*}[htp]
	\centering
	\ra{1.2}
	\footnotesize
	\begin{threeparttable}[b]
		\caption{Performance of detecting sudden degradation (SuddenD) anomalies.}
		\label{tab:simulation-norecovery}
		\begin{tabular}{lllllllllllllllll}
			\toprule
			
			\multirow{2}{*}{Approach}
			& \multirow{2}{*}{Technique}
			& \multicolumn{3}{c}{time-value features}
			& \phantom{}
			& \multicolumn{3}{c}{aggregated features}
			& \phantom{}
			& \multicolumn{3}{c}{histogram features}
			& \phantom{}
			& \multicolumn{3}{c}{frequency domain}
			\\\cmidrule{3-5}\cmidrule{7-9}\cmidrule{11-13} \cmidrule{15-17}
			
			& 
			& Prec.
			& Rec.
			& F1
			& 
			& Prec.
			& Rec.
			& F1
			& 
			& Prec.
			& Rec.
			& F1
			& 
			& Prec.
			& Rec.
			& F1
			\\\midrule
			
			Baseline & Threshold (Tab.~\ref{tab:threshold-config})	& 0.66 & 1.00 & 0.79\tnote{1} &	& 0.97 & 1.00 & 0.98\tnote{1} &	& 0.44 & 1.00 & 0.61\tnote{1} && - & - & - \\\midrule

			\multirow{6}{*}{Supervised} & LR	& 1.00 & 1.00 & \textbf{1.00}\tnote{1} &	& 1.00 & 1.00 & \textbf{1.00}\tnote{2} &	& 0.99 & 0.99 & 0.99\tnote{1} &	& 1.00 & 1.00 & \textbf{1.00}\tnote{1}  \\
			& encoder~+~LR	& 1.00 & 1.00 & \textbf{1.00}\tnote{6} &	& 1.00 & 1.00 & \textbf{1.00}\tnote{3} &	& 1.00 & 1.00 & \textbf{1.00}\tnote{6} &	& 1.00 & 1.00 & \textbf{1.00}\tnote{1}  \\\cmidrule{2-17}
			
			& RForest	& 1.00 & 1.00 & \textbf{1.00}\tnote{1} &	& 0.99 & 0.99 & 0.99\tnote{4} &	& 0.99 & 1.00 & \textbf{1.00}\tnote{4} &	& 1.00 & 1.00 & \textbf{1.00}\tnote{1}  \\
			& encoder~+~RForest	& 1.00 & 1.00 & \textbf{1.00}\tnote{1} &	& 1.00 & 1.00 & \textbf{1.00}\tnote{4} &	& 1.00 & 1.00 & \textbf{1.00}\tnote{4} &	& 1.00 & 1.00 & \textbf{1.00}\tnote{1}  \\\cmidrule{2-17}
			
			& SVM	& 1.00 & 1.00 & \textbf{1.00}\tnote{1,7} &	& 1.00 & 1.00 & \textbf{1.00}\tnote{4,7} &	& 0.99 & 1.00 & \textbf{1.00}\tnote{5,8} &	& 1.00 & 1.00 & \textbf{1.00}\tnote{1,7}  \\
			& encoder~+~SVM	& 1.00 & 1.00 & \textbf{1.00}\tnote{1,8} &	& 1.00 & 1.00 & \textbf{1.00}\tnote{4,8} &	& 1.00 & 1.00 & \textbf{1.00}\tnote{4,7} &	& 1.00 & 1.00 & \textbf{1.00}\tnote{1,7}  \\\midrule
			
			\multirow{6}{*}{Unsupervised} & LOF	& 0.36 & 0.53 & 0.43\tnote{1} &	& 0.51 & 0.38 & 0.43\tnote{6} &	& 0.88 & 0.67 & 0.76\tnote{4} &	& 1.00 & 0.20 & 0.33\tnote{5}  \\
			& encoder~+~LOF	& 0.85 & 0.25 & 0.38\tnote{3} &	& 0.65 & 0.16 & 0.26\tnote{4} &	& 0.65 & 0.19 & 0.29\tnote{1} &	& 0.59 & 0.19 & 0.29\tnote{2}  \\\cmidrule{2-17}
			
			& IForest	& 0.98 & 0.48 & 0.64\tnote{1} &	& 0.90 & 0.77 & 0.83\tnote{4} &	& 0.91 & 0.60 & 0.72\tnote{4} &	& 0.89 & 0.47 & 0.61\tnote{2}  \\
			& encoder~+~IForest	& 0.94 & 1.00 & \textbf{0.97}\tnote{3} &	& 0.89 & 0.86 & \textbf{0.88}\tnote{2} &	& 0.94 & 1.00 & \textbf{0.97}\tnote{3} &	& 0.94 & 0.99 & \textbf{0.97}\tnote{5}  \\\cmidrule{2-17}
			
			& OC-SVM	& 0.87 & 0.93 & \textbf{0.90}\tnote{4,8} &	& 0.81 & 0.86 & 0.83\tnote{1,8} &	& 0.94 & 0.99 & \textbf{0.96}\tnote{5,8} &	& 1.00 & 0.96 & \textbf{0.98}\tnote{2,7}  \\
			& encoder~+~OC-SVM	& 1.00 & 0.84 & \textbf{0.91}\tnote{3,7} &	& 0.99 & 0.93 & \textbf{0.96}\tnote{5,7} &	& 1.00 & 0.84 & \textbf{0.91}\tnote{1,7} &	& 0.98 & 0.99 & \textbf{0.99}\tnote{3,7}  \\

			\bottomrule
		\end{tabular}	
	\end{threeparttable}
\end{table*}
\begin{table*}[htp]
	\centering
	\ra{1.2}
	\footnotesize
	\begin{threeparttable}[b]
		\caption{Performance of detecting sudden degradation with recovery (SuddenR) anomalies.}
		\label{tab:simulation-recovery}
		\begin{tabular}{lllllllllllllllll}
			\toprule
			
			\multirow{2}{*}{Approach}
			& \multirow{2}{*}{Technique}
			& \multicolumn{3}{c}{time-value features}
			& \phantom{}
			& \multicolumn{3}{c}{aggregated features}
			& \phantom{}
			& \multicolumn{3}{c}{histogram features}
			& \phantom{}
			& \multicolumn{3}{c}{frequency domain}
			\\\cmidrule{3-5}\cmidrule{7-9}\cmidrule{11-13} \cmidrule{15-17}
			
			& 
			& Prec.
			& Rec.
			& F1
			& 
			& Prec.
			& Rec.
			& F1
			& 
			& Prec.
			& Rec.
			& F1
			& 
			& Prec.
			& Rec.
			& F1
			\\\midrule

			Baseline & Threshold (Tab.~\ref{tab:threshold-config})	& 0.66 & 1.00 & 0.79\tnote{1} &	& 0.97 & 0.97 & 0.97\tnote{1} &	& 0.44 & 1.00 & 0.61\tnote{1} & & - & - & - \\\midrule
			
			\multirow{6}{*}{Supervised} & LR	& 0.92 & 0.86 & 0.89\tnote{3} &	& 0.99 & 0.99 & \textbf{0.99}\tnote{3} &	& 1.00 & 0.98 & \textbf{0.99}\tnote{2} &	& 1.00 & 1.00 & \textbf{1.00}\tnote{2}  \\
			& encoder~+~LR	& 0.99 & 0.98 & \textbf{0.99}\tnote{2} &	& 0.99 & 0.99 & \textbf{0.99}\tnote{6} &	& 1.00 & 0.99 & \textbf{0.99}\tnote{2} &	& 1.00 & 1.00 & \textbf{1.00}\tnote{2}  \\\cmidrule{2-17}
			
			& RForest	& 0.96 & 0.96 & 0.96\tnote{3} &	& 0.99 & 0.99 & \textbf{0.99}\tnote{2} &	& 1.00 & 0.99 & \textbf{0.99}\tnote{5} &	& 0.99 & 0.99 & 0.99\tnote{5}  \\
			& encoder~+~RForest	& 0.99 & 0.99 & \textbf{0.99}\tnote{5} &	& 0.99 & 0.98 & \textbf{0.99}\tnote{6} &	& 1.00 & 0.99 & \textbf{0.99}\tnote{6} &	& 1.00 & 1.00 & \textbf{1.00}\tnote{1}  \\\cmidrule{2-17}
			
			& SVM	& 0.98 & 0.96 & 0.97\tnote{2,8} &	& 0.99 & 0.99 & \textbf{0.99}\tnote{5,8} &	& 0.99 & 0.99 & \textbf{0.99}\tnote{3,8} &	& 1.00 & 1.00 & \textbf{1.00}\tnote{1,7}  \\
			& encoder~+~SVM	& 0.99 & 0.98 & \textbf{0.99}\tnote{5,7} &	& 0.99 & 0.99 & \textbf{0.99}\tnote{6,8} &	& 1.00 & 0.99 & \textbf{0.99}\tnote{2,7} &	& 1.00 & 1.00 & \textbf{1.00}\tnote{2,8}  \\\midrule
			
			\multirow{6}{*}{Unsupervised} & LOF	& 0.88 & 0.99 & \textbf{0.93}\tnote{5} &	& 0.53 & 0.39 & 0.45\tnote{6} &	& 0.98 & 0.97 & \textbf{0.98}\tnote{2} &	& 1.00 & 0.39 & 0.56\tnote{5} \\
			& encoder~+~LOF	& 0.95 & 0.61 & 0.74\tnote{1} &	& 0.67 & 0.16 & 0.26\tnote{2} &	& 0.79 & 0.27 & 0.40\tnote{3} &	& 0.80 & 0.29 & 0.43\tnote{1}  \\\cmidrule{2-17}
			
			& IForest	& 0.48 & 0.20 & 0.28\tnote{1} &	& 0.95 & 0.62 & 0.75\tnote{1} &	& 0.99 & 0.26 & 0.41\tnote{1} &	& 0.94 & 0.49 & 0.64\tnote{6}  \\
			& encoder~+~IForest	& 0.98 & 0.97 & \textbf{0.98}\tnote{5} &	& 0.93 & 0.98 & \textbf{0.95}\tnote{2} &	& 0.95 & 0.96 & \textbf{0.95}\tnote{5} &	& 0.97 & 0.97 & \textbf{0.97}\tnote{3}  \\\cmidrule{2-17}
			
			& OC-SVM	& 0.92 & 0.98 & \textbf{0.95}\tnote{2,8} &	& 0.98 & 0.82 & \textbf{0.89}\tnote{2,7} &	& 0.93 & 0.95 & \textbf{0.94}\tnote{5,8} &	& 1.00 & 0.83 & \textbf{0.91}\tnote{2,7}  \\
			& encoder~+~OC-SVM	& 0.81 & 0.87 & 0.84\tnote{5,8} &	& 0.76 & 0.81 & \textbf{0.78}\tnote{2,8} &	& 0.74 & 0.96 & 0.83\tnote{5,7} &	& 0.73 & 0.98 & \textbf{0.84}\tnote{6,7} \\
			
			\bottomrule
		\end{tabular}	
	\end{threeparttable}
\end{table*}

\begin{table*}[htp]
	\centering
	\ra{1.2}
	\footnotesize
	\begin{threeparttable}[b]
		\caption{Performance of detecting spike (InstaD) anomalies.}
		\label{tab:simulation-spikes}
		\begin{tabular}{lllllllllllllllll}
			\toprule
			
			\multirow{2}{*}{Approach}
			& \multirow{2}{*}{Technique}
			& \multicolumn{3}{c}{time-value features}
			& \phantom{}
			& \multicolumn{3}{c}{aggregated features}
			& \phantom{}
			& \multicolumn{3}{c}{histogram features}
			& \phantom{}
			& \multicolumn{3}{c}{frequency domain}
			\\\cmidrule{3-5}\cmidrule{7-9}\cmidrule{11-13} \cmidrule{15-17}
			
			& 
			& Prec.
			& Rec.
			& F1
			& 
			& Prec.
			& Rec.
			& F1
			& 
			& Prec.
			& Rec.
			& F1
			& 
			& Prec.
			& Rec.
			& F1
			\\\midrule
			
			Baseline & Threshold (Tab.~\ref{tab:threshold-config})	& 0.64 & 0.97 & 0.77\tnote{1} &	& 0.94 & 0.67 & 0.78\tnote{1} &	& 0.42 & 1.00 & 0.60\tnote{1} && - & - & - \\\midrule
			
			\multirow{6}{*}{Supervised} & LR	& 0.92 & 0.87 & 0.89\tnote{1} &	& 0.96 & 0.99 & 0.97\tnote{2} &	& 0.95 & 0.95 & 0.95\tnote{1} &	& 0.97 & 0.91 & 0.94\tnote{1}  \\
			& encoder~+~LR	& 0.95 & 0.92 & \textbf{0.94}\tnote{4} &	& 0.98 & 0.98 & \textbf{0.98}\tnote{5} &	& 0.98 & 0.95 & \textbf{0.97}\tnote{3} &	& 0.98 & 0.94 & \textbf{0.96}\tnote{4}  \\\cmidrule{2-17}
			
			& RForest	& 0.98 & 0.86 & 0.91\tnote{3} &	& 0.98 & 0.96 & 0.97\tnote{6} &	& 0.98 & 0.95 & \textbf{0.96}\tnote{2} &	& 0.96 & 0.89 & 0.92\tnote{1}  \\
			& encoder~+~RForest	& 0.96 & 0.91 & \textbf{0.94}\tnote{4} &	& 0.97 & 0.97 & 0.97\tnote{6} &	& 0.98 & 0.95 & \textbf{0.96}\tnote{3} &	& 0.97 & 0.93 & \textbf{0.95}\tnote{4}  \\\cmidrule{2-17}
			
			& SVM	& 0.96 & 0.91 & \textbf{0.94}\tnote{3,8} &	& 0.97 & 0.98 & \textbf{0.98}\tnote{5,8} &	& 0.97 & 0.96 & \textbf{0.96}\tnote{6,8} &	& 0.97 & 0.91 & 0.94\tnote{1,8}  \\
			& encoder~+~SVM	& 0.95 & 0.93 & \textbf{0.94}\tnote{4,7} &	& 0.98 & 0.98 & \textbf{0.98}\tnote{4,7} &	& 0.98 & 0.95 & \textbf{0.97}\tnote{6,8} &	& 0.99 & 0.93 & \textbf{0.96}\tnote{4,8}  \\\midrule
			
			\multirow{6}{*}{Unsupervised} & LOF	& 0.79 & 0.86 & \textbf{0.82}\tnote{2} &	& 0.60 & 0.32 & 0.42\tnote{3} &	& 0.94 & 0.85 & \textbf{0.89}\tnote{5} &	& 0.90 & 0.25 & 0.39\tnote{4}  \\
			& encoder~+~LOF	& 0.89 & 0.28 & 0.43\tnote{5} &	& 0.58 & 0.26 & 0.36\tnote{6} &	& 0.65 & 0.27 & 0.38\tnote{3} &	& 0.45 & 0.11 & 0.17\tnote{3}  \\\cmidrule{2-17}
			
			& IForest	& 0.29 & 0.16 & 0.21\tnote{1} &	& 0.67 & 0.73 & 0.70\tnote{1} &	& 0.98 & 0.34 & 0.50\tnote{5} &	& 0.97 & 0.42 & 0.59\tnote{2}  \\
			& encoder~+~IForest	& 0.91 & 0.82 & \textbf{0.86}\tnote{2} &	& 0.87 & 0.97 & \textbf{0.92}\tnote{1} &	& 0.80 & 0.96 & \textbf{0.87}\tnote{1} &	& 0.82 & 0.97 & \textbf{0.89}\tnote{1}  \\\cmidrule{2-17}
			
			& OC-SVM	& 0.77 & 0.87 & \textbf{0.82}\tnote{5,8} &	& 0.99 & 0.82 & \textbf{0.90}\tnote{2,7} &	& 0.76 & 0.82 & 0.79\tnote{5,8} &	& 0.82 & 0.90 & 0.86\tnote{6,7}  \\
			& encoder~+~OC-SVM	& 0.76 & 0.85 & 0.80\tnote{3,8} &	& 0.71 & 0.91 & \textbf{0.80}\tnote{4,7} &	& 0.70 & 0.80 & 0.75\tnote{3,8} &	& 0.92 & 0.95 & \textbf{0.93}\tnote{1,7}  \\

			\bottomrule
		\end{tabular}
	\end{threeparttable}
\end{table*}
\begin{table*}[htp]
	\centering
	\ra{1.2}
	\footnotesize
	\begin{threeparttable}[b]
		\caption{Performance of detecting slow degradation (SlowD) anomalies.}
		\label{tab:simulation-slow}
		\begin{tabular}{lllllllllllllllll}
			\toprule
			
			\multirow{2}{*}{Approach}
			& \multirow{2}{*}{Technique}
			& \multicolumn{3}{c}{time-value features}
			& \phantom{}
			& \multicolumn{3}{c}{aggregated features}
			& \phantom{}
			& \multicolumn{3}{c}{histogram features}
			& \phantom{}
			& \multicolumn{3}{c}{frequency domain}
			\\\cmidrule{3-5}\cmidrule{7-9}\cmidrule{11-13} \cmidrule{15-17}
			
			& 
			& Prec.
			& Rec.
			& F1
			& 
			& Prec.
			& Rec.
			& F1
			& 
			& Prec.
			& Rec.
			& F1
			& 
			& Prec.
			& Rec.
			& F1
			\\\midrule
			
			Baseline & Threshold (Tab.~\ref{tab:threshold-config})	& 0.17 & 0.10 & 0.13\tnote{1} &	& 0.47 & 0.03 & 0.06\tnote{1} &	& 0.31 & 0.57 & 0.40\tnote{1} && - & - & - \\\midrule
			
			\multirow{6}{*}{Supervised} & LR	& 0.97 & 0.96 & 0.97\tnote{3} &	& 0.44 & 0.16 & 0.91\tnote{4} &	& 0.92 & 0.87 & 0.90\tnote{2} &	& 0.92 & 0.89 & 0.90\tnote{6}  \\
			& encoder~+~LR	& 0.99 & 0.98 & \textbf{0.99}\tnote{2} &	& 0.99 & 1.00 & \textbf{1.00}\tnote{3} &	& 1.00 & 1.00 & \textbf{1.00}\tnote{1} &	& 0.98 & 0.98 & \textbf{0.98}\tnote{4}  \\\cmidrule{2-17}
			
			& RForest	& 0.98 & 0.98 & \textbf{0.98}\tnote{5} &	& 0.98 & 0.99 & 0.99\tnote{4} &	& 0.99 & 0.99 & 0.99\tnote{6} &	& 0.92 & 0.92 & 0.92\tnote{3}  \\
			& encoder~+~RForest	& 0.98 & 0.98 & \textbf{0.98}\tnote{4} &	& 0.99 & 1.00 & \textbf{1.00}\tnote{3} &	& 1.00 & 1.00 & \textbf{1.00}\tnote{6} &	& 0.97 & 0.97 & \textbf{0.97}\tnote{4}  \\\cmidrule{2-17}
			
			& SVM	& 0.97 & 0.97 & 0.97\tnote{2,7} &	& 0.98 & 0.99 & 0.99\tnote{6,8} &	& 1.00 & 1.00 & \textbf{1.00}\tnote{4,8} &	& 0.93 & 0.94 & 0.94\tnote{2,8}  \\
			& encoder~+~SVM	& 0.98 & 0.99 & \textbf{0.99}\tnote{4,7} &	& 1.00 & 1.00 & \textbf{1.00}\tnote{3,7} &	& 1.00 & 1.00 & \textbf{1.00}\tnote{6,7} &	& 0.98 & 0.98 & \textbf{0.98}\tnote{4,7}  \\\midrule
			
			\multirow{6}{*}{Unsupervised} & LOF	& 0.36 & 0.37 & 0.36\tnote{3} &	& 0.28 & 0.23 & 0.26\tnote{4} &	& 0.32 & 0.26 & 0.29\tnote{5} &	& 0.43 & 0.02 & 0.04\tnote{2}  \\
			& encoder~+~LOF	& 0.59 & 0.12 & 0.20\tnote{4} &	& 0.36 & 0.18 & 0.24\tnote{4} &	& 0.29 & 0.20 & 0.23\tnote{3} &	& 0.59 & 0.11 & 0.18\tnote{1}  \\\cmidrule{2-17}	
			
			& IForest	& 0.29 & 0.20 & 0.24\tnote{1} &	& 0.74 & 0.55 & \textbf{0.63}\tnote{3} &	& 0.33 & 0.13 & 0.18\tnote{6} &	& 0.30 & 0.09 & 0.14\tnote{1}  \\
			& encoder~+~IForest	& 0.86 & 0.97 & \textbf{0.91}\tnote{4} &	& 0.40 & 0.41 & 0.40\tnote{6} &	& 0.49 & 0.58 & \textbf{0.53}\tnote{6} &	& 0.64 & 0.61 & \textbf{0.63}\tnote{1}  \\\cmidrule{2-17}
			
			& OC-SVM	& 0.46 & 0.81 & 0.59\tnote{5,7} &	& 0.41 & 0.92 & \textbf{0.56}\tnote{4,7} &	& 0.65 & 0.69 & \textbf{0.67}\tnote{1,7} &	& 0.69 & 0.73 & \textbf{0.71}\tnote{6,7}  \\
			& encoder~+~OC-SVM	& 0.71 & 0.76 & \textbf{0.73}\tnote{4,8} &	& 0.90 & 1.00 & \textbf{0.95}\tnote{4,7} &	& 0.77 & 1.00 & \textbf{0.87}\tnote{6,7} &	& 0.63 & 0.91 & \textbf{0.75}\tnote{5,7}  \\
			
			\bottomrule
			
		\end{tabular}	
	
		$^1$No-scaling $^2$Only mean scaling (by standard scaler)  $^3$Mean and deviation scaling (by standard scaler) $^4$Only mean scaling (by robust scaler with respect to values between Q25 and Q75) $^5$Mean and deviation scaling (by robust scaler with respect to values between Q25 and Q75)  $^6$Min-Max scaler  $^7$Linear kernel
		$^8$RBF kernel

	\end{threeparttable}
\end{table*}

\subsubsection{The influence of the representations on the performance of the learnt models}
The best performing results of the classification with respect to F1 score are presented in Table~\ref{tab:simulation-norecovery} for SuddenD, Table~\ref{tab:simulation-recovery} for SuddenR, Table~\ref{tab:simulation-spikes} for InstaD and Table~\ref{tab:simulation-slow} for SlowD. The first column of the tables lists the approach, the second column outlines the used ML techniques, while columns 3 to 6 list the results for time-value, aggregated, histogram and FFT representations, respectively.

The encoded representation introduced in Section~\ref{sec:representations}-e and employed according to the methodology in Section~\ref{sec:encodedRep} is inserted into the above-mentioned performance tables with the name of respective ML technique using the term "encoder". More precisely, referring to the rows corresponding to the ML technique, say IForest, the performance results are implemented for the four mentioned representations for the IForest ML technique. Additionally, at the row entitled "Encoder + IF", the numerical results refer to the IForest ML technique that is applied to the codes generated from the four representations, respectively. Finally, the superscripts identify the scaling methods utilized. The three highest F1 scores for supervised approaches and the three highest F1 scores for unsupervised approaches are delineated in bold font.

With respect to the data representations, from the results listed in Tables~\ref{tab:simulation-norecovery},~\ref{tab:simulation-recovery},~\ref{tab:simulation-spikes} and~\ref{tab:simulation-slow}, two high level observations are outlined as follows. 

\begin{itemize}
	\item None of the four manually generated features clearly dominates the remaining ones in terms of anomaly detection performance. 
	\item In most cases, automatically generated encoded data representation improves anomaly detection performance compared to the same non-encoded counterpart.
\end{itemize}

\paragraph*{SuddenD anomalies}
For \textit{SuddenD} observed in Table~\ref{tab:simulation-norecovery}, all representations produce nearly perfect F1 scores of above 0.99 with all supervised ML approaches. Moving to unsupervised approaches, it can be readily seen that the histogram representation works best with LOF, however the F1 score of $0.76$ is modest. The aggregated features with $F1=0.83$ work best with IForest followed by the histogram features with $F1=0.72$. The encoded representations surpass all non-encoded ones with this approach reaching F1 scores up to $0.97$. All but the manual aggregated features yield good F1 scores of above $0.9$ with OC-SVM, however the frequency representation dominates with F1 score of above 0.98. The encoded representations improve the anomaly detection performance in three of the four possible cases.

\paragraph*{SuddenR anomalies}
For \textit{SuddenR} observed in Table~\ref{tab:simulation-recovery}, almost all representations produce high F1 scores of above $0.9$ with all supervised ML approaches. The time-value representation is slightly inferior to the other manual and autoencoded representations, producing $0.89$ F1 score with LR, $0.96$ with RForest and $0.97$ with SVM. 

For unsupervised approaches, unlike in the case of SuddenD, the time-series and histogram representations work best with LOF, with high F1 scores of above $0.93$. Similarly, the aggregated features with $F1=0.75$ work best with IForest followed by the frequency representation with $F1=0.64$ for SuddenD anomaly. The encoded representations surpass all non-encoded ones with this approach reaching F1 scores up to $0.98$. The manual features yield good scores of above $0.89$ with OC-SVM, however the time-value and histogram representations dominate with F1 score of above $0.94$. The encoded representations do not improve the anomaly detection performance for this anomaly type using OC-SVM.

\paragraph*{InstaD anomalies}
For \textit{InstaD} observed in Table~\ref{tab:simulation-spikes}, almost all representations produce high F1 scores of above $0.9$ with all supervised ML approaches. The time-value representation is slightly inferior to the other manual and autoencoded representations, producing $0.89$ F1 score with LR, $0.91$ with RForest and $0.94$ with SVM. While for the previous SuddenD and SuddenR the remaining three representations yielded comparable F1 scores with all ML approaches, for InstaD anomaly, frequency domain representation is less suitable when compared to histogram, and histogram features are less suitable than the aggregated features in terms of the anomaly detection performance. 

Considering unsupervised approaches, the more arbitrary the anomaly becomes, so the effect of the representation on the results. The time-value representation and histogram work best with LOF with F1 up to $0.89$ while the encoded representation provides no additional benefit. The manual representations work poorly with RForests while the encoded ones yield F1 scores of up to 0.92. The aggregated features and encoded frequency domain representations work best with OC-SVM with $F1=0.9$ and $F1=0.93$, respectively.

\paragraph*{SlowD anomalies}
For \textit{SlowD} observed in Table~\ref{tab:simulation-slow}, all representations produce high F1 scores of above $0.9$ with all supervised ML approaches. The time-value representation performs best with LR yielding $F1=0.97$, while all time-value, aggregated and histogram features work well with RForest and SVM yielding an F1 score of above $0.97$. This anomaly type is relatively more difficult to be detected using frequency representation when supervised approaches are considered. 

For unsupervised approaches, no representation works well with LOF while all manual representations perform modestly with F1 scores of up to $0.71$. However, in some specific cases, the encoded representation achieves higher detection performance. For instance, time-value encoded with IForest yields an F1 score of $0.91$, while aggregated encoded yields an F1 score of $0.95$ with OC-SVM. All encoded representations perform better with OC-SVM compared to their non-encoded counterparts.

\subsection{Performance analyses of ML approaches}\label{sub:MLappr}
We now analyse the detection performance of the ML approaches described in Section~\ref{sec:approaches} on all the anomaly types proposed in Section~\ref{sec:anomaly}. By using Tables~\ref{tab:simulation-norecovery},~\ref{tab:simulation-recovery},~\ref{tab:simulation-spikes} and~\ref{tab:simulation-slow} we perform an analysis across rows, unlike the cross-column analysis performed in Section~\ref{sec:perf-repr} for data representations. While in Section~\ref{sec:perf-repr} we already explained, as an example, how the LR approach works on our anomaly dataset, this section elaborates, as an exemplifying analysis, on what the tree based ensemble learns. We selected the tree based ensemble as it is also easily explainable and tractable similar to LR. For the start, we remark the following major observations.

\begin{itemize}
	\item For a given anomaly type, there is no major difference between the three selected supervised approaches. 
	\item Among the unsupervised approaches, OC-SVM performs the best F1 scores, closely followed by IForest, whereas LOF typically performs the worst F1 scores.
\end{itemize}

\paragraph*{SuddenD anomalies}

\begin{figure}[tbp]
	\centering
	\includegraphics[width=\linewidth]{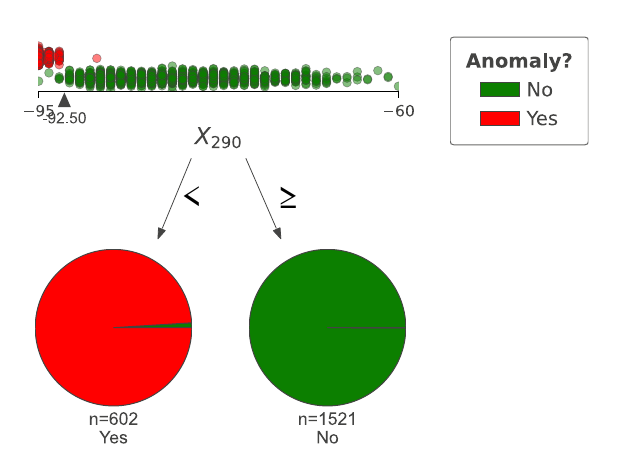}
	\caption{An exemplifying decision tree for the detection process of SuddenD anomaly.}
	\label{fig:dtree-explain-suddend}
\end{figure}

According to Table~\ref{tab:simulation-norecovery}, the supervised models are able to detect SuddenD anomalies more accurately than the unsupervised models. All three supervised models have achieved near perfect F1 score of $0.99$ on all data representations. 

The tree based ensemble models, such as supervised RForest and unsupervised IForest, learn a set of trees and subsequently use a voting mechanism on the decision of each individual tree to determine the final class. A tree that is the fundamental part of the two ensemble models also learns which features are the most important ones. The feature with the highest discrimination power (weight) is situated at the root of the tree, then on the left and right nodes, as it can be exemplified in Figure~\ref{fig:dtree-explain-suddend}, the next two important features are placed and the process follows until a certain stopping criterion is met. In our specific case, the trees learn the thresholds for particular values in the feature vector. For instance, depicted in Figure~\ref{fig:dtree-explain-suddend}, it can be seen that if the value at position 290 in the time-value representation, denoted by $X_{290}$, is below $-92.5$, then the link is anomalous, otherwise it is a normal link. This simple rule is able to correctly detect $n=596$ anomalous links and $n=1520$ normal links while only misclassifying 7 links, thus the performance of that tree alone is $F1=0.99$. 

The SVM models are more complex and difficult to visualize when a feature vector has more than 3 dimensions as it is the case with all manual and autoencoded representations used in this paper. SVMs essentially compute a hyperplane that attempts to separate the N-dimensional feature vector according to a criterion, such as the labels.

Among the unsupervised approaches, OC-SVM is able to achieve F1 scores close to the supervised approaches, for instance $0.98$, $0.96$ and $0.90$ on FFT, histogram and time-value representations, respectively. For OC-SVM model, with the aid of autoencoder the time-value representation is transformed to an important summary of the data by removing the noise and repetitions, leading to a performance increase from $F1=0.83$ to $F1=0.96$. Next, IForest achieved a lower performance with an F1 score between $0.61$ and $0.83$, the latter on the aggregated representation $0.83$ while the LOF performance reached $0.76$ on one occasion.

\paragraph*{SuddenR anomalies}
Compared to SuddenD, SuddenR gains a steep recovery slope, while the duration and occurrence are more random. The results in Table~\ref{tab:simulation-recovery} show that supervised models are able to detect SuddenR more accurately than the unsupervised models. F1 score of supervised models ranges from $0.89$ with LR on time-value representation to near perfect F1 score for remaining supervised approaches. Using encoded representation of the time-values improves the performance also in the case of LR to $0.99$, which corresponds to an about $11$\% improvement. For the LR case, as discussed in Section~\ref{sec:perf-repr} and depicted in Figure~\ref{fig:learnt:step-recovery}, the most important features are the ones that attempt to capture the random drops between packets 25 and 275.

\begin{figure}[tbp]
	\centering
	\includegraphics[trim=0 2300 1200 0,clip,width=1.2\linewidth]{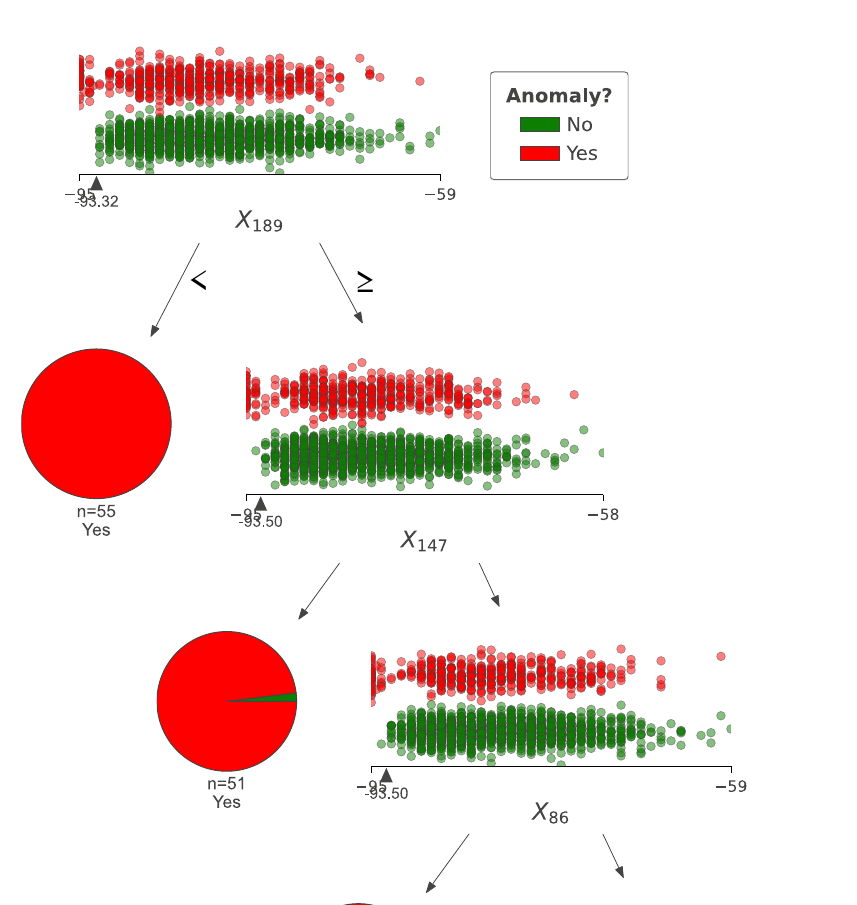}
	\caption{A part of the decision tree while detecting SuddenR anomaly over time-value representation.}
	\label{fig:dtree-explain-suddenr}
\end{figure}

\begin{figure}[tbp]
	\centering
	\includegraphics[width=0.9\linewidth]{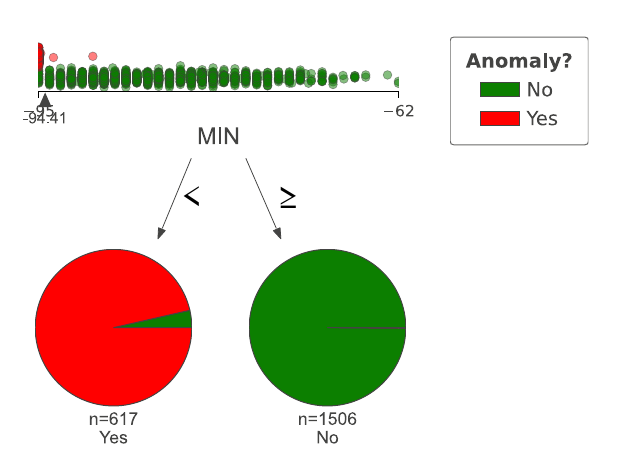}
	\caption {A part of the decision tree while detecting SuddenR anomaly over aggregated data representation.}
	\label{fig:dtree-aggr-explain-suddenr}
\end{figure}

\begin{figure*}[tbp]
	\centering
	\includegraphics[clip,trim=30 0 30 0,width=0.8\linewidth]{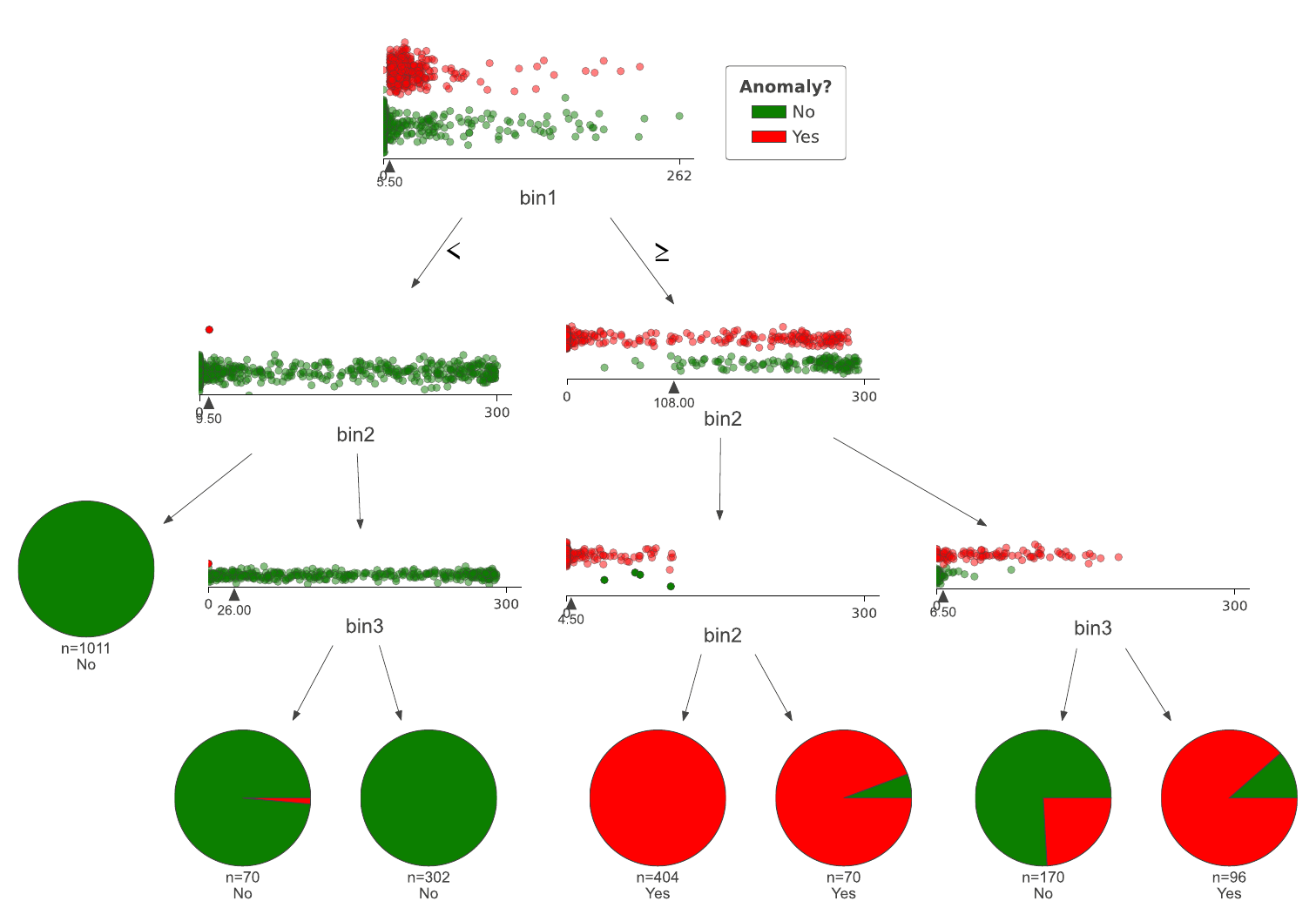}
	\caption{A part of the decision tree while detecting SuddenR anomaly over histogram representation.}
	\label{fig:dtree-hist-explain-suddenr}
\end{figure*}

\begin{figure}[tbp]
	\centering
	\includegraphics[width=0.9\linewidth]{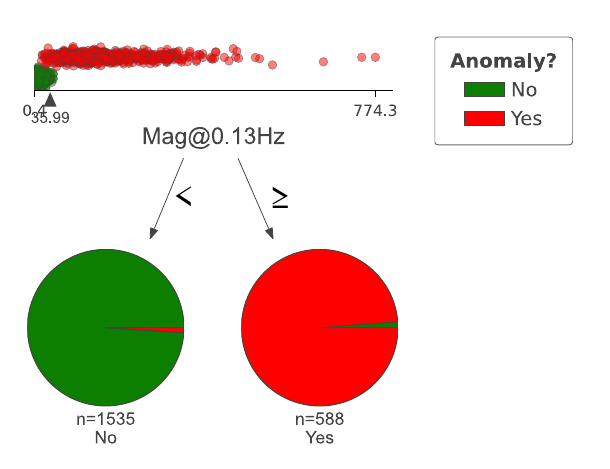}
	\caption{Decision tree for detecting SuddenR anomaly using FFT representation.}
	\label{fig:dtree-fft-explain-suddenr}
\end{figure}

A decision tree representing RForest and IForest ensembles is portrayed in Figure~\ref{fig:dtree-explain-suddenr} for the time-value representation of the SuddenR anomaly. It can be seen that the most discriminative data points are $X_{189}$, $X_{147}$, $X_{86}$ with $-93.5$~dBm RSSI threshold. The tree can grow very deep, eventually over-fitting the data, however, as discussed in Section~\ref{sec:meth}, we undertook standard methods for avoiding that in the experimental design. Figure~\ref{fig:dtree-aggr-explain-suddenr} presents an example tree learnt on aggregated feature representation. Similar to the tree in Figure~\ref{fig:dtree-explain-suddend}, it is simple and effective, where it compares minimal RSSI to $-94.407$~dBm threshold to decide whether it is anomaly or not. Figure~\ref{fig:dtree-hist-explain-suddenr} shows a tree learnt using the histogram representation as input. While performance is similar to the previous representation, we see that using aggregated representation requires less number of decisions, i.e., depth of tree, for effective anomaly detection. Similar observations can be made for the tree learnt on fft representation for this anomaly type depicted in Figure~\ref{fig:dtree-fft-explain-suddenr}.

Among the unsupervised approaches, OC-SVM, without encoded representation, is able to achieve an F1 score of around $0.90$ on average through all four representations, which is almost on par with supervised approaches. IForest, on the other hand, performs much better with encoded representations, where the most significant improvement is presented on time-value representation ramping its F1 score from $0.21$ to $0.86$. Since SuddenR is limited in duration and thus affecting less number of features, LOF is able to pull ahead in time-value and histogram representations, where it reaches an F1 score of above $0.93$.

\paragraph*{InstaD anomalies}

\begin{figure}[tbp]
	\centering
	\includegraphics[clip,trim=980 1200 350 0,width=\linewidth]{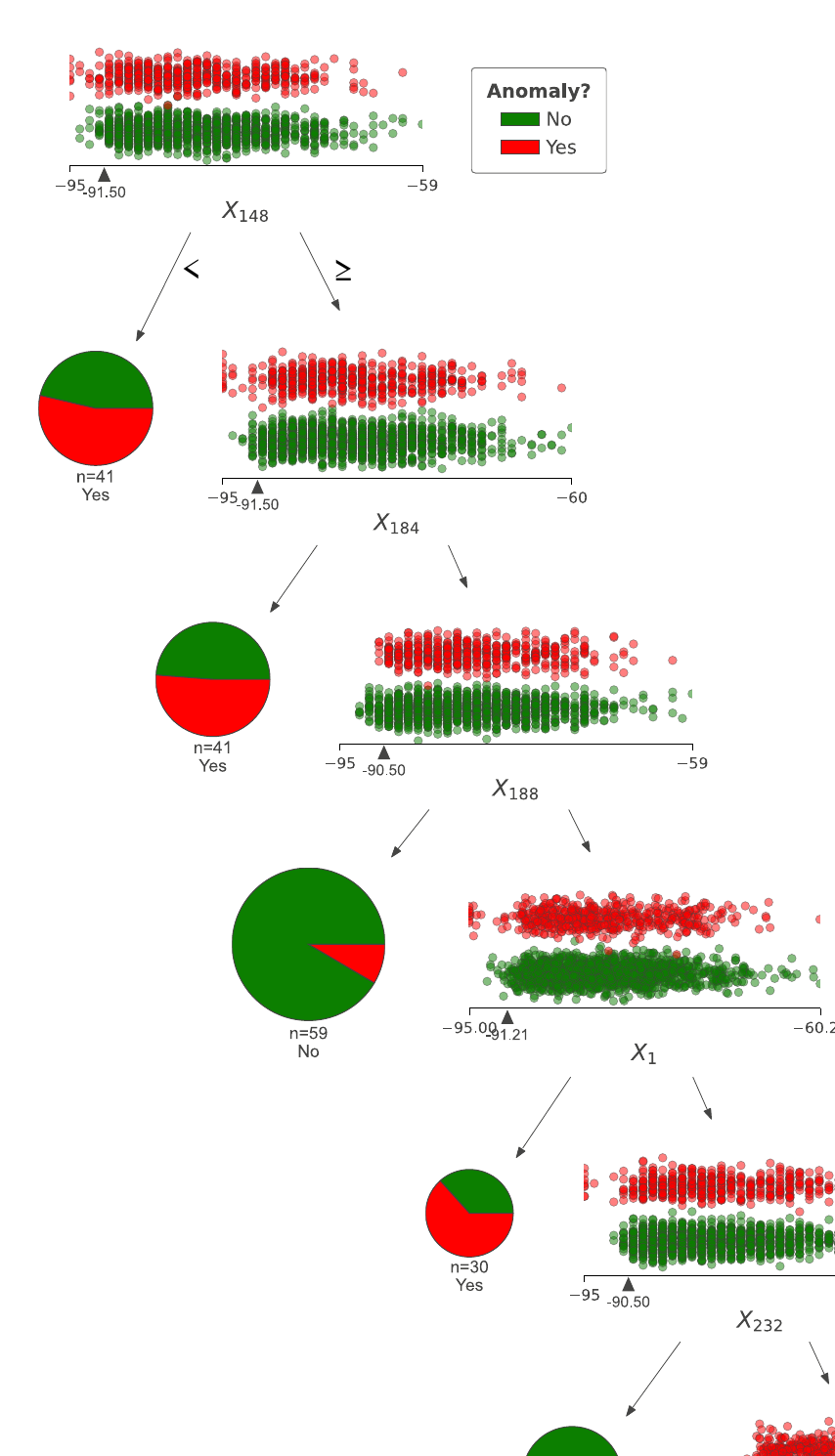}
	\caption{A part of the decision tree while detecting InstaD anomaly over time-value representation.}
	\label{fig:dtree-explain-instad}
\end{figure}

In contrast to SuddenD and SuddenR, InstanD appears as an anomaly with extremely short duration (pulse). The results in Table~\ref{tab:simulation-spikes} show that supervised approaches are slightly better at InstaD classification. F1 performance score of supervised approaches is slightly worse (up to 0.98) from what we have seen for SuddenD or SuddenR detection performance. Due to the arbitrary characteristics of this anomaly type, the F1 score is diminished further when the supervised approaches are trained with the time-value and frequency domain representations as outlined in Table~\ref{tab:simulation-spikes}.

To better understand decision making on classifying the InstaD anomaly, we examine a decision tree representing RForest and IForest ensembles  as depicted in Figure~\ref{fig:dtree-explain-instad}. Due to the random nature of this anomaly, the tree selects random points and verifies their value against a learnt threshold. For this particular tree, feature $X_{148}$ that is compared to $-91.50$~dBm RSSI threshold is selected in the root. Then, it follows with the comparisons of the features in order of $X_{184}$ and $X_{188}$ that are compared to $-91.50$~dBm and $-90.50$~dBm, respectively and this process terminates when the final depth of the three is reached. For this anomaly type, time-series and FFT domain may not be the optimal data representations for the sake of developing a reliable and non-overfitting model.

Among the unsupervised approaches, there is no clear best approach. The top five performing models are OC-SVM using encoded FFT with $0.93$ F1 score, IForest using encoded aggregated features with an F1 score of $0.92$, OC-SVM using aggregated representation with an F1 score of $0.90$, and LOF using histogram representation and IForest using encoded FFT, both achieving an F1 score of $0.89$.

\paragraph*{SlowD anomalies}

\begin{figure*}[tbp]
	\centering
	\includegraphics[clip,trim=0 1500 1500 0, width=0.7\linewidth]{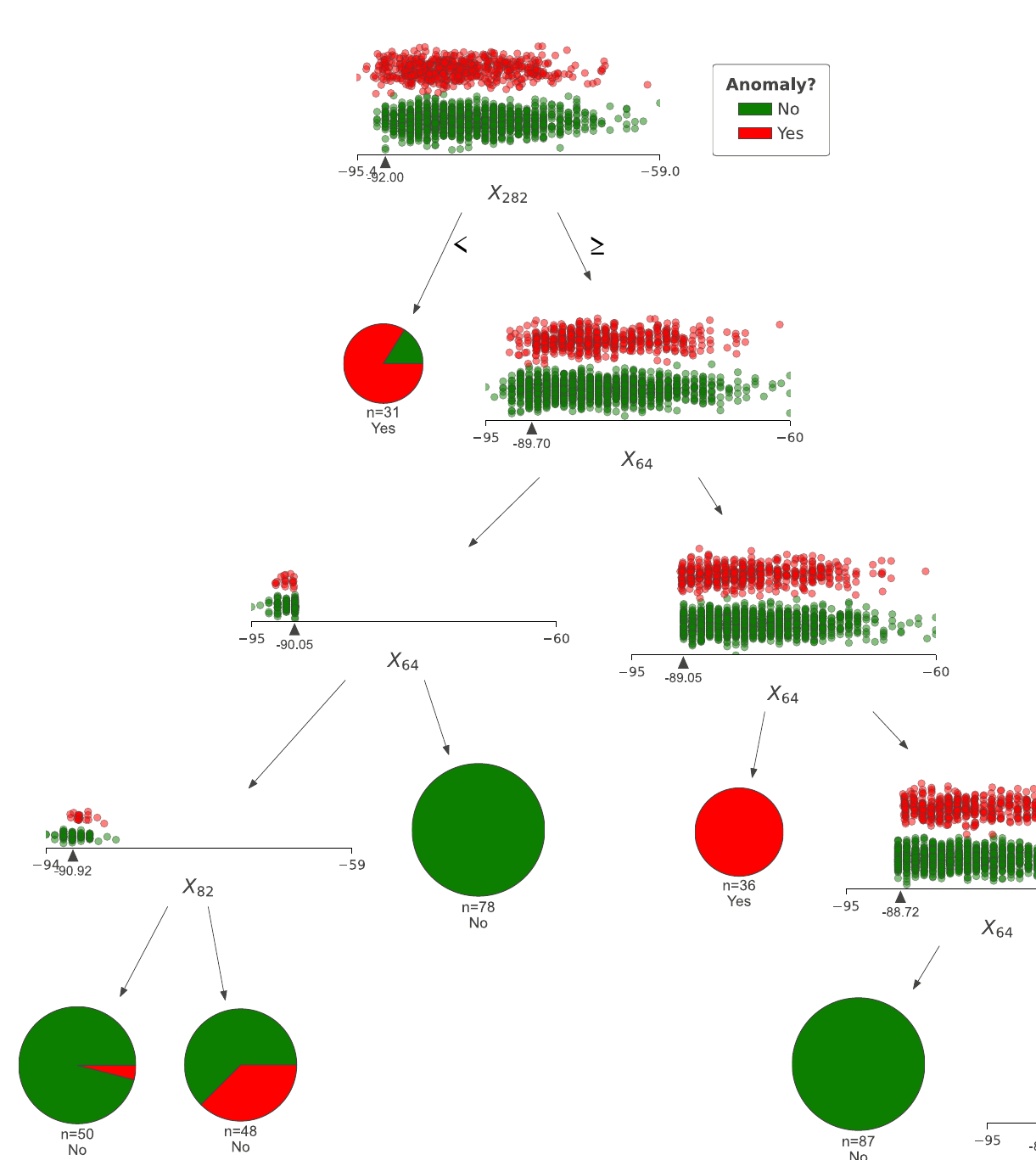}
	\caption{A part of the decision tree while detecting SlowD anomaly over time-value representation.}
	\label{fig:dtree-explain-slowd}
\end{figure*}

In contrast to SuddenD, SuddenR and InstaD, SlowD does not appear instantly, but rather gradually with random slope. The results in Table~\ref{tab:simulation-slow} show that supervised approaches are still superior to unsupervised ones. For supervised approaches, the average F1 score, ranging between $0.90$ and the perfect score, is slightly better than InstaD, but slightly worse than SuddenD and SuddenR. The most notable drop in performance is observed with LR approach over aggregated, histogram and frequency representations.

To better understand the underlying reasons behind the detection performance, we visualized in Figure~\ref{fig:dtree-explain-slowd} a typical decision tree learnt on the time-value representation of this anomaly. It can be seen from the figure that the tree commences with a comparison of feature $X_{282}$ (282nd item in time-series) to the threshold of $-92$~dBm. By doing so, it tries to distil anomalous samples at the end of the series, since samples with SlowD anomaly are suppose to have lower value towards the end of the time-series. However, as the first pie-chart reveals, this is not always the case, since some of the fully functioning non-anomalous (normal) links in the dataset have average RSSI close to that threshold, which leads to a high misclassification rate. In the second step of decision making, the process is repeated by comparing an earlier feature $X_{64}$ against $-89.70$~dBm threshold. The tree continues to learn according to this pattern until a stopping criterion is met. 

Among the unsupervised approaches, according to Table~\ref{tab:simulation-slow}, the best approach is OC-SVM with best F1 scores from 0.71 to 0.95, followed by IForest with best F1 scores from 0.63 to 0.91. LOF, as an alternative unsupervised candidate, has poorly performed over all scenarios.

\subsection{Limitations}\label{sec:limitation}
We identify three main limitations that apply to this treatise, and to the best of our understanding also to most of the other related works in wireless network and IoT anomaly data that do not target real-world application data, such as measurements.

\begin{figure*}
	\subfloat[N4 $\rightarrow$ N26\label{fig:validation:n4-n26}]{\includegraphics[width=0.33\linewidth]{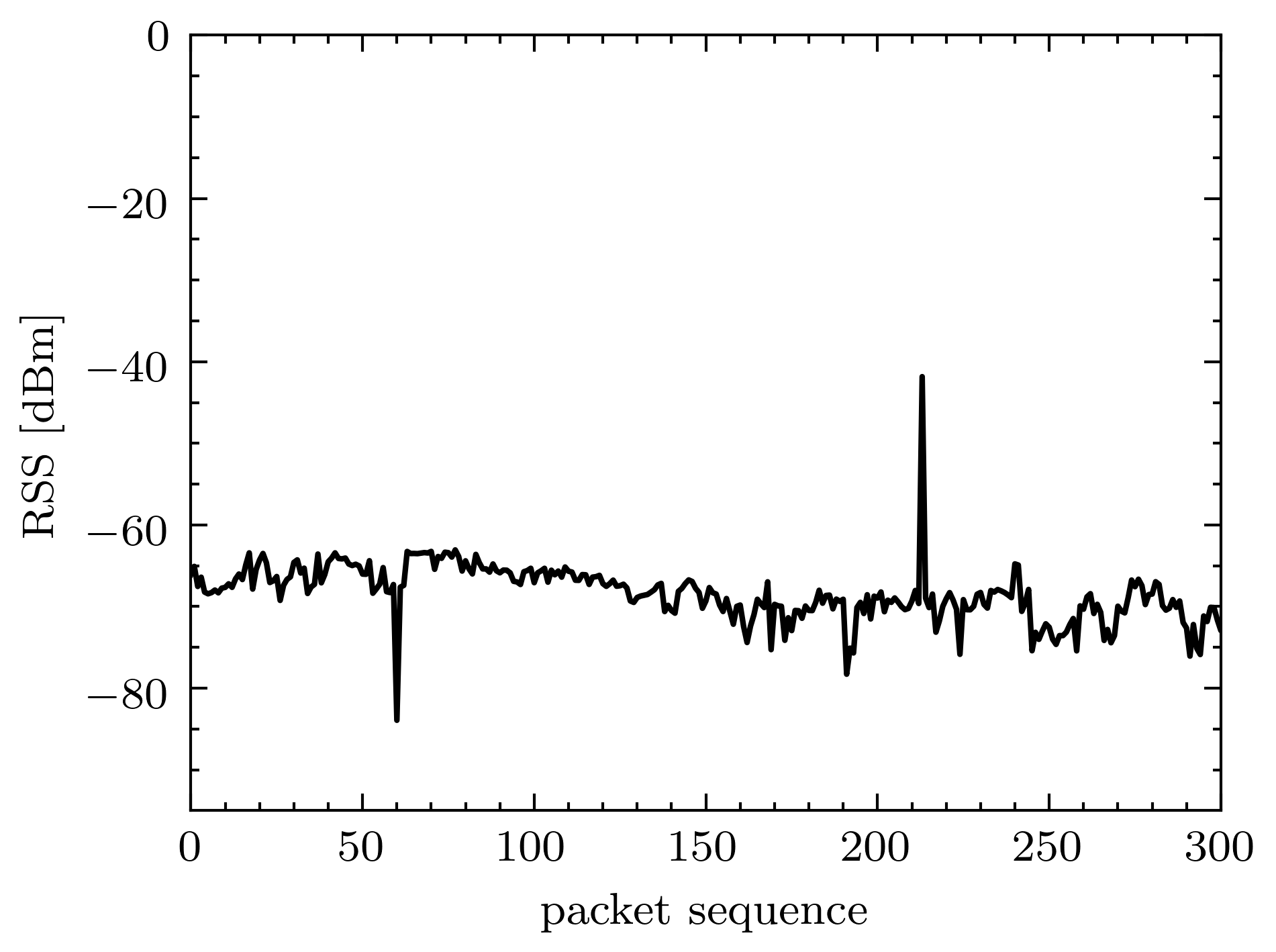}}
	\subfloat[N26 $\rightarrow$ N4\label{fig:validation:n26-n4}]{\includegraphics[width=0.33\linewidth]{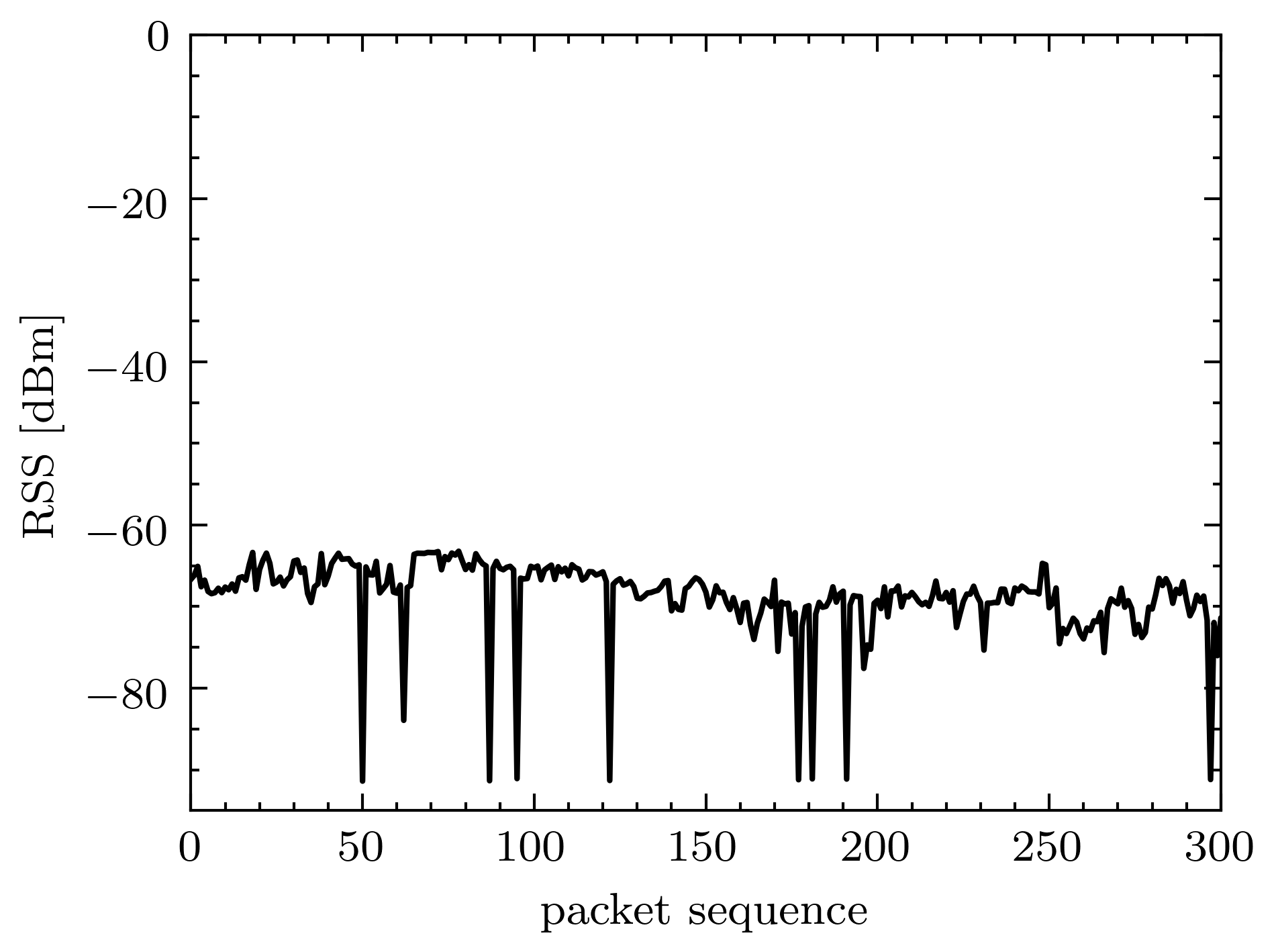}}
	\subfloat[N5 $\rightarrow$ N3 \label{fig:validation:n5-n3}]{\includegraphics[width=0.33\linewidth]{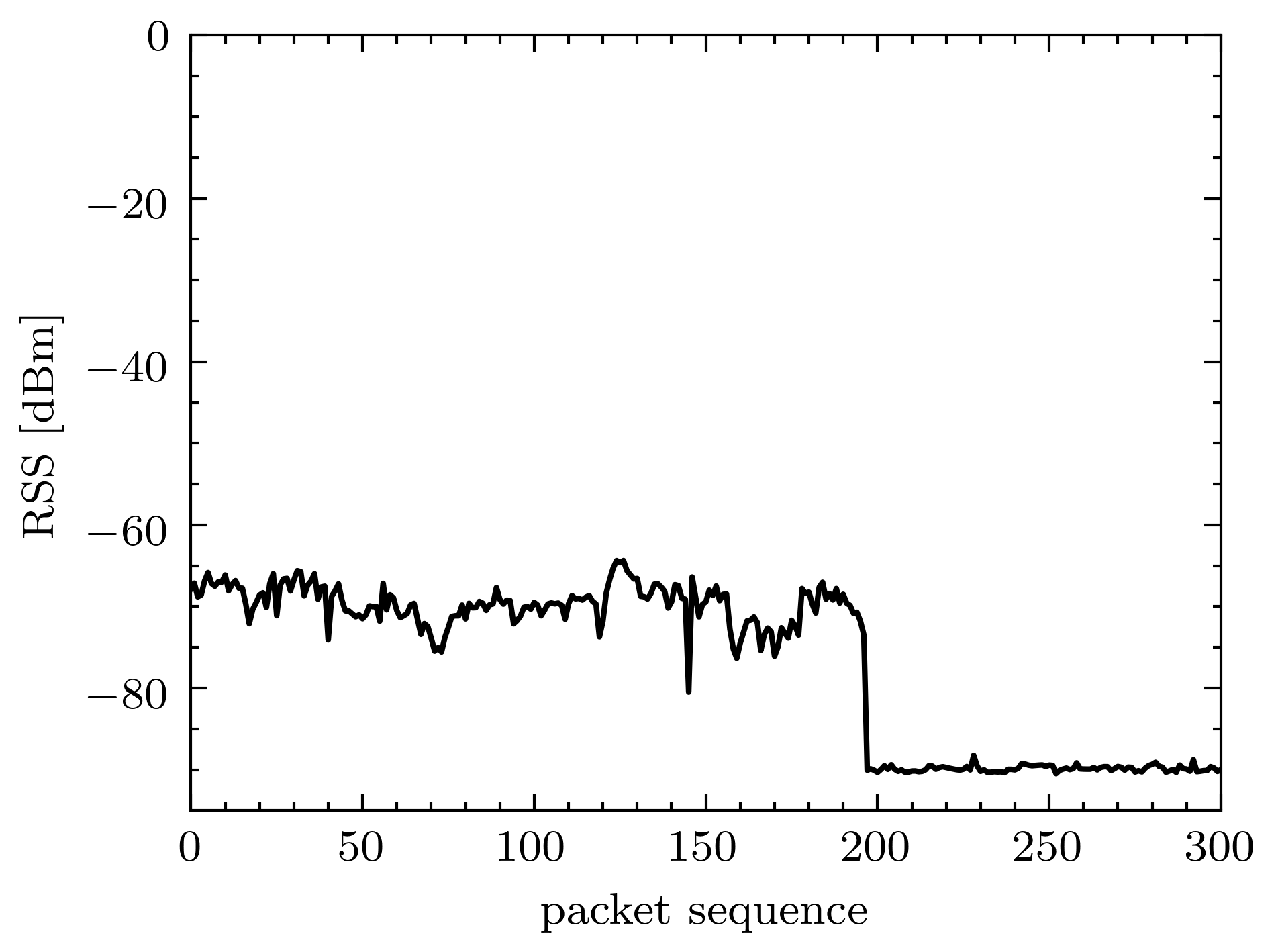}}
	
	\subfloat[N17 $\rightarrow$ N16 \label{fig:validation:n17-n16}]{\includegraphics[width=0.33\linewidth]{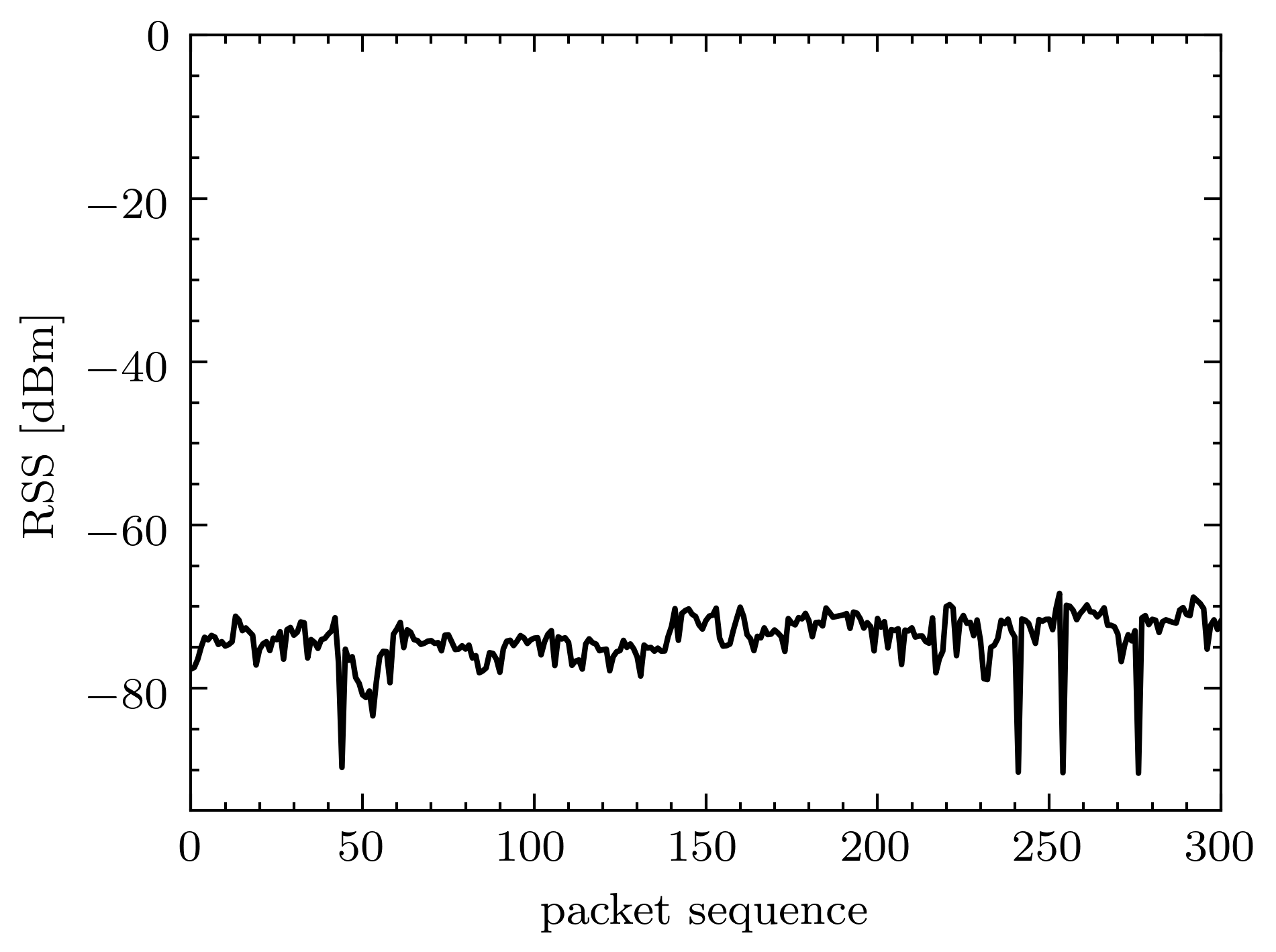}}
	\subfloat[N2 $\rightarrow$ N25 \label{fig:validation:n2-n25}]{\includegraphics[width=0.33\linewidth]{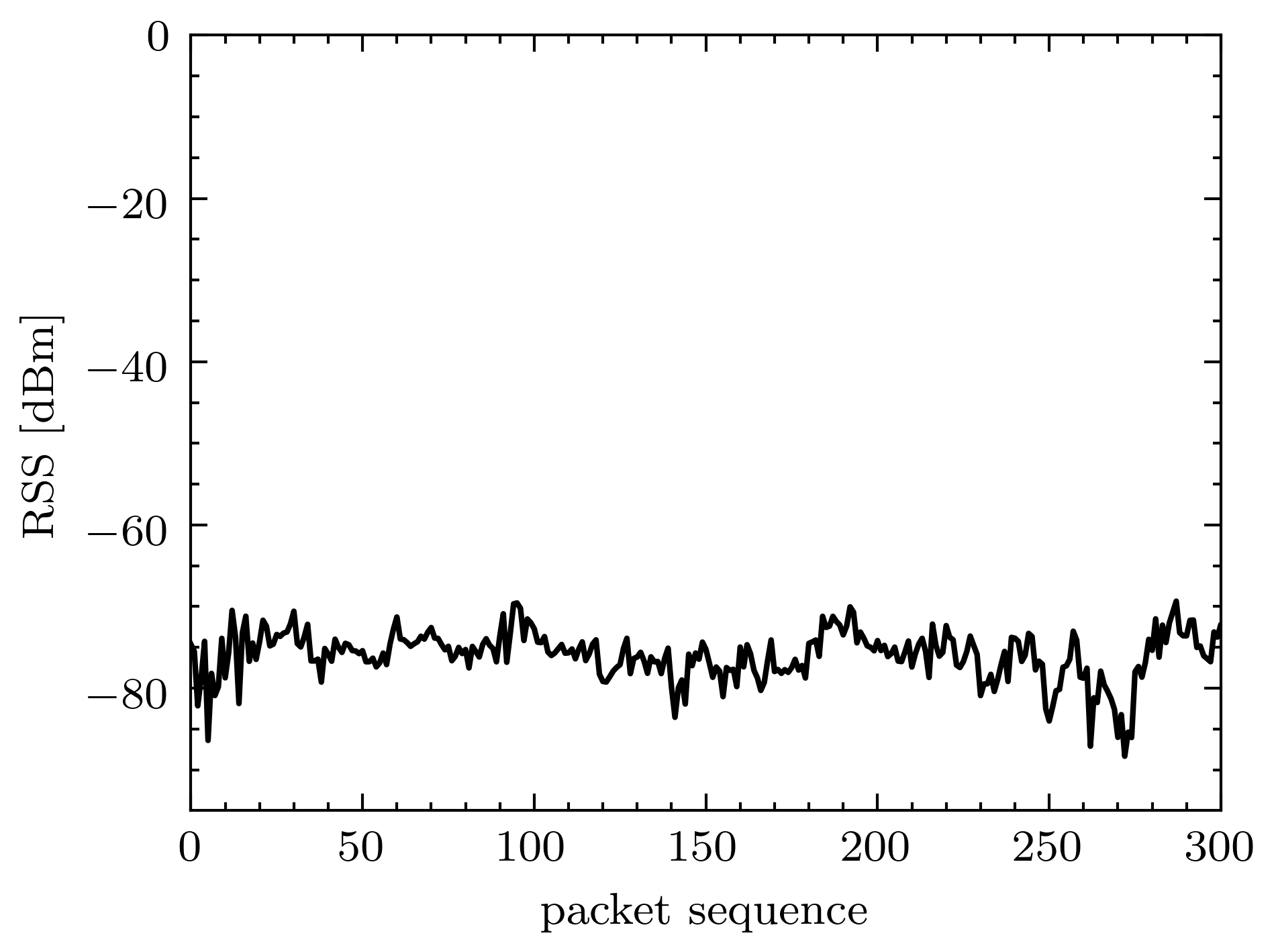}}
	\subfloat[N25 $\rightarrow$ N2\label{fig:validation:n25-n2}]{\includegraphics[width=0.33\linewidth]{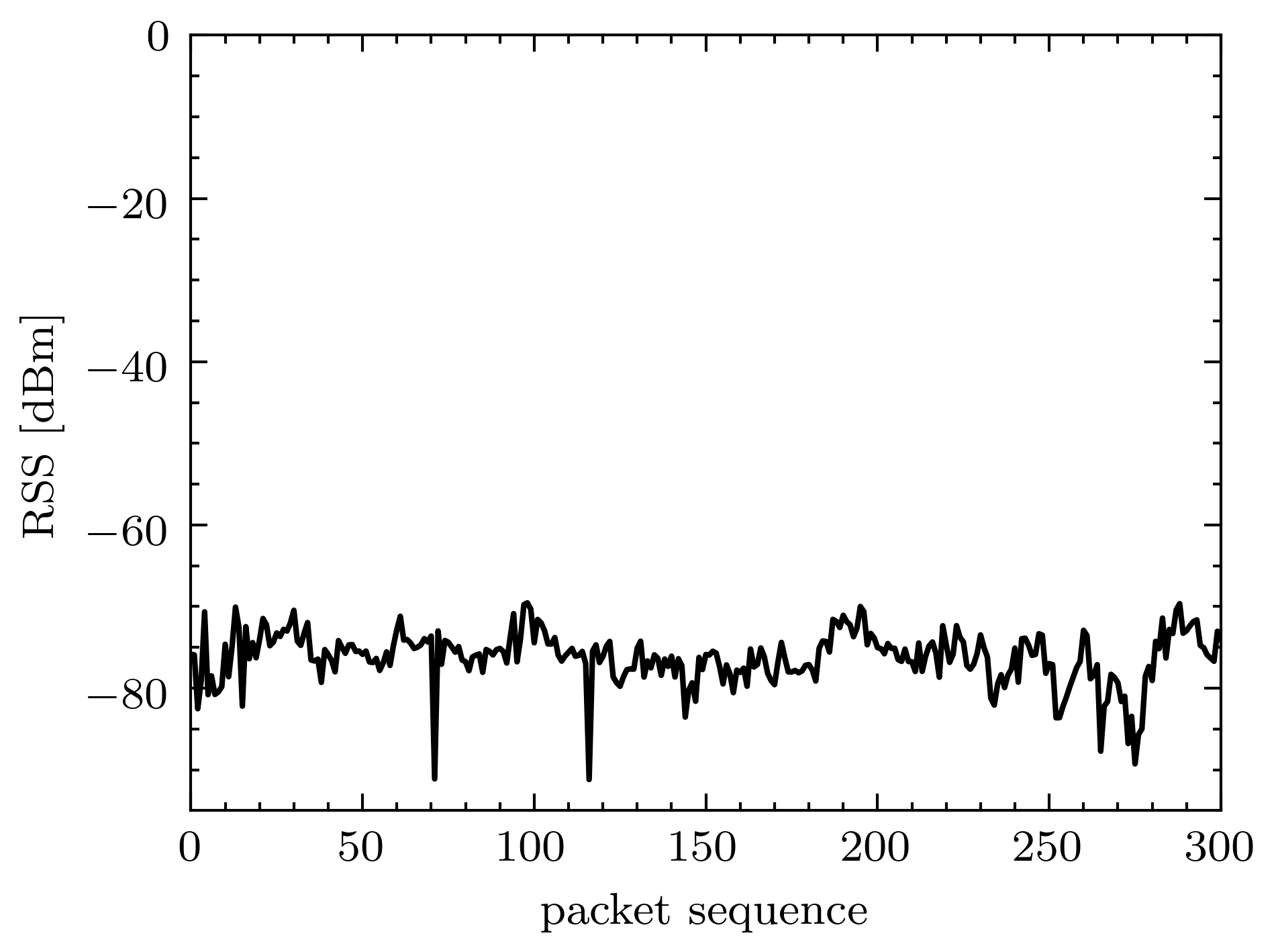}}
	
	\subfloat[N16 $\rightarrow$ N17\label{fig:validation:n16-n17}]{\includegraphics[width=0.33\linewidth]{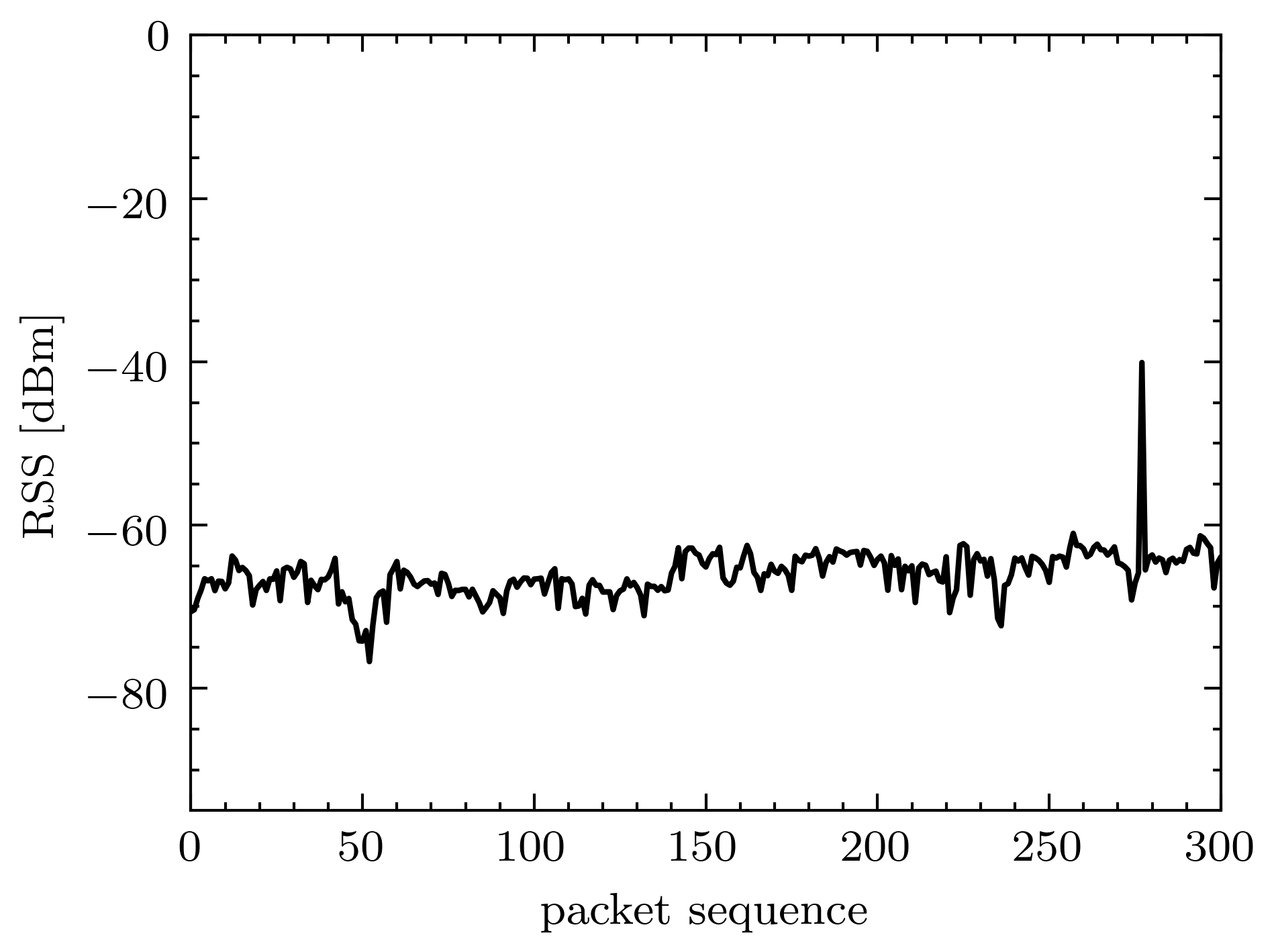}}
	\subfloat[N15 $\rightarrow$ N12\label{fig:validation:n15-n12}]{\includegraphics[width=0.33\linewidth]{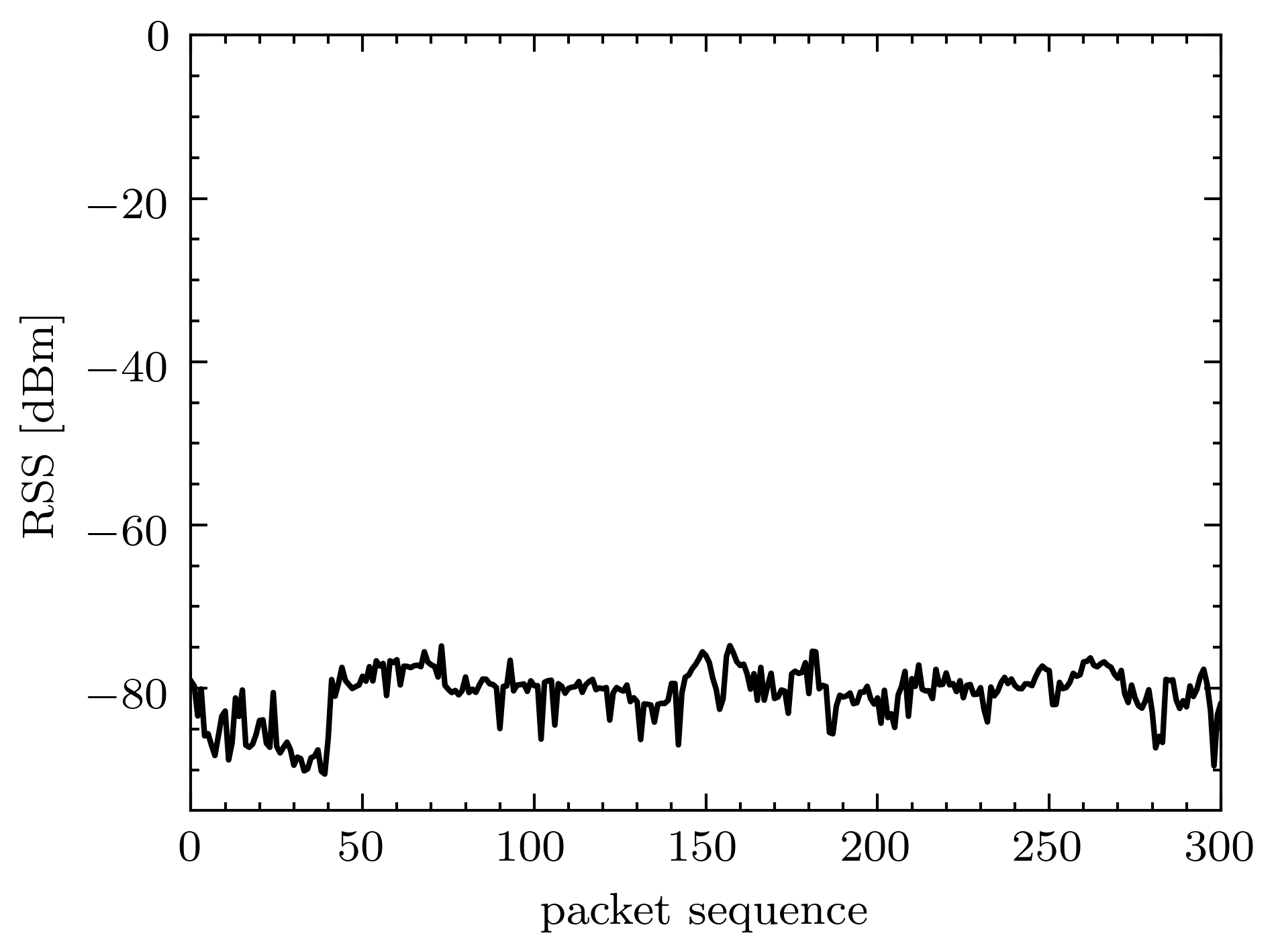}}
	\subfloat[N6 $\rightarrow$ N13 \label{fig:validation:n6-n13}]{\includegraphics[width=0.33\linewidth]{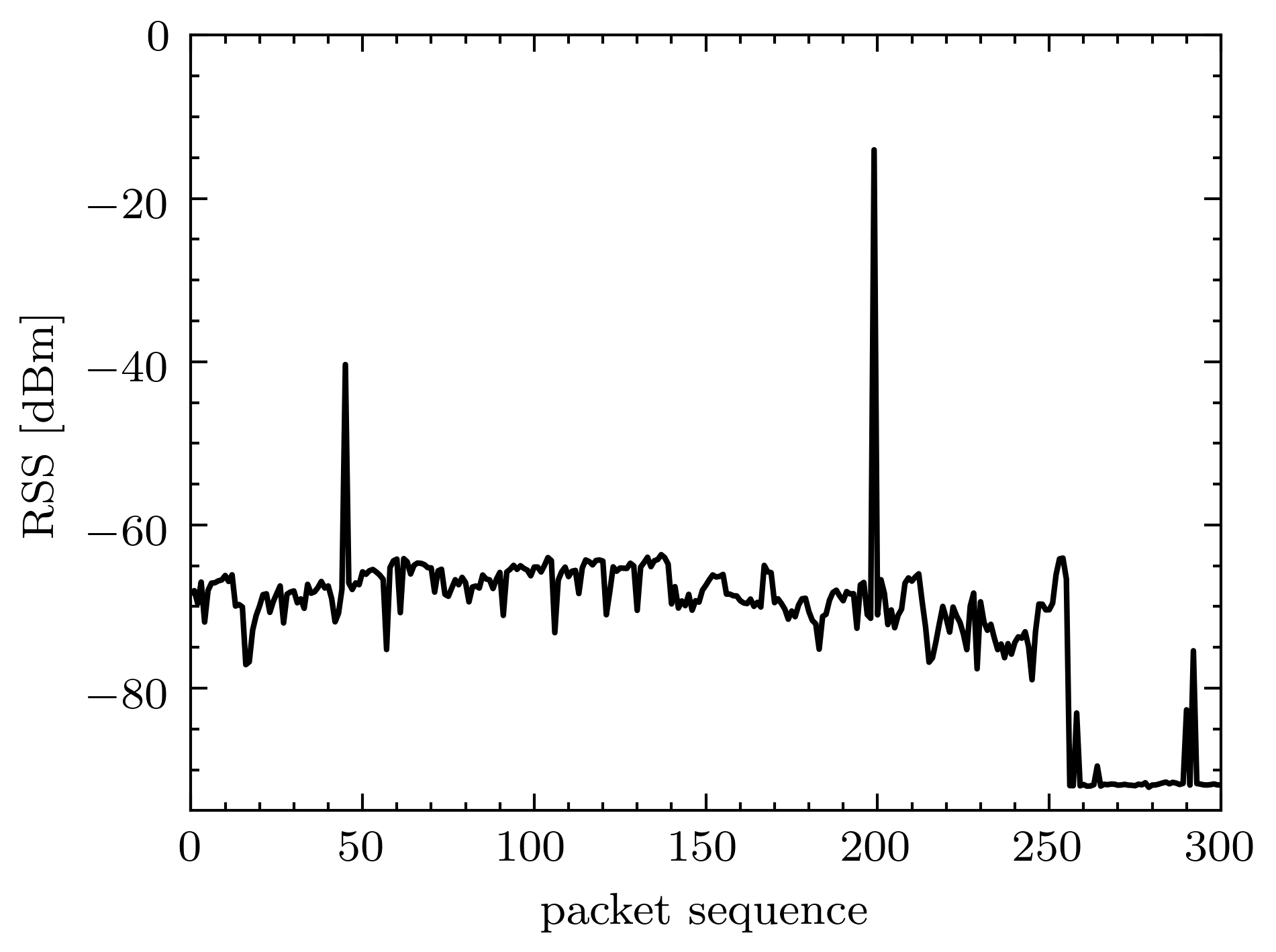}}

	\subfloat[N12 $\rightarrow$ N15\label{fig:validation:n12-n15}]{\includegraphics[width=0.33\linewidth]{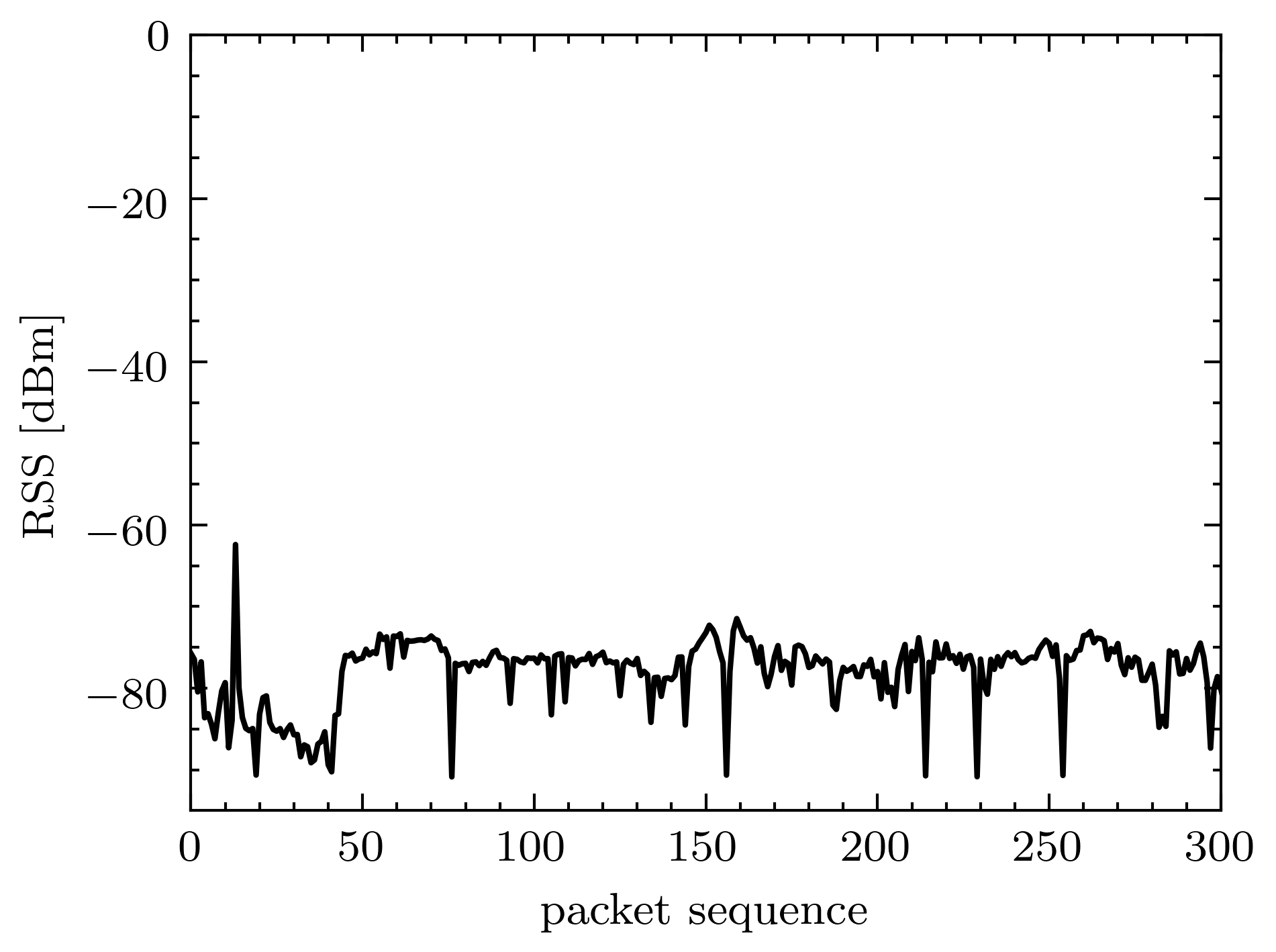}}
	
	\caption{Anomaly detection validation test employed over real-world measurements gleaned from the LOG-a-TEC testbed, where for example, as in (g) N16$\rightarrow$N17 indicates a communication link between nodes 16 and 17.}
	\label{fig:validation}
\end{figure*}

\paragraph*{Limitation 1} Every ML-based tool needs sufficient data for training and evaluation. Quantifying "sufficient" is difficult but in general it means that the model needs to see enough training examples to be able to accurately approximate the underlying distribution. Intuition would say that the data that is "sufficient" to learn a normal distribution would be smaller in size than the data needed to learn an exponential distribution. While synthetic data is useful to develop a proof of concept, for anything more than that real data is required. To the best of our knowledge, only few related works consider real-world data~\cite{yin2020anomaly} and none of them considers link layer traces. 

\begin{table}\label{tab:predAno}
	\begin{centering} \renewcommand{\arraystretch}{1.2}
		\begin{tabular}{|l|c|}
			\hline 
			Model & Predicted anomalies\tabularnewline
			\hline 
			\hline 
			LR SuddenD & Figures~\ref{fig:validation:n5-n3} and~\ref{fig:validation:n6-n13}\tabularnewline
			\hline 
			LR SuddenR & Figures~\ref{fig:validation:n5-n3} and~\ref{fig:validation:n25-n2}\tabularnewline
			\hline 
			LR InstaD & Figures~\ref{fig:validation:n5-n3} and~\ref{fig:validation:n6-n13}\tabularnewline
			\hline 
			LR SlowD & Figures~\ref{fig:validation:n26-n4},~\ref{fig:validation:n5-n3},~\ref{fig:validation:n17-n16},~\ref{fig:validation:n2-n25},~\ref{fig:validation:n25-n2} and~\ref{fig:validation:n6-n13}\tabularnewline
			\hline 
		\end{tabular}
		\par\end{centering}
	\caption{Predicted anomalies on validation data, as illustrated in Figure~\ref{fig:validation}.}
\end{table}

In this study, we developed the ML models using IEEE~802.11 traces available from a public dataset as the motivation data from LOG-a-TEC contains only 11 IEEE~802.15.4 traces all depicted in Figure~\ref{fig:validation}. Table~\ref{tab:predAno} shows how the LR model developed on IEEE~802.11 traces performs on the IEEE~802.15.4 traces. The first column of the table lists the LR model corresponding to the anomalies defined in this paper while the second includes the subfigures with links that were classified as having the respective anomalies. It can be seen from the first row of the table that the SuddenD degradations in the IEEE~802.15.4 traces are detected correctly and they appear in the links represented in Figures~\ref{fig:validation:n5-n3} and~\ref{fig:validation:n6-n13}, while for the other degradations the models seem to generate false positives. 

According to the second row of Table~\ref{tab:predAno}, it can be seen that the links represented in Figures~\ref{fig:validation:n5-n3} and~\ref{fig:validation:n25-n2} have been classified as SuddenR. However, when visually inspecting the links in those respective subfigures it can be seen that they are both classified as false positives. It is hard to determine the reason for misclassification since none of the classified traces even remotely resemble SuddenR anomaly presented in Figure~\ref{fig:synthetic:step-recovery:ts}.

As per to the third row of Table~\ref{tab:predAno}, we observe that the two links that are detected as having InstaD anomaly are false positives. As also discussed in Section~\ref{sec:perf-repr} and Figure~\ref{fig:LG_spikes_ts}, the weights change dynamically and arbitrarily for such anomalies, and thus no distinct pattern can be readily detected. 

Finally, the last row of the table shows that a large number of 802.15.4 links are falsely classified as having SlowD anomalies. While we can see that the trace in Figure~\ref{fig:validation:n26-n4} contains a slightly descending slope predicted to be SlowD anomaly, this model produces false positives over the other traces in Figure~\ref{fig:validation}. As discussed in Section~\ref{sec:perf-repr}, the discriminative importance of the features for the detection of SlowD is sought in the last part of the signal trace. This is why Figures~\ref{fig:validation:n5-n3} and~\ref{fig:validation:n6-n13}, and to some extent Figures~\ref{fig:validation:n2-n25} and~\ref{fig:validation:n25-n2} are inevitably misclassified,  since they contain lower values in the last portion of the trace.

As a conclusion, the learnt models on the relatively limited IEEE 802.11 traces are not directly and reliably transferable to the IEEE 802.15.4 traces, which indicates that the developed models cannot be readily generalized across various technologies and possibly for distinct applications.

\paragraph*{Limitation 2} The architecture of the autoencoder that learns the encoded features has been selected for a small number of candidates as a result of the trial-and-error method. Having more data would enable training an autoencoder, which then can be better generalized for even unseen examples. Autoencoder optimization and end-to-end deep learning for the proposed anomaly types might bring further insights into developing better performing and more reliable anomaly detection models. However, as hyperparameter search in deep learning is challenging and needs a large amount of training data, we leave such optimization for the future work. 

\paragraph*{Limitation 3} In this study, we only developed offline models that would need to be periodically retrained in real-world applications in order to account for the dynamically changing environments, which are the inherent characteristics of wireless networks. This leads us to online models that can learn from continuous incoming (streaming) data. Roughly speaking, offline models outperform online counterpart models in terms of the required computational power, albeit online models are able to rapidly adapt to the changes within the application environment in an automated way thus simplify the detection system that would otherwise need to periodically re-train and update the offline models.

\section{Conclusions}\label{sec:conclusions} 
In this paper, we introduce four types of anomalies that can be present in wireless links and are useful for being detected in real-world operational IoT deployments. We demonstrated that these anomalies were exposed on a real-world IoT deployment, namely the LOG-a-TEC testbed, and they significantly affected the expected operations of the testbed. Motivated by this, we develop detection models for each type of anomaly by considering five different data representations and six different ML techniques. We performed an extensive relative evaluation of the models from data representations and ML models perspective, and the limitations of our models are discussed. The resulting tool-set for anomaly injection, feature generation and model development are made publicly available for reproducibility.

Our study reveals that \textit{with respect to the data representations}; i) none of the four manually generated features clearly dominates the remaining ones in terms of anomaly detection performance, and ii) in most cases, automatically generated encoded data representations improve anomaly detection performance by up to 40\% compared to their non-encoded counterparts. 

\textit{With respect to the selected ML approach}, our results demonstrate that; i) there is no major difference among the selected supervised ML approaches, where all are capable of detecting anomalies with F1 scores of above $0.98$, and ii) the unsupervised approaches are also able to detect anomalies with F1 scores of, on average, about $0.90$ and OC-SVM outperforms all the other unsupervised ones reaching at F1 scores of $0.99$ for SuddenD, $0.95$ for SuddenR, $0.93$ for InstaD and $0.95$ for SlowD. 

\section*{Acknowledgment}
The authors would like to recognize Toma\v{z} \v{S}olc, one of the core developers of the LOG-a-TEC testbed for his contribution to the motivation of this work.

\bibliographystyle{IEEEtran}
\bibliography{anomaly}

\end{document}